%% file: XVPxDES.tex
\documentclass[useAMS,usenatbib,iop,numberedappendix]{mn2e} 

\usepackage{mathptmx}
\usepackage{amsmath,amsfonts,amssymb}
\usepackage{makeidx}
\usepackage{graphicx}
\usepackage{epstopdf}
\usepackage[colorlinks=True]{hyperref}
\usepackage{multirow}
\usepackage{rotating}
\usepackage{color}
\usepackage{tablefootnote}
\usepackage{lscape}
\usepackage{subcaption}
\usepackage{tablefootnote}
\usepackage[T1]{fontenc}
\usepackage{aecompl}
\usepackage[normalem]{ulem}
\usepackage{siunitx}
\usepackage{eso-pic}

\AddToShipoutPictureBG*{%
  \AtPageUpperLeft{%
    \hspace{0.75\paperwidth}%
    \raisebox{-3.5\baselineskip}{%
      \makebox[0pt][l]{\textnormal{DES 2017-0259}}
}}}%

\AddToShipoutPictureBG*{%
  \AtPageUpperLeft{%
    \hspace{0.75\paperwidth}%
    \raisebox{-4.5\baselineskip}{%
      \makebox[0pt][l]{\textnormal{FERMILAB-PUB-17-442-AE}}
}}}%

\definecolor{inon}{rgb}{1.00,0.27,0.00}

\input{authorlist}

\newcommand{\LCDM}{\ensuremath{\Lambda\textrm{CDM}}}
\newcommand{\OmegaM}{\ensuremath{\Omega_{\mathrm{M}}}}

\newcommand{\Hnow}{\ensuremath{H_{0}}}

\newcommand{\Msun}{\ensuremath{\mathrm{M}_{\odot}}}

\newcommand{\Rfiveoo}{\ensuremath{R_{500}}}
\newcommand{\Mfiveoo}{\ensuremath{M_{500}}}
\newcommand{\Mfiveoosub}{\ensuremath{M_{\mathrm{sub}}}}

\newcommand{\redshift}{\ensuremath{z}}
\newcommand{\dif}{\ensuremath{\mathrm{d}}}
\newcommand{\angstrom}{\textup{\AA}}

\newcommand{\CHANDRA}{\emph{Chandra}}

\newcommand{\Spitzer}{\emph{Spitzer}}
\newcommand{\WISE}{\emph{WISE}}

\newcommand{\PLANCK}{\emph{Planck}}

\newcommand{\XMMBCS}{XMM-BCS}

\newcommand{\zp}{\ensuremath{z_{\mathrm{p}}}}
\newcommand{\zd}{\ensuremath{z_{\mathrm{d}}}}
\newcommand{\zs}{\ensuremath{z_{\mathrm{s}}}}

\newcommand{\Mstar}{\ensuremath{M_{\star}}}
\newcommand{\MBCG}{\ensuremath{M_{\star,\mathrm{BCG}}}}

\newcommand{\Lx}{\ensuremath{L_{\mathrm{X}}}}
\newcommand{\Tx}{\ensuremath{T_{\mathrm{X}}}}
\newcommand{\Mgas}{\ensuremath{M_{\mathrm{ICM}}}}
\newcommand{\Mbary}{\ensuremath{M_{\mathrm{b}}}}
\newcommand{\Mhalo}{\ensuremath{M_{\mathrm{h}}}}
\newcommand{\Yx}{\ensuremath{Y_{\mathrm{X}}}}
\newcommand{\mstarchar}{\ensuremath{m_{\star, \mathrm{char}}}}
\newcommand{\mstar}{\ensuremath{m_{\star}}}
\newcommand{\IRACone}{\ensuremath{[3.6]}}
\newcommand{\IRACtwo}{\ensuremath{[4.5]}}
\newcommand{\Wone}{\ensuremath{W1}}
\newcommand{\Wtwo}{\ensuremath{W2}}

\newcommand{\MPIV}{\ensuremath{M_{\mathrm{piv}}}}
\newcommand{\ZPIV}{\ensuremath{z_{\mathrm{piv}}}}

\newcommand{\Agas}{\ensuremath{A_{\mathrm{ICM}}}}
\newcommand{\Bgas}{\ensuremath{B_{\mathrm{ICM}}}}
\newcommand{\Cgas}{\ensuremath{C_{\mathrm{ICM}}}}

\newcommand{\Abary}{\ensuremath{A_{\mathrm{b}}}}
\newcommand{\Bbary}{\ensuremath{B_{\mathrm{b}}}}
\newcommand{\Cbary}{\ensuremath{C_{\mathrm{b}}}}

\newcommand{\Acold}{\ensuremath{A_{\mathrm{c}}}}
\newcommand{\Bcold}{\ensuremath{B_{\mathrm{c}}}}
\newcommand{\Ccold}{\ensuremath{C_{\mathrm{c}}}}

\newcommand{\Astar}{\ensuremath{A_{\star}}}
\newcommand{\Bstar}{\ensuremath{B_{\star}}}
\newcommand{\Cstar}{\ensuremath{C_{\star}}}

\newcommand{\Asz}{\ensuremath{A_{\mathrm{SZ}}}}
\newcommand{\Bsz}{\ensuremath{B_{\mathrm{SZ}}}}
\newcommand{\Csz}{\ensuremath{C_{\mathrm{SZ}}}}
\newcommand{\Dsz}{\ensuremath{D_{\mathrm{SZ}}}}
\newcommand{\Dszcom}{\ensuremath{\sigma_{\ln \zeta|M_{500}}}}

\newcommand{\fstar}{\ensuremath{f_{\star}}}
\newcommand{\fgas}{\ensuremath{f_{\mathrm{ICM}}}}
\newcommand{\fcold}{\ensuremath{f_{\mathrm{c}}}}
\newcommand{\fbary}{\ensuremath{f_{\mathrm{b}}}}

\newcommand{\percent}{\ensuremath{\%}}

\newcommand{\numAstarez}{\ensuremath{4.00\pm0.13}}
\newcommand{\numBstarez}{\ensuremath{0.80\pm0.12}}
\newcommand{\numCstarez}{\ensuremath{0.03\pm0.26}}
\newcommand{\numDstarez}{\ensuremath{0.22\pm0.03}}

\newcommand{\numAstaronez}{\ensuremath{4.00\pm0.13}}
\newcommand{\numBstaronez}{\ensuremath{0.80\pm0.12}}
\newcommand{\numCstaronez}{\ensuremath{0.05\pm0.25}}
\newcommand{\numDstaronez}{\ensuremath{0.22\pm0.03}}
\newcommand{\numAstaronezsys}{\ensuremath{4.00\pm0.28}}
\newcommand{\numBstaronezsys}{\ensuremath{0.80\pm0.12}}
\newcommand{\numCstaronezsys}{\ensuremath{0.05\pm0.27}}
\newcommand{\numDstaronezsys}{\ensuremath{0.22\pm0.03}}

\newcommand{\numAgasez}{\ensuremath{5.70\pm0.11}}
\newcommand{\numBgasez}{\ensuremath{1.32\pm0.07}}
\newcommand{\numCgasez}{\ensuremath{-0.17\pm0.15}}
\newcommand{\numDgasez}{\ensuremath{0.12\pm0.02}}

\newcommand{\numAgasonez}{\ensuremath{5.69\pm0.11}}
\newcommand{\numBgasonez}{\ensuremath{1.33\pm0.07}}
\newcommand{\numCgasonez}{\ensuremath{-0.15\pm0.14}}
\newcommand{\numDgasonez}{\ensuremath{0.11\pm0.02}}
\newcommand{\numAgasonezsys}{\ensuremath{5.69\pm0.62}}
\newcommand{\numBgasonezsys}{\ensuremath{1.33\pm0.09}}
\newcommand{\numCgasonezsys}{\ensuremath{-0.15\pm0.22}}
\newcommand{\numDgasonezsys}{\ensuremath{0.11\pm0.04}}

\newcommand{\numAbaryez}{\ensuremath{6.16\pm0.12}}
\newcommand{\numBbaryez}{\ensuremath{1.28\pm0.08}}
\newcommand{\numCbaryez}{\ensuremath{-0.17\pm0.16}}
\newcommand{\numDbaryez}{\ensuremath{0.12\pm0.02}}
\newcommand{\numAbaryonez}{\ensuremath{6.17\pm0.12}}
\newcommand{\numBbaryonez}{\ensuremath{1.29\pm0.07}}
\newcommand{\numCbaryonez}{\ensuremath{-0.16\pm0.15}}
\newcommand{\numDbaryonez}{\ensuremath{0.12\pm0.02}}
\newcommand{\numAbaryonezsys}{\ensuremath{6.17\pm0.62}}
\newcommand{\numBbaryonezsys}{\ensuremath{1.29\pm0.09}}
\newcommand{\numCbaryonezsys}{\ensuremath{-0.16\pm0.23}}
\newcommand{\numDbaryonezsys}{\ensuremath{0.12\pm0.04}}

\newcommand{\numAcoldez}{\ensuremath{6.79\pm0.22}}
\newcommand{\numBcoldez}{\ensuremath{-0.53\pm0.12}}
\newcommand{\numCcoldez}{\ensuremath{0.05\pm0.27}}
\newcommand{\numDcoldez}{\ensuremath{0.24\pm0.03}}

\newcommand{\numAcoldonez}{\ensuremath{6.78\pm0.22}}
\newcommand{\numBcoldonez}{\ensuremath{-0.51\pm0.12}}
\newcommand{\numCcoldonez}{\ensuremath{0.08\pm0.24}}
\newcommand{\numDcoldonez}{\ensuremath{0.24\pm0.03}}
\newcommand{\numAcoldonezsys}{\ensuremath{6.78\pm0.36}}
\newcommand{\numBcoldonezsys}{\ensuremath{-0.51\pm0.12}}
\newcommand{\numCcoldonezsys}{\ensuremath{0.08\pm0.26}}
\newcommand{\numDcoldonezsys}{\ensuremath{0.24\pm0.03}}

\newcommand{\appropto}{\mathrel{\vcenter{
  \offinterlineskip\halign{\hfil$##$\cr
    \propto\cr\noalign{\kern2pt}\sim\cr\noalign{\kern-2pt}}}}}

\title[Baryon Content of Galaxy Clusters]{
Baryon Content in a Sample of 91
Galaxy Clusters Selected by the South Pole Telescope at $0.2 < \redshift < 1.25$
}

\begin{document}
\pdfpageheight 11.7in
\pdfpagewidth 8.3in

%
%

\maketitle 

%
%

\begin{abstract}

We estimate total mass (\Mfiveoo), intracluster medium (ICM) mass (\Mgas) and stellar mass (\Mstar) in a Sunyaev-Zel'dovich effect (SZE) selected sample of 91 galaxy clusters with masses $\Mfiveoo\gtrsim2.5\times10^{14}\Msun$ and redshift $0.2 < \redshift < 1.25$ from the 2500~$\deg^2$ South Pole Telescope SPT-SZ survey.  The total masses \Mfiveoo\ are estimated from the SZE observable, the ICM masses \Mgas\ are obtained from the analysis of \CHANDRA\ X-ray observations, and the stellar masses \Mstar\ are derived by fitting spectral energy distribution templates to Dark Energy Survey (DES) $griz$ optical photometry and \WISE\ or \Spitzer\ near-infrared photometry.  We study trends in the stellar mass, the ICM mass, the total baryonic mass and the cold baryonic fraction with cluster halo mass and redshift.  We find significant departures from self-similarity in the mass scaling for all quantities, while the redshift trends are all statistically consistent with zero, indicating that the baryon content of clusters at fixed mass has changed remarkably little over the past $\approx9$~Gyr. We compare our results to the mean baryon fraction (and the stellar mass fraction) in the field, finding that these values lie above (below) those in cluster virial regions in all but the most massive clusters at low redshift.  Using a simple model of the matter assembly of clusters from infalling groups with lower masses and from infalling material from the low density environment or field surrounding the parent halos, we show that the 
measured mass trends without strong redshift trends
in the stellar mass scaling relation could be explained by
a mass and redshift dependent fractional contribution from field material.  Similar analyses of the ICM and baryon mass scaling relations provide evidence for the so-called ``missing baryons'' outside cluster virial regions.

\end{abstract}

%
%


\clearpage

%
%

\section{Introduction}
\label{sec:introduction}

Galaxy clusters originate from the peaks of primordial fluctuations of the density field in the early Universe, and their growth contains a wealth of information about structure formation.  Of particular interest are scaling relations---the relation between cluster halo mass and other physical properties of the cluster---because these relations enable a link between the cluster observables and the underlying true halo mass.  This link then enables the use of galaxy cluster samples for the measurement of cosmological parameters and studies of the cosmic acceleration and of structure formation \citep{haiman01,holder01b,carlstrom02}.
In addition, energy feedback from star formation, Active Galactic Nuclei or other sources during cluster formation can leave an imprint on these scaling relations, affecting their mass or redshift dependence and providing an observational handle to inform studies of cluster astrophysics.

Over the last few decades, the scaling relations of galaxy clusters have been intensely studied using X-ray observables
\citep{mohr97,mohr99,arnaud99,reiprich02,ohara06,arnaud07,sun09,vikhlinin09a,pratt09,mantz16b}, populations of cluster galaxies \citep{lin03b,lin04a,rozo09,saro13,mulroy14}, or a combination of them \citep{zhang10,lin12,rozo14c}; in addition, scaling relations have been studied using hydrodynamical simulations \citep[e.g.,][]{evrard97,bryan98,nagai07,stanek10,truong16,barnes17,pillepich17}.

Observations indicate that the ensemble properties of the baryonic components of galaxy clusters correlate well with the halo mass.
For example, the detailed way in which the mass of intracluster medium (ICM) systematically trends with total cluster mass, and the scatter about that mean behavior, shed light on the thermodynamic history of massive cosmic halos \citep[e.g.,][]{ponman99,mohr99,pratt10,young11}.
To date, the bulk of observational results have been obtained using cluster samples selected at low redshift ($\redshift\le0.6$).  Studying scaling relations at high redshift remains difficult due to the lack of sizable cluster samples and/or the adequately deep datasets to extract physical properties from the clusters.

Enabled by the Sunyaev-Zel'dovich Effect \citep[SZE;][]{sunyaev70b,sunyaev72}---a signature on the Cosmic Microwave Background (CMB) that is caused by the inverse Compton scattering between the CMB photons and hot ICM---teams of scientists have developed novel instrumentation and have used it to search for galaxy clusters out to a redshift $z\approx1.8$.   These large SZE surveys, carried out with the South Pole Telescope \citep[SPT;][]{carlstrom11}, the Atacama Cosmology Telescope \citep{fowler07}, and the \PLANCK\ mission \citep{planck06}, have delivered large cluster samples and enabled studies of meaningful ensembles of clusters to high redshift \citep{high10,menanteau10b,andersson11,planck11_stat_sr,semler12,sifon13,mcdonald13,mcdonald16b}. 
Further breakthroughs in the area of wide-and-deep optical and NIR surveys, such as the Blanco Cosmology Survey \citep[][]{desai12}, the \Spitzer\ South Pole Telescope Deep Field \citep{ashby13a}, the Hyper Suprime-Cam survey \citep[][]{miyazaki12,aihara17}, and the Dark Energy Survey \citep[DES;][]{DES05,des2016morethande} have helped in delivering the needed optical data to study the galaxy populations of these SZE selected samples.

Recent studies of the baryon content of galaxy clusters or groups show strong mass trends of the observable to halo mass scaling relations but no significant redshift trends out to $\redshift\approx1.3$ \citep{chiu16a, chiu16c}.  That is, the stellar and ICM mass fractions vary rapidly with cluster mass but have similar values at fixed mass, regardless of cosmic time.  The combination of strong mass trends and weak redshift trends in the context of hierarchical structure formation implies that halos must accrete a significant amount of material that lies outside the dense virial regions of halos and that has values of the stellar mass fraction or IGM fraction that are closer to the cosmic mean.  A mixture of infall from 
lower mass halos and material outside the 
dense virial regions
would then allow for the stellar and ICM mass fractions to vary 
weakly over cosmic time.
However, it is important to note that current constraints on the redshift trends of scaling relations suffer from significant systematics introduced by comparing heterogeneous cluster samples analyzed in different ways \citep{chiu16a}. 
To overcome these systematic uncertainties, one needs to use a large sample with a well-understood selection function and---most importantly---employ an unbiased method on homogeneous datasets to determine the masses consistently across the mass and redshift range of interest.

In this study, we aim to analyze the baryon content of massive galaxy clusters selected by their SZE signatures in the 2500~$\deg^2$ South Pole Telescope SZE (SPT-SZ) survey.  We have focused on a sample of 91 galaxy clusters that have X-ray observations from \CHANDRA, optical imaging from DES, and near-infrared (NIR) data from \Spitzer\ and \WISE.  This sample is selected to lie above a detection significance $\xi>6.8$ and spans a broad redshift range $0.25\lesssim\redshift\lesssim1.25$. 
This sample is currently the largest, approximately mass-limited sample of galaxy clusters extending to high redshift with the required uniform, multi-wavelength datasets needed to carry out this analysis.  Moreover, we adopt self-consistent methodologies to estimate the ICM, stellar and total masses of each galaxy cluster in our sample; this dramatically minimizes the potential systematics that could bias the observed mass and redshift trends in the scaling relations.

This paper is organized as follows. 
The cluster sample and data are described in Section~\ref{sec:xvp_sample_and_data}, while the determinations of cluster mass \Mfiveoo, ICM mass \Mgas\ and the stellar mass \Mstar\ are given in Section~\ref{sec:various_mass_estimation}.  
We describe our fitting procedure and the method for estimating both statistical and systematic uncertainties on the scaling relation parameters in Section~\ref{sec:method}, and we
present the results of power law fits to the observed scaling relations in Section~\ref{sec:scalingrelation}.
We then discuss our results and quantify the potential systematics in Section~\ref{sec:discussion}. 
Conclusions are given in Section~\ref{sec:xvp_conclusion}.
Throughout this paper, we adopt the flat \LCDM\ cosmology with the fiducial cosmological parameters $(\OmegaM, \Hnow) = (0.304, 68~\mathrm{km}~\mathrm{s}^{-1}~\mathrm{Mpc}^{-1})$, which constitute the most recent cosmological constraints from the SPT collaboration \citep{dehaan16}.    Unless otherwise stated, the uncertainties indicate the $1\sigma$ confidence regions, the cluster halo mass \Mfiveoo\ is estimated at the overdensity of 500 with respect to the critical density $\rho_{\mathrm{crit}}$ at the cluster redshift \zd, the cluster radius \Rfiveoo\ is calculated using \Mfiveoo\footnote{$\Rfiveoo = \left( \Mfiveoo / \left( \frac{4\pi}{3} \rho_{\mathrm{crit}}(\zd) 500\right) \right)^{1 / 3}$}, and the photometry is in the AB magnitude system.

%
%

\section{Cluster Sample and Data}
\label{sec:xvp_sample_and_data}
\begin{figure}
\vskip-0.25in
\centering
\resizebox{!}{0.5\textwidth}{
\includegraphics[scale=1.0]{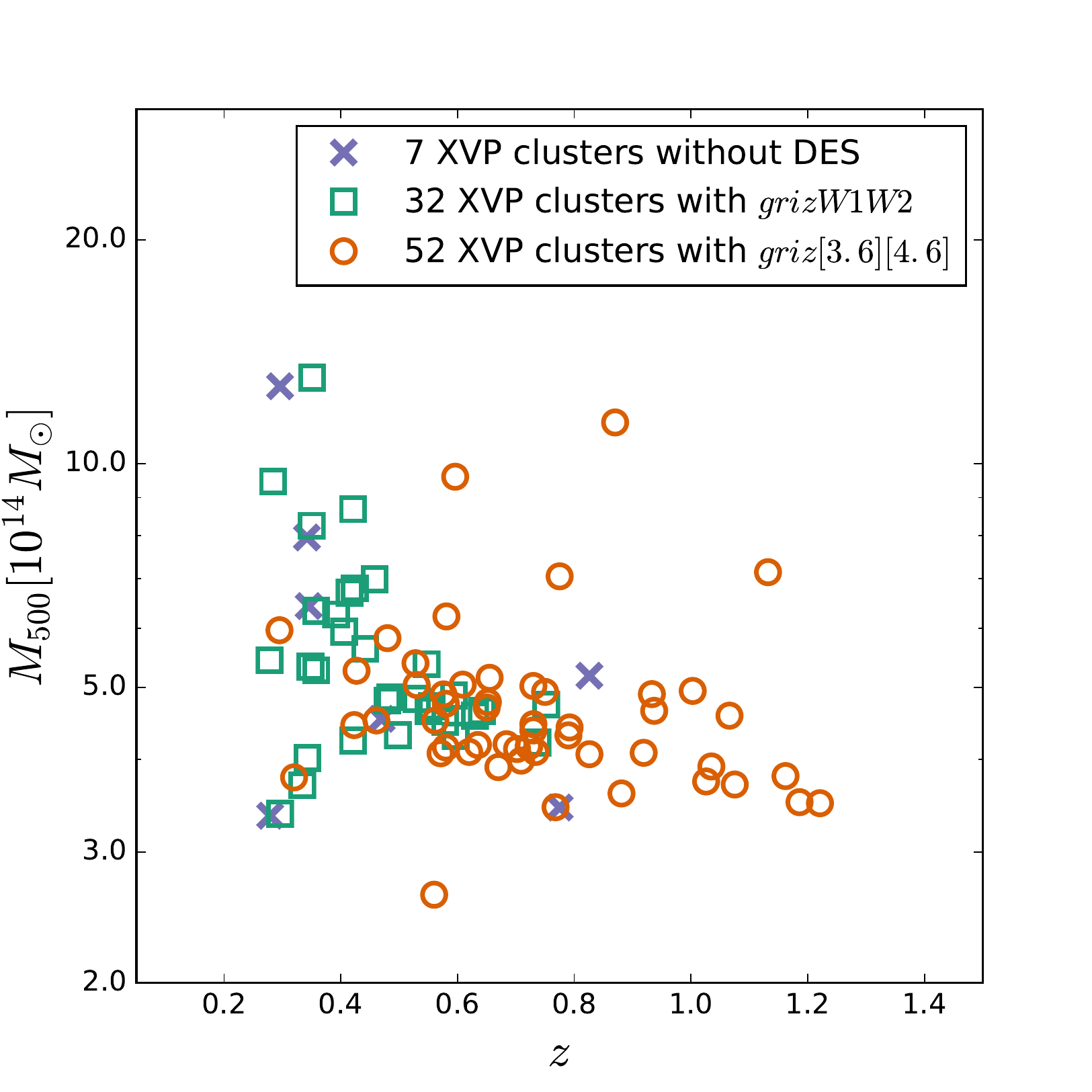}
}
\vskip-0.15in
\caption{
The SZE selected sample of galaxy clusters from the SPT-SZ survey that we use in this work is plotted in mass and redshift.
The subset of 32 clusters with combined DES $griz$ and \WISE\ $\Wone\Wtwo$ photometry is shown with green squares, and the subset of 52 clusters with combined DES $griz$ and \Spitzer\ $\IRACone\IRACtwo$ photometry is shown with red circles.
The sample of 7 clusters currently without DES $griz$ photometry is marked with blue crosses.  These systems are excluded from the stellar mass analysis.
}
\label{fig:sample}
\end{figure}

\subsection{Cluster sample}
\label{sec:xvp_sample}

The cluster sample used in this work is selected from the SPT-SZ 2500 $\deg^2$ survey \citep{bleem15}.
A subset of 80 SPT-selected clusters at $\redshift>0.4$ with SZE detection significance $\xi>6.8$ has been followed up by the \CHANDRA\ X-ray Observatory through an X-ray Visionary Project (hereafter XVP, PI Benson).
We extend this 80 cluster sample by including other SPT selected clusters at redshift $\redshift>0.2$ that have also been observed by \CHANDRA\ through previous proposals from SPT, the Atacama Cosmology Telescope \citep{marriage11b} collaboration or the \PLANCK\ \citep{planck06} consortium. 
The final sample consists of 91 galaxy clusters at redshifts $0.25<\redshift<1.25$, all SPT-SZ systems with an associated SZE significance $\xi$ that allows us to estimate the cluster masses with an uncertainty of $\approx20$~percent \citep{bocquet15}.

The redshifts of a subset of 61 clusters in our sample have been determined spectroscopically \citep{ruel14,bayliss16}. 
For the rest, we adopt photometric redshifts that are estimated by using the Composite Stellar Population (hereafter CSP) of the Bruzual and Charlot \cite[BC03;][]{bruzual03} model with formation redshift  ${\redshift}_{\mathrm{f}}=3$ and an exponentially-decaying star formation rate with the $e$-folding timescale $\tau = 0.4$~Gyr.
This CSP model is built by running \texttt{EzGal} \citep{mancone12b} and is calibrated using the red sequence of the Coma cluster and by using six different metallicities associated with different luminosities \citep[see more details in][]{song12a}.
The resulting CSP model has been demonstrated to provide accurate and precise measurements of the photometric redshifts of galaxy clusters with an 
accuracy of $\Delta\redshift / (1 + \redshift)\lesssim0.025$ through comparison with spectroscopic redshifts \citep{song12a,song12b,liu15b,bleem15}.  
This CSP model is also used to estimate the cluster characteristic magnitude \mstarchar\ that will be used to define the magnitude cut of each cluster in a consistent manner across the wide redshift range (see Section~\ref{sec:selection_of_cluster_galaxies}).
The characteristic magnitude \mstarchar\ is a parameter in the Schechter luminosity function that marks the transition magnitude between the exponential cutoff and the power-law components of the galaxy luminosity function.
In the following work we neglect these uncertainties in photo-z.  Fig.~\ref{fig:sample} contains a plot of the mass and redshift distribution of the sample, and the basic properties of each cluster are listed in Table~\ref{tab:sample}.

\begin{figure}
\vskip-0.20in
\centering
\resizebox{!}{0.5\textwidth}{
\includegraphics[scale=1.0]{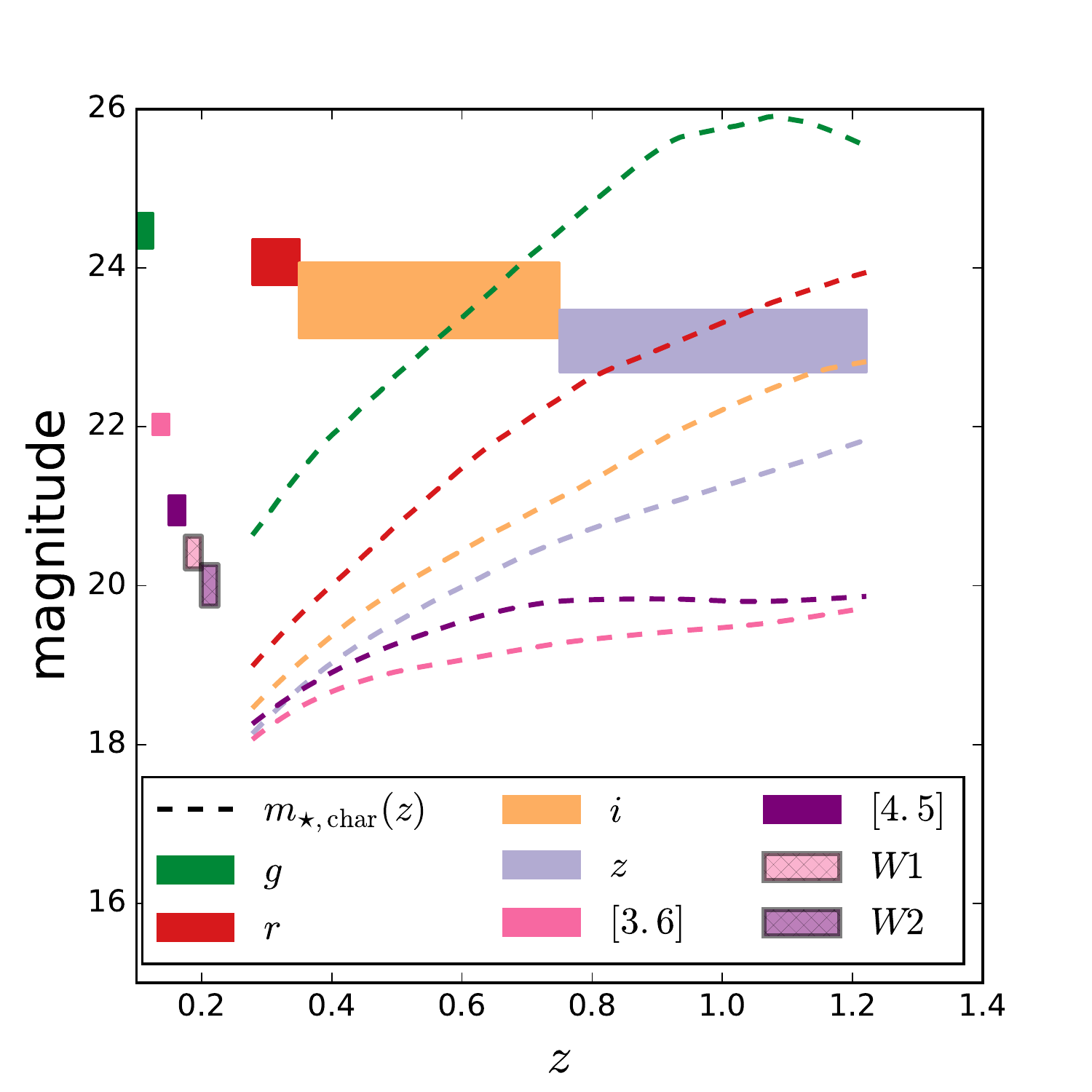}
}
\vskip-0.15in
\caption{
A comparison of photometric depths (bars of constant magnitude) and the redshift variation of the characteristic magnitude \mstarchar\ (dashed lines) over the relevant redshift range.  
The median and the $1\sigma$~variation of the 50\percent\ completeness for the DES $griz$ bands are shown with green, red, orange and purple bars, respectively. 
The $10\sigma$ depth of \IRACone\ and \Wone\ (\IRACtwo\ and \Wtwo) are shown in pink (dark purple), where the \WISE\ passbands are marked with black boundaries.
The redshift ranges of the $riz$ bands stand for the redshift ranges over which we use those bands to apply the magnitude cut (see Section~\ref{sec:selection_of_cluster_galaxies}), while the limits in the other bands (i.e., $g\IRACone\IRACtwo\Wone\Wtwo$) all appear on the left.
}
\label{fig:desdepth}
\end{figure}
\begin{table}
\centering
\caption{
For the full sample, we list the cluster name, the redshift, the Right Ascension ($\alpha_{\mathrm{X}}$) and Declination ($\delta_{\mathrm{X}}$) inferred from the X-ray imaging, the Right Ascension ($\alpha_{\mathrm{BCG}}$) and Declination ($\delta_{\mathrm{BCG}}$) of the BCG, and the optical/NIR datasets used in the SED fitting (see Section~\ref{sec:xvp_optical_nir_data}).
}
\label{tab:sample}
\resizebox{!}{0.45\textheight}{
\begin{tabular}{ccccccc}
\\ \hline\hline
Name     &\redshift\    &$\alpha_{\mathrm{X}}$    &$\delta_{\mathrm{X}}$    &$\alpha_{\mathrm{BCG}}$    &$\delta_{\mathrm{BCG}}$    & Optical + NIR datasets \\ \hline\hline
\input{xvp.master}
\hline\hline
\end{tabular}
}
\end{table}

\subsection{X-ray data}
\label{sec:xvp_xray_data}

All the clusters in our sample have been observed with the \CHANDRA\ X-ray Observatory.
The X-ray data were largely motivated by the need to determine an X-ray mass proxy to support the cosmological analysis \citep{andersson11,benson13, bocquet15,dehaan16}; observing times were tuned to obtain $\approx2000$ source photons per cluster.
With these X-ray data, we are able to estimate the total luminosity \Lx, temperature \Tx\ and ICM masses \Mgas\ as well as the mass proxy $\Yx\equiv\Tx\Mgas$ for each cluster.  These cluster parameters have been used in several previous works \citep{benson13, mcdonald13,mcdonald14b,bocquet15, dehaan16,chiu16a}.
In this work, we use only the \Mgas\ measurement (see Section~\ref{sec:icmmass}), while the total cluster masses are estimated from the SPT observables (see Section~\ref{sec:totalmass}).  Following \cite{chiu16a}, we adopt the X-ray center as the cluster center for our analysis (see Table~\ref{tab:sample}).
More details of the X-ray data acquisition, reduction and analysis are described elsewhere \citep{andersson11,benson13,mcdonald13}.

\subsection{Optical and NIR data}
\label{sec:xvp_optical_nir_data}

To estimate the stellar mass of each cluster galaxy in our sample, we use optical photometry in the $griz$ bands observed by the Dark Energy Survey \citep{DES05,des2016morethande} together with near-infrared (NIR) photometry obtained with either the Wide-field Infrared Survey Explorer \citep[\WISE;][]{wright10} or the Infrared Array Camera (IRAC, \citealt{fazio04}) of the \Spitzer\ telescope. 
The combination of DES, \WISE, and \Spitzer\ datasets enables us to estimate the stellar masses of cluster galaxies by constraining their Spectral Energy Distributions (SED; see Section~\ref{sec:stellarmass}).

The \Spitzer\ observations originate from an SPT follow-up program (see Section~\ref{sec:spitzer_data_sets}) that was designed to aid in the cluster confirmation at high redshift ($\redshift \gtrsim 0.4$).  These data are much deeper than the \WISE\ observations, which we have acquired from the public archive (see Section~\ref{sec:wise_data_sets}).  The field of view of our \Spitzer\ imaging is small and can only sufficiently cover the angular area out to \Rfiveoo\ for clusters at $\redshift\gtrsim0.4$, while we specifically re-process the \WISE\ data such that each resulting coadd image is centered on the cluster with coverage of $\approx40\arcmin\times40\arcmin$ (see Section~\ref{sec:wise_data_sets}).
In addition, the \Spitzer\ imaging has better angular resolution, making it more appropriate for galaxy studies in crowded environments---especially cluster cores---at higher redshift.  Thus, when we combine the optical data from DES with the additional NIR datasets, we choose to use \Spitzer\ observations whenever available.
These shallower \WISE\ observations are used only for the lower redshift clusters.

There are 84 out of the 91 clusters covered by the footprints of the Science Verification, Year One and Year Two of the DES datasets, and the remaining 7 clusters are or will be imaged by the continuing efforts from the DES. 
Therefore, we do not have the stellar mass measurements for the 7 clusters in our sample.
In Fig.~\ref{fig:sample} the sample is shown, color coded according to whether \WISE\ or \Spitzer\ imaging was used. We describe the details of each dataset in Sections~\ref{sec:des_data_sets} to~\ref{sec:wise_data_sets}, and then we present the procedure for combining these datasets in Section~\ref{sec:deblending}.

\subsubsection{Optical dataset}
\label{sec:des_data_sets}

For the optical data, the Science Verification, Year One and Year Two of the DES datasets \citep{diehl16} are used to obtain the $griz$ photometry. 
For each cluster, we build Point Spread Function (PSF)-homogenized coadd images for the $griz$ bands with the field of view of $\approx1\deg^{2}$ centered on the cluster; this avoids the edge effects that are typically seen in wide field surveys.  The optical imaging is processed by the CosmoDM pipeline \citep{mohr12}, and the full description of data reduction, source extraction and photometric calibration is given elsewhere \citep{desai12,liu15b,hennig17}.  These DES catalogs and images have been specifically processed for studying the SPT clusters, and the excellent photometric quality has been presented elsewhere \citep{hennig17,klein17}.  We increase the flux uncertainties by a factor of 2 based on tests of photometric repeatability on faint sources, which crudely accounts for contributions to the photometric noise from sources that are not tracked in the image weight maps.  These include, for example, cataloging noise and uncertainties in photometric calibration.  Through these efforts the catalogs of $griz$ photometry are available for 84 of the 91 clusters.  

Following the procedures of previous studies \citep{zenteno11, chiu16b, hennig17}, we estimate the completeness of the photometric catalogs by comparing the observed number counts to those estimated from the deeper COSMOS field \citep{capak2007, ilbert2008, laigle16}, where the source catalogs are complete down to $\gtrsim25.5$~mag in the $griz$ bands. Specifically, we first estimate the logarithmic slope of the source count--magnitude relation of the COSMOS field over a range of magnitudes, assuming that it follows a power law.  We then compare the histogram of the source counts---which are observed in the cluster field and are away from the cluster center at projected separations $>3\Rfiveoo$---to the derived power law model with the slope fixed to the best-fit value of the COSMOS and the normalization that is fitted to the source counts observed between $19.5$~mag and $21$~mag in the cluster field.  Finally, we fit an error function to the ratio of DES galaxy counts to those predicted by the power law model from COSMOS to obtain the completeness function.

The procedure above is carried out for each cluster, and the resulting 50~percent completeness depths are shown in Fig.~\ref{fig:desdepth}, where the median and root mean square variation of the completeness are plotted with horizontal bars. For clarity, we only show the results of the $riz$ bands where we will perform the magnitude cut on our galaxy samples in the following analysis (see Section~\ref{sec:selection_of_cluster_galaxies}); the median $50$\percent\ completeness of the $g$ band is $24.51\pm0.02$~mag. 
In Fig.~\ref{fig:desdepth}, we also plot the characteristic magnitude $\mstarchar\left(\redshift\right)$ as a function of redshift for the $griz$ bands as predicted by the CSP model (see Section~\ref{sec:xvp_sample}). Overall, the $50$\percent\ completeness of the $griz$ bands is deeper than $\mstarchar\left(\redshift\right)$ by $>2$~mag ($\approx1.5$~mag for $\redshift\gtrsim1.1$).  This suggests that the depth of DES optical data is sufficiently deep to detect the cluster galaxies that are dominating the stellar content of our sample, allowing us to estimate the total stellar mass of each cluster. The incompleteness corrections applied in the following analysis (see Section~\ref{sec:statistical_background_subtraction}) are based on these derived completeness functions.

\subsubsection{Spitzer dataset}
\label{sec:spitzer_data_sets}

The \Spitzer\ observations are obtained in IRAC channels $3.6$~\micron\ and $4.5$~\micron\ with the Program IDs (PI Brodwin) 60099, 70053 and 80012, resulting in photometry of \IRACone\ and \IRACtwo, respectively.  The data acquisition, processing and photometric calibration of the \Spitzer\ observations are fully described in \cite{ashby13a}, to which we defer the reader for more details. 
By design, the depths of IRAC observations are sufficient to image the cluster galaxies, which are brighter than \mstarchar(\redshift) + 1~mag in \IRACone\ and \IRACtwo, as predicted by the CSP model (see Section~\ref{sec:xvp_sample}) out to redshift $\redshift\approx1.5$ with more than $90$\percent\ completeness.  The $10\sigma$ depth of \IRACone\ and \IRACtwo\ are $22.00$ and $20.92$~mag with $1\sigma$ variation of $0.13$ and $0.18$~mag, respectively, as shown in Fig.~\ref{fig:desdepth}. 
The field of view of the \Spitzer\ mosaics is $\approx7\arcmin\times7\arcmin$ (and is $\approx5\arcmin\times5\arcmin$ for the full depth), which as already mentioned is sufficient to cover the \Rfiveoo\ region of the clusters in our sample at redshift $\redshift\gtrsim0.4$.  Among the 84 clusters imaged by DES in this work, there are 52 clusters that are also imaged by the \Spitzer\ telescope.  For these systems we use the photometry of $griz\IRACone\IRACtwo$ (see Section~\ref{sec:deblending} for SED fitting.

\subsubsection{WISE dataset}
\label{sec:wise_data_sets}

For each cluster, we acquire the NIR imaging observed by the \textit{Wide-field Infrared Survey Explorer} (\WISE) through the \WISE\ all-sky survey and the project of Near-Earth Object Wide-field Infrared Survey Explorer \citep[NEOWISE;][]{mainzer14}.
The NEOWISE project is part of the primary \WISE\ survey, which started in 2009 with the goal of imaging the full sky in four bands ($3.4$, $4.6$, $12$ and $22$~$\micron$, denoted by \Wone, \Wtwo, $W3$ and $W4$, respectively).  The main mission was followed by the NEOWISE Reactivation Mission---beginning in 2013---with the goal of continuing to find and characterize near-Earth objects using passbands \Wone\ and \Wtwo.  The data taken by the primary \WISE\ survey, the NEOWISE project and its Reactivation Mission  have been regularly released since 2011, providing a valuable legacy dataset to the community.

In this work, we collect the imaging of the 91 galaxy clusters in our sample in the passbands of \Wone\ and \Wtwo\ from the ALLWISE data release combined with the release from the NEOWISE Reactivation Mission through 2016, which enables us to detect fainter sources than the official catalogs from the ALLWISE data release.  After acquiring the single-exposure images centered on each cluster,  we coadd them with the area weighting by the Image Coaddition with Optional Resolution Enhancement \citep[ICORE; ][]{masci09} in the WISE/NEOWISE Coadder \footnote{http://irsa.ipac.caltech.edu/applications/ICORE/docs/instructions.html}. The photometric zeropoint is calibrated on the basis of each single-exposure image using the measurements of a network of calibration standard stars near the ecliptic poles \citep[see more details in][]{wright10,jarrett11}, and we properly calculate the final zero point of the coadd image in the reduction process.
The $10\sigma$ depth of \Wone\ and \Wtwo\ are $20.39$ and $19.98$~mag with $1\sigma$ variation of $0.20$ and $0.26$~mag, respectively, as shown in Fig.~\ref{fig:desdepth}. 
We create the coadd images with footprints of $40\arcmin\times40\arcmin$ centered on each cluster, allowing us to completely cover a region extending to $\gtrsim4\Rfiveoo$ in our cluster sample. 

\subsubsection{Combining DES and NIR datasets}
\label{sec:deblending}

Our goal is to construct a photometric catalog (either $griz\Wone\Wtwo$ or $griz\IRACone\IRACtwo$) for each cluster. 
However, one of the greatest challenges in combining these multi-wavelength datasets is that the source blending varies from band to band due to the variation in the PSF size.  Many methods have been proposed to solve or alleviate the blending problem \citep[e.g., ][]{laidler07,desantis2007,mancone13,joseph2016,laigle16}, and good performance has been demonstrated when one adopts priors based on the images with better resolution. 

In this work, we use the \texttt{T-PHOT} package \citep{merlin15}---a state-of-the-art PSF-matching technique---to deblend the NIR fluxes on the NIR images observed by either the \WISE\ or \Spitzer\ telescope based on priors from the DES optical imaging.
The coadd images are used in the following procedures.
Here we list the steps. 
\begin{enumerate}
\item We first prepare the NIR images in the native pixel scale ($0.263\arcsec$) of the 
DECam using \texttt{SWarp} \citep{bertin02}.
\item We then run \texttt{SExtractor} \citep{bertin96} on these swarped images to produce the background-subtracted images. The background-subtracted DES images will be used to construct the ``real 2-d profiles'' \citep[see details in][]{merlin15} as the inputs of \texttt{T-PHOT}.
\item The PSFs of the NIR and optical images are then derived by stacking a few tens of stars that are selected in the DES catalog. Specifically, these stars are selected in $i$-band to have (1) $\|\mathtt{spread\_model}\| \leq 0.002$, (2) $17~\mathrm{mag} < \mathtt{mag\_auto} < 22~\mathrm{mag}$, (3) $\mathtt{FLAGS} = 0$ in \texttt{SExtractor} flags indicating no blending, and then these stars are subjected to 3$\sigma$ iterative clipping in \texttt{FLUX\_RADIUS} to exclude any outliers. 
\item Once the PSFs of NIR and optical images are obtained, the kernel connecting them is derived by a method similar to that used in \cite{galametz13}, where a low passband filter is applied to suppress the high-frequency noise in the Fourier space. 
\item We run \texttt{T-PHOT} on each source in the optical catalog using the priors of ``real 2-d profiles'', convolving them with the derived kernel, and obtaining the best-fit fluxes of the corresponding sources observed on the NIR images. The $i$-band images are good as optical priors because they are deep and have good seeing.
\item The second round of \texttt{T-PHOT} is run with locally registered kernels to account for small offsets in astrometry and/or position-dependent variation of the kernel. The final deblended NIR fluxes are obtained.
\end{enumerate}

Following this process for each cluster with available \Spitzer\ or \WISE\ data, we extract PSF-matched NIR photometry (\IRACone\IRACtwo\ or \Wone\Wtwo) for each source in the DES $griz$ catalogs.  As previously stated, we use the $griz\IRACone\IRACtwo$ catalogs whenever available.  An implication is that all sources in our photometric catalogs are constructed based on optical detection.
Although the wavelength coverage of $\IRACone\IRACtwo$ and $\Wone\Wtwo$ is similar, we observe a small systematic offset in stellar masses extracted using the two sets of photometry: $griz\Wone\Wtwo$ and $griz\IRACone\IRACtwo$. 
We quantify this systematic in Section~\ref{sec:sedfitting}, apply a correction to the stellar masses in Section~\ref{sec:smf}.\

\begin{table}
\centering
\caption{Cluster parameters include, from left to right, the cluster name, the cluster radius, the halo mass, the ICM mass, the stellar mass, the total baryon mass and the cold baryon fraction.}
\label{tab:measurements}
\resizebox{!}{0.43\textheight}{
\begin{tabular}{lccccccc}
\\ \hline\hline
    &\Rfiveoo    &$\Mfiveoo$    &$\Mgas$    &$\Mstar$    &$\Mbary$    &  \\ 
Name     &[arcmin]    &$[10^{14}\Msun]$    &$[10^{13}\Msun]$    &$[10^{12}\Msun]$    &$[10^{13}\Msun]$    &$\fcold$  \\ \hline\hline
\input{xvp.measurement}
\hline\hline
\end{tabular}
}
\end{table}
%

%
%

\section{Cluster mass estimation}
\label{sec:various_mass_estimation}

In the subsections below we describe in turn how the total cluster masses, ICM masses and stellar masses are measured.
The estimates of the ICM masses and stellar masses assume spherical symmetry for the galaxy clusters.

\subsection{Halo masses}
\label{sec:totalmass}
We use the latest SZE scaling relation from the SPT collaboration (i.e., Table~3 in \citealt{dehaan16}) to estimate the halo mass or total mass \Mfiveoo\ for each cluster.
The best fit scaling relation parameters were determined using the number counts of the SPT galaxy clusters  together with external information from 
big-bang nucleosynthesis (BBN) calculations \citep{cooke14}, direct measurement of the Hubble parameter \citep{riess11},
and $Y_X$ measurements for a subset of the clusters. 
The resulting \Mfiveoo\ of each cluster is listed in the Table~\ref{tab:measurements}.

The details of the mass determination are given in \cite{bocquet15}, to which we refer the readers for more details.
We briefly describe the method as follows.  For each cluster with the SZE signal to noise $\xi$ observed in the SPT-SZ survey, we estimate the cluster total mass \Mfiveoo\ using the $\zeta$-mass relation,
\begin{equation}
\label{eq:zeta2mass}
\zeta = \Asz 
\left( \frac{\Mfiveoo}{3\times10^{14}h^{-1}\Msun}  \right)^{\Bsz}
\left( \frac{E(\redshift)}{E(\redshift = 0.6)} \right)^{\Csz} \, 
\end{equation}
with log-normal intrinsic scatter $\Dsz\equiv\Dszcom$, where we connect the observable $\xi$---a biased estimator of the cluster SZE signature---to the unbiased SZE observable $\zeta$ by its ensemble behavior that can be described by a normal distribution with unit width,
\begin{equation}
\label{eq:xi2zeta}
\left\langle \xi \right\rangle = \sqrt{\zeta^2 + 3} \, .
\end{equation}
The biases in $\xi$ arise because there are three degrees of freedom adopted in the cluster detection:  the cluster coordinates on the sky and the scale or core radius of the matched filter employed in the detection.  For the mass estimates used here we adopt for these scaling relation parameters 
\begin{eqnarray}
\label{eq:params_sz}
r_{\mathrm{SZ}} &\equiv &\left( \Asz, \Bsz, \Csz, \Dsz \right) \, \nonumber \\
 &= &\left( 4.84, 1.66, 0.55, 0.20 \right) \,
\end{eqnarray}
along with their associated uncertainties \citep{dehaan16}.  When we calculate the SZE-derived masses \Mfiveoo, we account for the Malmquist bias, which arises from the intrinsic scatter and measurement uncertainty coupling with the selection function in the SPT-SZ survey, and the Eddington bias, which comes from the intrinsic scatter and measurement uncertainties together with the steeply falling behavior of the mass function on the high-mass end.  

Our mass measurements are inferred from the SZE observable $\xi$ using the parameters of the SZE observable to mass relation that have been calibrated self-consistently using information from (1) the X-ray mass proxy $Y_{\mathrm{X}}$ that is externally calibrated through weak lensing information, and (2) the observed distribution of the SPT-SZ cluster sample in $\xi$ and \redshift, which is connected to the underlying mass function through the observable to mass relation.  The mass calibration information coming from the mass function itself is substantial---especially when external cosmological constraints are adopted as priors (see Section~\ref{sec:sys_mass} for more discussions of systematics in mass estimation).  
With these external constraints within a flat \LCDM\ context, the cosmological parameters are already so well constrained that there is very little freedom in the underlying halo mass function.  Thus, the observable to mass relation that describes the mapping to the observed cluster distribution in $\xi$ and \redshift\ is tightly constrained.

Note that the mass calibration obtained in \cite{dehaan16} and used in our analysis is statistically consistent with the direct mass calibration through weak lensing using 32 SPT-SZ clusters \citep{dietrich17} and dynamical analyses of 110 SPT-SZ clusters \citep{capasso17}.   The equivalent mass offset measurement relative to \citet{dehaan16} is ($-9\pm 21)$~\percent\ in the former analyses and $(+12\pm12)$~\percent\ in the latter analysis.  These analyses involve clusters over the full redshift range of interest to our analysis $0.2 \lesssim \redshift \lesssim 1.2$.

We stress that in this analysis we account for both statistical and systematic halo mass uncertainties.  We consider the intrinsic scatter in $\zeta$ at fixed mass together with the measurement noise in $\xi$ as a representation of the underlying $\zeta$ as statistical components of the uncertainties, because they are independent from cluster to cluster.  The impact of this statistical component of the uncertainty on the results we present here can be reduced by enlarging the sample we study. We consider the uncertainties in the $\zeta$-mass scaling relation parameters $r_{\mathrm{SZ}}$ to be systematic uncertainties, because a shift in one of those parameters, for example the normalization parameter $A_\mathrm{SZ}$, systematically shifts the halo masses of the entire cluster ensemble.  Moreover, these systematic uncertainties can only be reduced through an improved mass calibration of the sample.

To account for both statistical and systematic halo mass uncertainties, we adopt a two step process.  We first fix the cosmological parameters and scaling relation parameters $r_{\mathrm{SZ}}$
rather than marginalizing over the full posterior parameter distributions from \citet{dehaan16} when we estimate the total cluster mass \Mfiveoo.   In this first step, the uncertainties of the cluster masses only reflect the measurement uncertainties and intrinsic scatter in our SZE halo masses,  but do not include the systematic uncertainties due to the uncertainties in the cosmological and SZE $\zeta$--mass scaling relation parameters.  Characteristically, these statistical mass uncertainties are at the level of $\approx20$~percent for a $\xi=5$ cluster.
In a second step, described in Section~\ref{sec:sys_mass}, we quantify the impact of the systematic uncertainties on the best fit baryonic scaling relation parameters that arise due to the uncertainties on the SZE $\zeta$--mass scaling relation parameters $r_{\mathrm{SZ}}$ presented in \citet{dehaan16}.  Marginalizing over the uncertainties in the scaling relation parameters $r_{\mathrm{SZ}}$ corresponds characteristically to a  $\approx15$~percent systematic uncertainty on the cluster halo mass for our sample.  Adding these two components in quadrature leads to a $\approx25$~percent total characteristic uncertainty on a single cluster halo mass.

Unless otherwise stated, the baryonic scaling relation parameter uncertainties presented in the paper are the quadrature sum of the statistical and systematic uncertainties.  Note that both statistical and systematic uncertainties are presented separately in Table~\ref{tab:sr_params}.

\subsection{ICM masses}
\label{sec:icmmass}

We estimate the ICM mass \Mgas\ of each cluster by fitting the X-ray surface brightness profile. The resulting \Mgas\ of each cluster is listed in the Table~\ref{tab:measurements}. We briefly summarize the procedures below, and we defer the readers to \cite{mcdonald13} for more details.

After the reduction of the \CHANDRA\ X-ray data, the surface brightness profile is extracted in the energy range of $0.7-2.0$~keV out to $1.5\Rfiveoo$.  The surface brightness profile is further corrected for the spatial variation of the ICM properties (e.g., temperature) as well as the telescope effective area.  We then project the modified $\beta$-model \citep{vikhlinin06} along the line of sight to fit the observed surface brightness profile.  In the end, the ICM mass \Mgas\ is obtained by integrating the best-fit modified $\beta$-model to the radius of \Rfiveoo, which comes directly from the SZE based halo mass.

We stress that the clusters used in this work have been imaged uniformly by the \CHANDRA\ X-ray telescope in the XVP program with the goal of obtaining $\approx2000$ source counts, which is sufficient signal to allow us to measure \Mgas\ with an uncertainty $\lesssim5$\percent\ for each cluster.

\subsection{Stellar masses}
\label{sec:stellarmass}

To estimate the stellar mass \Mstar\ of each cluster, we carry out SED fitting of individual galaxies using six band photometry---either $griz\IRACone\IRACtwo$ or $griz\Wone\Wtwo$ (see Section~\ref{sec:xvp_optical_nir_data} for more information of constructing the photometric catalogs).
In this work, we have made no attempt to measure the intra cluster light because of the limiting depth of the available imaging.
We describe the procedure of deriving stellar mass in the following subsections.

\subsubsection{SED fitting}
\label{sec:sedfitting}

After constructing the catalogs (see Section~\ref{sec:deblending}), we use \texttt{Le~Phare} \citep{arnouts99, ilbert06} to perform the $\chi^2$-based SED fitting on the galaxies that lie in each cluster field.
We first compile a template library using the BC03 code with various input parameters. 
The parameters and their ranges include (1) metallicities $Z = 0.02, 0.008$, (2) star formation rates that are exponentially decaying with e-folding timescales of $\tau = 1.0,~2.0,~3.0,~5.0,~10.0,~15.0,~30.0$~Gyr, (3) 40 ages logarithmically increasing from 0.01~Gyr to 13.5~Gyr, (4) redshift ranging from 0 to 3.0 with a step of 0.02 and (4) the \citet{calzetti00} extinction law with reddening $E\left(B-V\right) = 0.0,~0.1,~0.2,~0.3$.
This template library is constructed with the goal of sampling the wide range of physical characteristics expected for stellar populations in galaxies in and near clusters.
The \citet{chabrier03} initial mass function is used in constructing the library.  Then, we run \texttt{Le Phare} on each galaxy that lies within the observed footprint to estimate the stellar mass and photometric redshift (photo-\redshift\ or \zp) simultaneously.  During the fitting, we interpolate the templates among the redshift steps.
The SED fitting is performed by comparing predicted and observed fluxes in each of the six bands.

For estimating the stellar mass \MBCG\ of the Brightest Cluster Galaxy (BCG), we run the same SED fitting pipeline on the BCG of each cluster with its redshift fixed to the cluster redshift.  The BCGs have been visually identified and studied in \cite{mcdonald15}, and their sky coordinates are listed in Table~\ref{tab:sample}.  When calculating the stellar masses of non-BCG galaxies, we use the stellar masses estimated based on the photo-\redshift, regardless of whether there is a spec-\redshift\ available, in the interest of uniformity.

\begin{figure}
\vskip-0.15in
\centering
\resizebox{!}{0.7\textwidth}{
\includegraphics[scale=1.0]{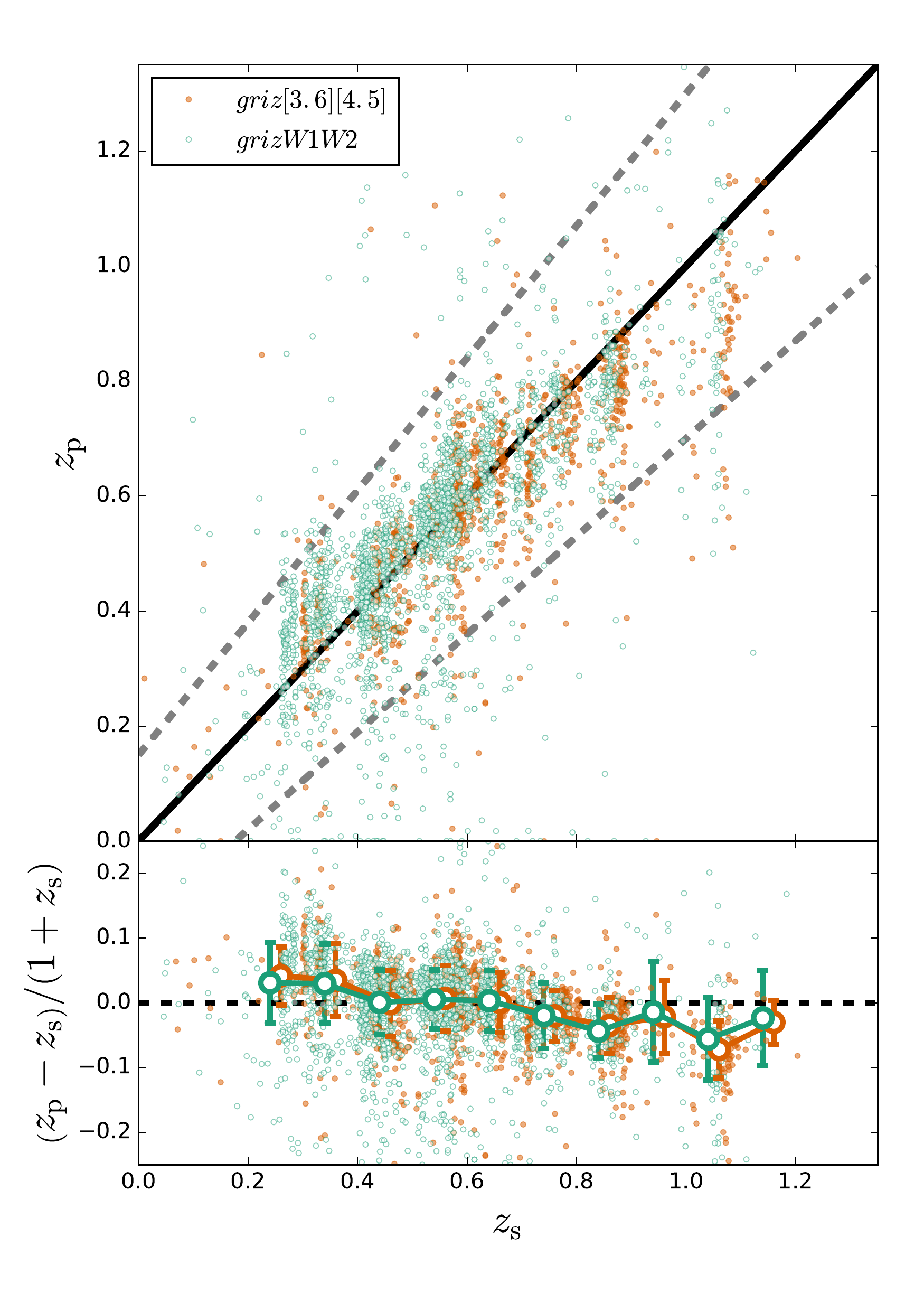}
}
\vskip-0.20in
\caption{
The plot of photo-\redshift\ versus spec-\redshift\ (top) and of photo-\redshift\ mean error and RMS scatter (bottom) for the galaxy sample.
Photo-\redshift's are measured using SED fitting with either $griz\IRACone\IRACtwo$ (red) or $griz\Wone\Wtwo$ (green) photometry.  The dashed lines in the upper panel bracket the region with $\| \Delta\redshift \| / (1 + \redshift) < 0.15$.  For clarity, these two samples are plotted with an offset of 0.02 on the x-axis.
}
\label{fig:zpzs}
\end{figure}

We use the sample of galaxies with spectroscopic redshifts (spec-\redshift\ or \zs) to gauge the accuracy of our photo-\redshift, and also to quantify the systematics in deriving the stellar masses.  In the following comparison, we discard stars (see star/galaxy separation discussion in Section~\ref{sec:selection_of_cluster_galaxies}).
The measurements of spec-\redshift\ are taken from the previous SPT spectroscopic follow-up programs \citep{ruel14,bayliss16}, where a subset of $\approx100$ SPT selected clusters is targeted with the goal of obtaining $\approx25-35$ spectra of galaxies per cluster for the purposes of cluster mass calibration \citep{saro13,bocquet15}.
It is worth mentioning that we include all galaxies with spec-\redshift\ measurements in the comparison even if they are in the fore- or background of the clusters.
The spec-\redshift\ galaxies are selected to sample the cluster red sequence, and we expect this strategy to provide an excellent sample for the purposes of this work, given that the member galaxies of our clusters are indeed dominated by passively evolving populations \citep{hennig17}.
We also note that we re-process the \WISE\ datasets for the full sample (including the ones with \Spitzer\ follow-up observations); therefore, we are able to (1) increase the statistics in our study of the SED fitting performance on $griz\Wone\Wtwo$ and (2) cross-compare the results obtained between $griz\IRACone\IRACtwo$ and $griz\Wone\Wtwo$ for those 52 clusters with both \WISE\ and \Spitzer\ data. 
In the cluster fields there are 2,149 galaxies in total with spec-\redshift\ measurements for the clusters with $griz\Wone\Wtwo$ photometry; for the clusters with the IRAC photometry there are 999 galaxies. 

A comparison between the photo-\redshift's and the spectroscopic redshifts (spec-\redshift\ or \zs) for this sample is contained in Fig.~\ref{fig:zpzs}.
When using the estimator \texttt{Z\_BEST}\footnote{The best estimate of photometric redshift from the maximum likelihood estimation.} for the photo-\redshift,
we measure the mean bias $\Delta\redshift \equiv \left( \zp - \zs \right) / \left( 1 + \zs \right)$ and the root-mean-square variation about the mean as a function of redshift for the sample (bottom panel).  These values are in good agreement, regardless of which NIR photometry is used ($griz\IRACone\IRACtwo$ in red and $griz\Wone\Wtwo$ in green).
Although the mean bias of the photo-\redshift's is statistically consistent with no bias at each redshift bin separately, we do observe mild differences between the photo-\redshift's and spec-\redshift's, and these translate into systematics in our stellar mass estimation.
In at least one previous work \citep{vdb15}, a photo-\redshift\ correction was applied to reduce this systematic.
In this work we account for the uncertainties of the photo-\redshift's by employing the information from the full probability distribution of the photo-\redshift\ (instead of only using a photo-\redshift\ point estimator).

\begin{figure}
\vskip-0.05in
\centering
\resizebox{!}{0.5\textwidth}{
\includegraphics[scale=1.0]{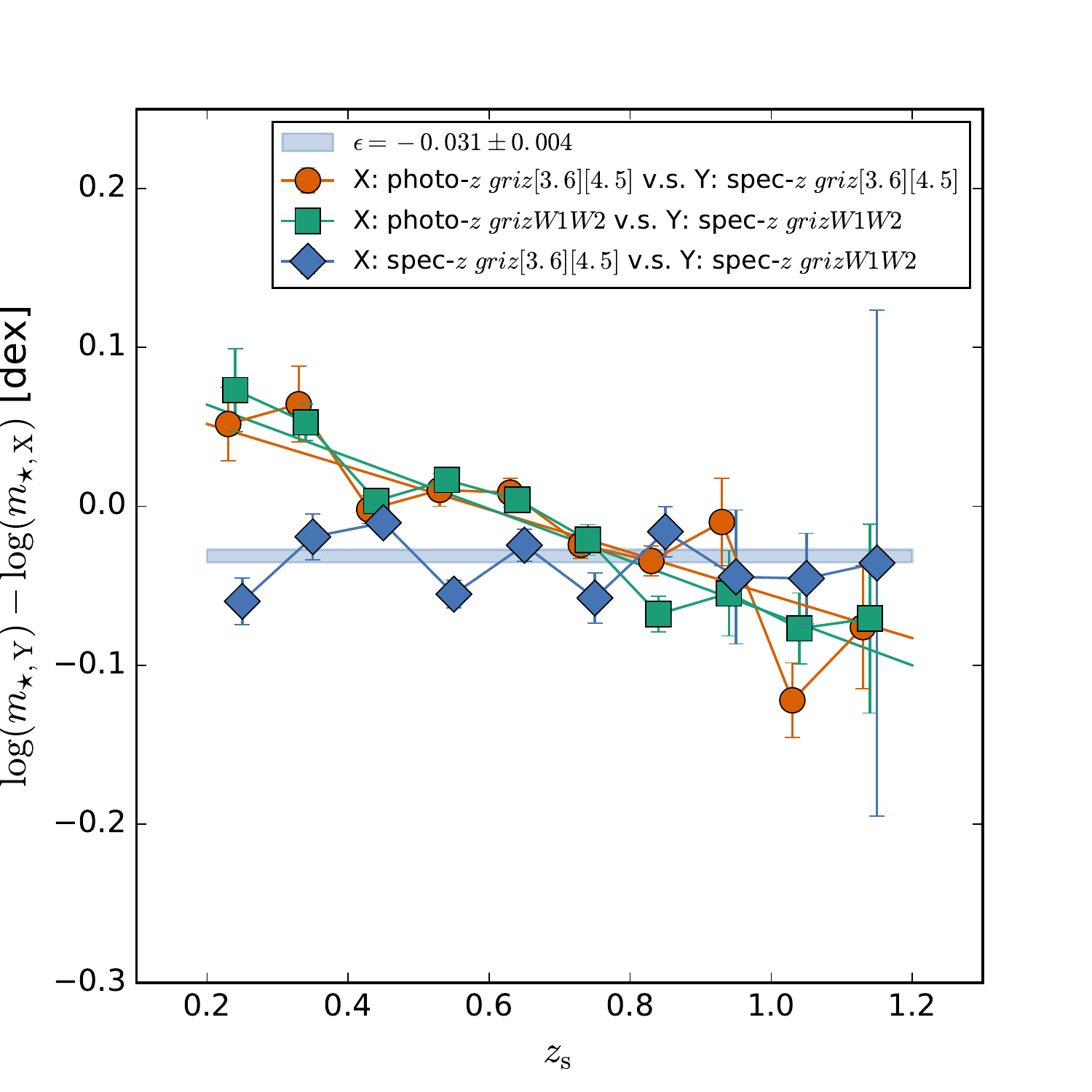}
}
\vskip-0.15in
\caption{
The comparison of derived stellar masses $m_{\star}$ among the cases of using photo-\redshift, spec-\redshift, and photometric catalogs.
These comparisons are made based on the galaxies with available spec-\redshift\ measurements (see Section~\ref{sec:sedfitting}).
The comparison of stellar masses $m_{\star}$ estimated with photometry of $griz\IRACone\IRACtwo$ between photo-\redshift\ and spec-\redshift\ are in red circles, while the same comparison but with photometry of $griz\Wone\Wtwo$ is marked by the green squares.
Additionally, the comparison using the spec-\redshift\ between the photometry of $griz\IRACone\IRACtwo$ and $griz\Wone\Wtwo$ is shown by the blue diamonds.
For clarity, these samples are plotted with an offset 0.02 in the x-axis.
}
\label{fig:dm}
\end{figure}

To quantify any resulting systematics in the derived stellar masses that arise from the accuracy of the photo-\redshift's, we compare the derived stellar masses using the same SED fitting on each galaxy when adopting the photo-\redshift\ and the spec-\redshift.  The results are shown in Fig.~\ref{fig:dm}.
We find that the stellar masses $m_{\star}$ of galaxies show a mild bias as a function of redshift if photo-\redshift's are used:  the derived $m_{\star}$ is biased high by $\approx0.05-0.07$~dex at low redshift and trends to a bias of $\approx-0.07$~dex as the redshift increases to $\approx1.2$.  This fractional bias in stellar mass---denoted as $\delta(\redshift)$---is present for both photometric datasets.
To correct for this bias, we fit a linear function $\delta_{\star}(\redshift)$ to the observed $\delta(\redshift)$, and we use this model to correct the derived stellar masses of each cluster.

We repeat the exercise while using spectroscopic redshifts \zs\ with the two different sets of photometry.
The results are shown by the blue diamonds in Fig.~\ref{fig:dm}.  This comparison reveals any systematic offsets between the stellar masses when using the different datasets, and this appears to be well described by a redshift independent factor.  Namely,  the stellar masses estimated with $griz\Wone\Wtwo$ are systematically lower than those obtained with $griz\IRACone\IRACtwo$ by $\approx0.031$~dex ($\approx7\percent$).  We apply the correction $\epsilon \equiv \log(m_{\star, griz\Wone\Wtwo}) - \log(m_{\star, griz\IRACone\IRACtwo}) = -0.031$~dex to the stellar masses estimated from  $griz\Wone\Wtwo$ to bring them into consistency with those obtained using $griz\IRACone\IRACtwo$ (see Section~\ref{sec:statistical_background_subtraction}).
In summary, the correction $\epsilon$ accounts for the redshift-independent offset in results from the two datasets, while the correction $\delta_{\star}(\redshift)$ accounts for the redshift-dependent discrepancy in the derived stellar masses introduced by using photo-\redshift's.

\subsubsection{Selection of cluster galaxies}
\label{sec:selection_of_cluster_galaxies}

After the SED fitting, we select the galaxies that are used in this work by carrying out (1) the star/galaxy separation, (2) the photo-\redshift\ selection, and (3) the magnitude cut.

For the star/galaxy separation, the parameter \texttt{spread\_model} provides a robust identification of stars down to $i$-band magnitude of $\approx22$~mag in the DES data \citep{hennig17}; therefore we exclude the stars in $i$ band exhibiting $\|\mathtt{spread\_model}\| \le 2\times10^{-3}$ with magnitudes brighter than 22~mag.  We also discard any objects with $\mathtt{spread\_model} \le -2\times10^{-3}$; these consist mainly of defects or unreliable detections.  The remaining faint stars ($i \geq 22$~mag) are excluded by the statistical fore/background subtraction as described below.

After discarding the stars, we reject the galaxies (1) whose photo-\redshift\ probability distributions are inconsistent with the cluster redshift \zd\ at the $3\sigma$ or greater level, and 
(2) whose photo-\redshift\ point estimators satisfy $\| \left( \zp - \zd \right) / \left( 1 + \zd \right) \| \geq 0.15$.  Note that for the photo-\redshift\ point estimator we use a conservative threshold  (i.e., $0.15$) that is $\gtrsim3$ times the RMS scatter of $\| \left( \zp - \zd \right) / \left( 1 + \zd \right) \|$
we observed in the \zp-\zs\ relation.  The purpose of the photo-\redshift\ selection is to remove the galaxies that are certainly outside the cluster, obtaining a highly complete sample of cluster members with lower purity as a trade-off.  We ultimately remove the contamination from fore/background galaxies leaking into our sample by conducting a statistical background subtraction (see Section~\ref{sec:statistical_background_subtraction}).

In the end, we apply a magnitude cut to select the galaxies brighter than \mstarchar\ + 2~mag in the band that is just redder than the 4000~\angstrom\ break in the observed frame.
Specifically, we only select galaxies with $\mathtt{MAG\_AUTO} \le \mstarchar + 2$ in the $r$ ($i$, $z$) band for clusters at $\redshift \le 0.35$ ($0.35 < \redshift \le 0.75$, $\redshift > 0.75$), where the \mstarchar\ is predicted by the CSP model at the cluster redshift \zd\ (see Section~\ref{sec:xvp_sample}).  By employing the selections above, we ensure that we study and select the galaxy populations in a consistent manner across the whole redshift range of the cluster sample.

\subsubsection{Statistical background subtraction}
\label{sec:statistical_background_subtraction}

To eliminate the contamination of (1) faint stars that are not discarded by the cut in \texttt{spread\_model} and (2) non-cluster galaxies 
due to
the photo-\redshift\ scatter, we perform statistical background subtraction.  Specifically, we select the footprint with the field of view of $\approx1\deg^{2}$ located at the center of the COSMOS field \citep{capak2007, ilbert2008} as the background field, because this region is also observed by DES and lies within the \Spitzer\ Large Area Survey with Hyper-Suprime-Cam \citep[SPLASH, ][]{capak12}, with the same wavelength coverage as our cluster fields.
In the COSMOS field, we only use the passbands $griz\IRACone\IRACtwo$ to ensure the uniformity between the optical and NIR datasets available for the SPT clusters.
Moreover, we stress that (1) this region is free from any cluster that is as massive as the SPT clusters, (2) we specifically build this background field by coadding the single exposures observed by the DES to reach comparable depth in the $griz$ bands as we have in the cluster fields, even though the combined DES data in the COSMOS field would be much deeper, and (3) the photometric catalogs of the $griz$ bands are also processed and cataloged using the CosmoDM system.  
For the photometry of \IRACone\ and \IRACtwo\ used in the background field, we match our optical catalog to the \texttt{COSMOS2015} catalog released in \cite{laigle16}, using a matching radius of $1\arcsec$ to obtain the magnitudes and fluxes observed by the SPLASH survey.
We note that the photometry of \IRACone\ and \IRACtwo\ in the \texttt{COSMOS2015} catalog has been properly deblended; therefore, the number of ambiguous pairs in matching is negligible.  

In this way, the photometric catalog of the background field is constructed using the CosmoDM system and is based on the optical detections in the same manner as the cluster fields.
After constructing the photometric catalog of the background field, we perform the same SED fitting and galaxy selection (e.g., the spread model, photo-\redshift\ and magnitude cuts) to obtain the background properties for each cluster.  In other words, we have the stellar mass estimates of the galaxy populations selected and analyzed in the same way on each cluster field and in a corresponding background field.
We randomly draw multiple background apertures with the same size as the cluster \Rfiveoo\ (typically $\approx20$ independent apertures, depending on cluster size), and then adopt the ensemble behavior of the stellar masses among these apertures for use as the background model of the stellar masses toward each galaxy cluster (see Section~\ref{sec:smf}).

\subsubsection{Modelling stellar mass functions}
\label{sec:smf}

We obtain the stellar mass of each cluster by integrating the stellar mass function (SMF).  In the process of modelling the SMF, we exclude the BCG because the luminosity function of the BCGs appears to follow a Gaussian function separately from the satellite galaxies \citep[e.g.,][]{hansen05,hansen09}.  The details of modelling stellar mass function are described as follows.

First, we create the histograms of stellar mass $\mathcal{M}_{\star} \equiv \log(\mstar)$ 
after performing the galaxy selection (see Section~\ref{sec:selection_of_cluster_galaxies}) for both cluster and background fields using stellar mass binning between 9~dex and 13~dex with an equal step of 0.2~dex.  We use the \Rfiveoo\ for each cluster to define the region of interest in the cluster and background fields.  If there are non-observed portions of the \Rfiveoo\ region in the cluster field, then we modify the radii of the background apertures such that their areas match those of the cluster field.  

Second, we model the stellar mass function using a \citet{schechter76} function,
\[
\phi(\mathcal{M}_{\star}) = 
\phi_{\star}~
10^{
(\mathcal{M}_{\star} - m_{\mathrm{char}})
(\alpha + 1)
} 
e^{ (-10^{[\mathcal{M}_{\star} - m_{\mathrm{char}}] }) } \, ,
\]
for each cluster where $\phi_{\star}$ is the normalization, $m_\mathrm{char}$ is the characteristic mass scale denoting the transition between the exponential cutoff and the power-law componentns of the SMF, and $\alpha$ is the faint end slope.  We employ \citet{cash79} statistics to properly deal with observations in the Poisson regime.  Namely, we maximize the log-likelihood
\begin{equation}
\label{eq:cstat}
\ln\left( L_{\mathrm{cstat}} \right) = -\Sigma_{i}\left( M_{i} - D_{i} + D_{i}(\ln(D_{i}) - \ln(M_{i}) ) \right) \, ,
\end{equation}
where $i$ runs over the stellar mass bins, $D_i$ is the observed number of galaxies in the $i$-th bin of the stellar mass histogram observed on the cluster field (which includes the cluster members and background), and $M_i$ is the value of the model stellar mass function in the $i$-th bin.
We construct the model $M_i$ as
\[
M_i = f_{\mathrm{incmp, clu}}({\mstar}_{i}) \phi({\mstar}_{i}) + f_{\mathrm{incmp, bkg}}({\mstar}_{i}) B_i  \, ,
\]
where $B_i$ is the mean number of galaxies in the $i$-th bin of the stellar mass histograms among the apertures that are randomly drawn from the COSMOS field, and the uncertainty of the mean serves as the background uncertainty.
The incompleteness at the low-mass end is accounted for by 
boosting 
the number of sources---for both cluster and background fields (denoted by $f_{\mathrm{incmp, clu}}$ and $f_{\mathrm{incmp, bkg}}$, respectively)---based on the completeness functions in magnitude (see Section~\ref{sec:xvp_optical_nir_data}) derived in the band used for the magnitude cut (see Section~\ref{sec:selection_of_cluster_galaxies}).
Specifically, we bin the galaxies in magnitude space and randomly draw galaxies in each magnitude bin to meet the number count required by the measured completeness function (i.e., the original number multiplied by a factor of $1/f_{\mathrm{incmp}}(m)$).  In this way, we use the completeness function in magnitude to derive a completeness correction for each stellar mass bin.

Finally, we use \texttt{emcee} \citep{foreman13} to explore the likelihood space of $(\phi_{\star}, m_{\mathrm{char}}, \alpha)$.
We begin with flat and largely uninformative priors on these three parameters, and we find that they are ill-constrained on a single cluster basis.
The mean values of $m_{\mathrm{char}}$ and $\alpha$ among the cluster sample are $10.89$~dex and $-0.47$, respectively, both with scatter of $0.25$.
Moreover, the ensemble behavior of $m_{\mathrm{char}}$ and $\alpha$ show no trends with cluster mass \Mfiveoo\ or redshift.
Motivated by the data, we therefore apply a Gaussian prior with mean of $10.89$~dex and width of 0.25 (mean of $-0.47$ and width of 0.25) on $m_{\mathrm{char}}$ ($\alpha$) for modelling the stellar mass function of each cluster.  The main impact of this informative prior is a reduction in the uncertainty of the cluster stellar masses; specifically, 
a return to the flat priors would increase the stellar mass uncertainties by $\approx(5\pm20)\percent$ over the cluster sample.
Once the parameter constraints are in hand for each cluster, we derive the integrated stellar mass $M_{\star, \mathrm{sat}}$ of each cluster by integrating over the stellar mass function from a lower mass limit of $\log\left(\mstar/\Msun\right) = 10$, where we are $\gtrsim80\percent$ complete for all but seven clusters.  
Extrapolating our best-fit stellar mass function to lower stellar masses increases the cluster stellar mass by $\approx3.5\percent$.
We assess the uncertainty due to the cosmic variance in our background estimation that cannot be captured by the solid angle COSMOS survey.
Specifically, we use the analytic function derived in \citet{driver10} to calculate the cosmic variance of the galaxy population in the COSMOS field at the cluster redshift $\redshift_{\mathrm{d}}$ with the line-of-sight length enclosed by the redshifts\footnote{We use half bin width of $0.3$ because it is our typical photo-\redshift\ uncertainty for individual galaxies.} of $\redshift_{\mathrm{d}} - 0.3$ and $\redshift_{\mathrm{d}} + 0.3$; the resulting uncertainties due to cosmic variance are $11\percent, 8.7\percent, 7.8\percent$ and $7.5\percent$ at $\redshift_{\mathrm{d}}=0.3, 0.6, 0.9$ and $1.2$, respectively.
We do not include the cosmic variance in the error budget of the stellar mass estimation. 

There are three corrections that we need to apply to the integrated stellar mass---the masking correction, the correction for systematics of the SED fit and a deprojection correction.
First, due to the insufficient field of view of the \Spitzer\ follow-up observations, we apply the masking correction $f_{\mathrm{mask}}$ to the integrated stellar mass estimates.
The masking correction $f_{\mathrm{mask}}$ is obtained by calculating the weighted ratio of geometric areas of the cluster footprint ($\pi\Rfiveoo^2$) to the observed footprint.
The weighting factor is derived based on a projected NFW profile $\Sigma(r)$ with concentration of $C_{500}=2$ \citep{lin04a,burg14,chiu16b,hennig17} to account for the radial distribution of the cluster galaxies (e.g., the number densities of galaxies drop significantly at large radii).
For example, the weighting factor for the area $A$ is derived as 
$\int_{0}^{r_{\mathrm{eff}}}\Sigma(x)\dif x$,
where $r_{\mathrm{eff}}$ is the effective radius such that $A = \pi r_{\mathrm{eff}}^2$.
By construction, $f_{\mathrm{mask}} = 1$ for the case of using $griz\WISE$ photometry because our \WISE\ imaging is wide enough to cover the whole cluster footprint.

Second, we apply the correction to account for the systematics in the SED fitting (see Section~\ref{sec:sedfitting}).
We apply the correction to account for the systematics caused by the use of photometric redshifts and a correction to take into account the systematic differences in stellar masses when measured using the two different NIR datasets ($griz\Wone\Wtwo$ and $griz\IRACone\IRACtwo$).
We explicitly express the stellar mass estimate \Mstar\ within \Rfiveoo\ of each cluster as follows.
\begin{equation}
\label{eq:stellar_mass}
\Mstar = 10^{-\epsilon} \times \left( \MBCG\ + D_{\mathrm{dpj}}~f_{\mathrm{mask}}~\delta_{\star}(\redshift)~M_{\star, \mathrm{sat}} \right) \,  , 
\end{equation}
where $M_{\star, \mathrm{sat}}$ is the integrated stellar masses of non-BCG cluster galaxies that lie within the cluster \Rfiveoo; 
$\delta_{\star}(\redshift)$ is the linear model as a function of cluster redshift \redshift\ taking the photo-\redshift\ bias into account (see Section~\ref{sec:sedfitting}); 
$f_{\mathrm{mask}}$ is the masking correction due to the unobserved area in the footprint; 
$D_{\mathrm{dpj}} = 0.71$ is the deprojection factor \citep{lin03b} converting the galaxy distribution from the volume of a cylinder to a sphere by assuming a NFW model with concentration $C_{500} = 2$; $\epsilon = -0.031$ is the correction for the systematic between different NIR datasets (see Section~\ref{sec:sedfitting}); 
we apply the correction $10^{-\epsilon}$ to bring the stellar masses from the basis of $griz\Wone\Wtwo$ to $griz\IRACone\IRACtwo$---therefore---by definition, $\epsilon = 0$ for the case of using \Spitzer\ datasets.
Using a higher concentration $C_{500} = 3.2$, which would be correct if the galaxy populations in clusters were completely dominated by passively evolving galaxies \citep{hennig17}, would result in a higher deprojection factor of $D_{\mathrm{dpj}} = 0.76$.  
Therefore, we estimate that there is an associated deprojection systematic in the stellar mass estimates that is at the level of $\approx5\percent$.
The resulting \Mstar\ of each cluster is listed in the Table~\ref{tab:measurements}.

\begin{figure*}
\centering
\resizebox{0.47\textwidth}{!}{
\includegraphics[scale=1.0]{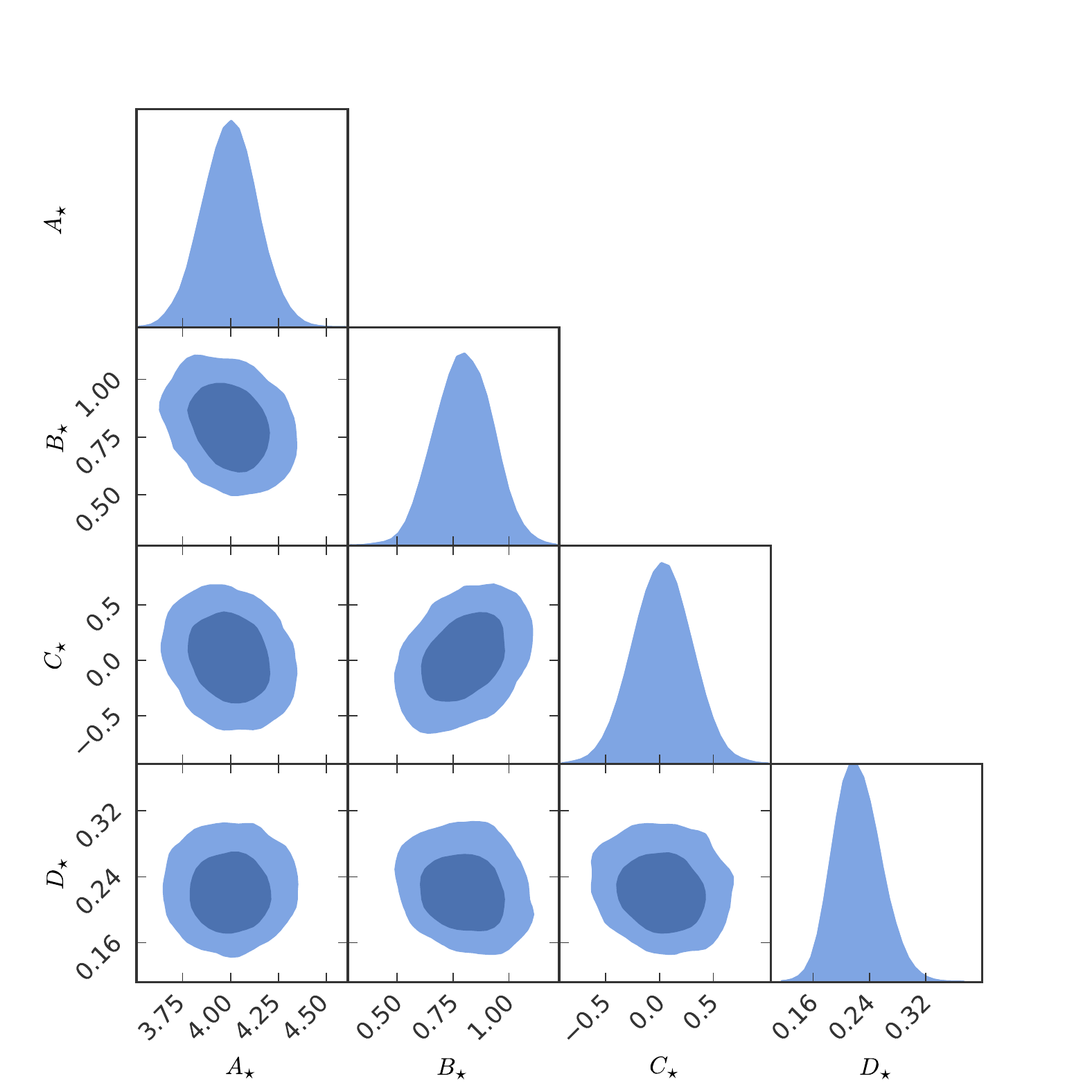}
}
\resizebox{0.47\textwidth}{!}{
\includegraphics[scale=1.0]{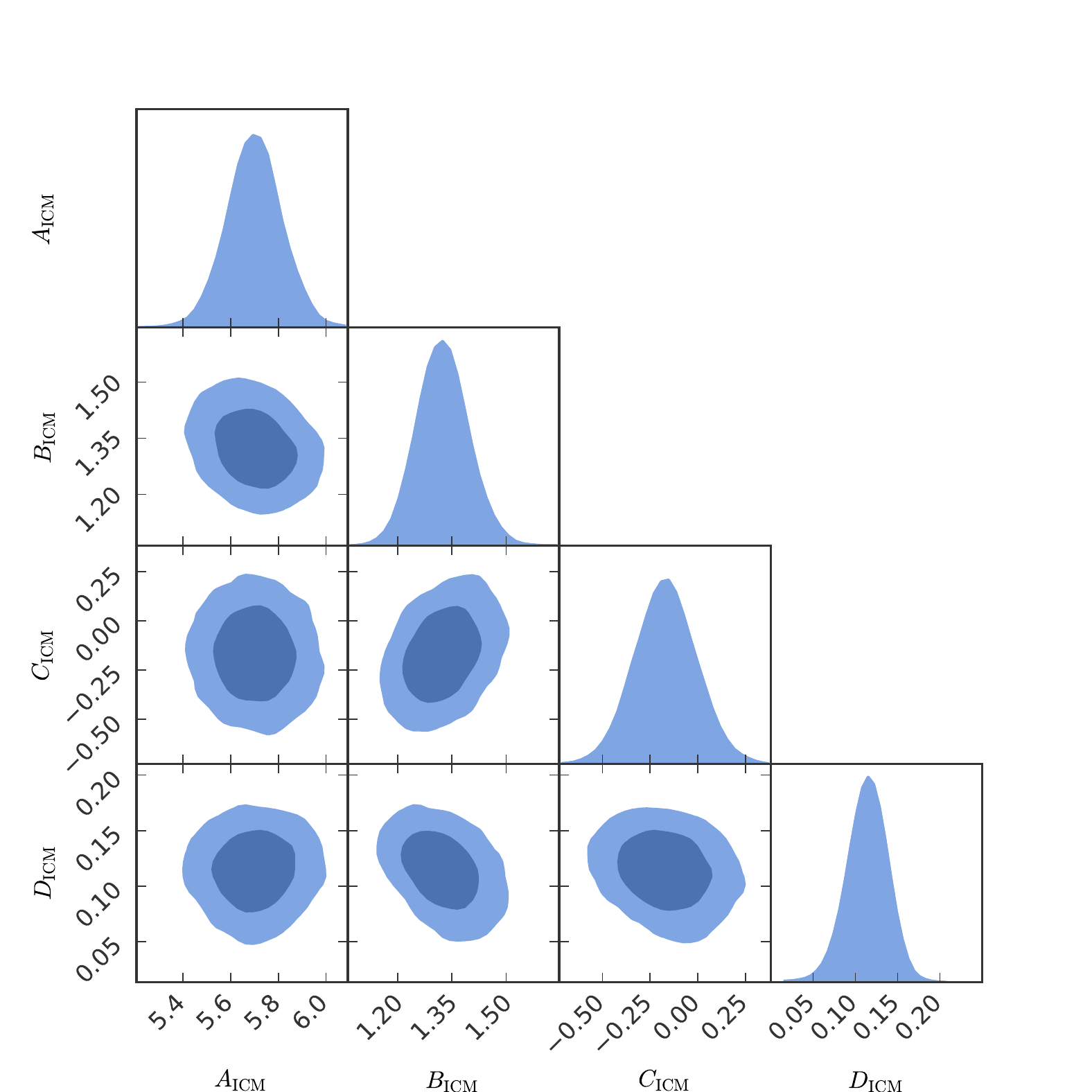}
}
\resizebox{0.47\textwidth}{!}{
\includegraphics[scale=1.0]{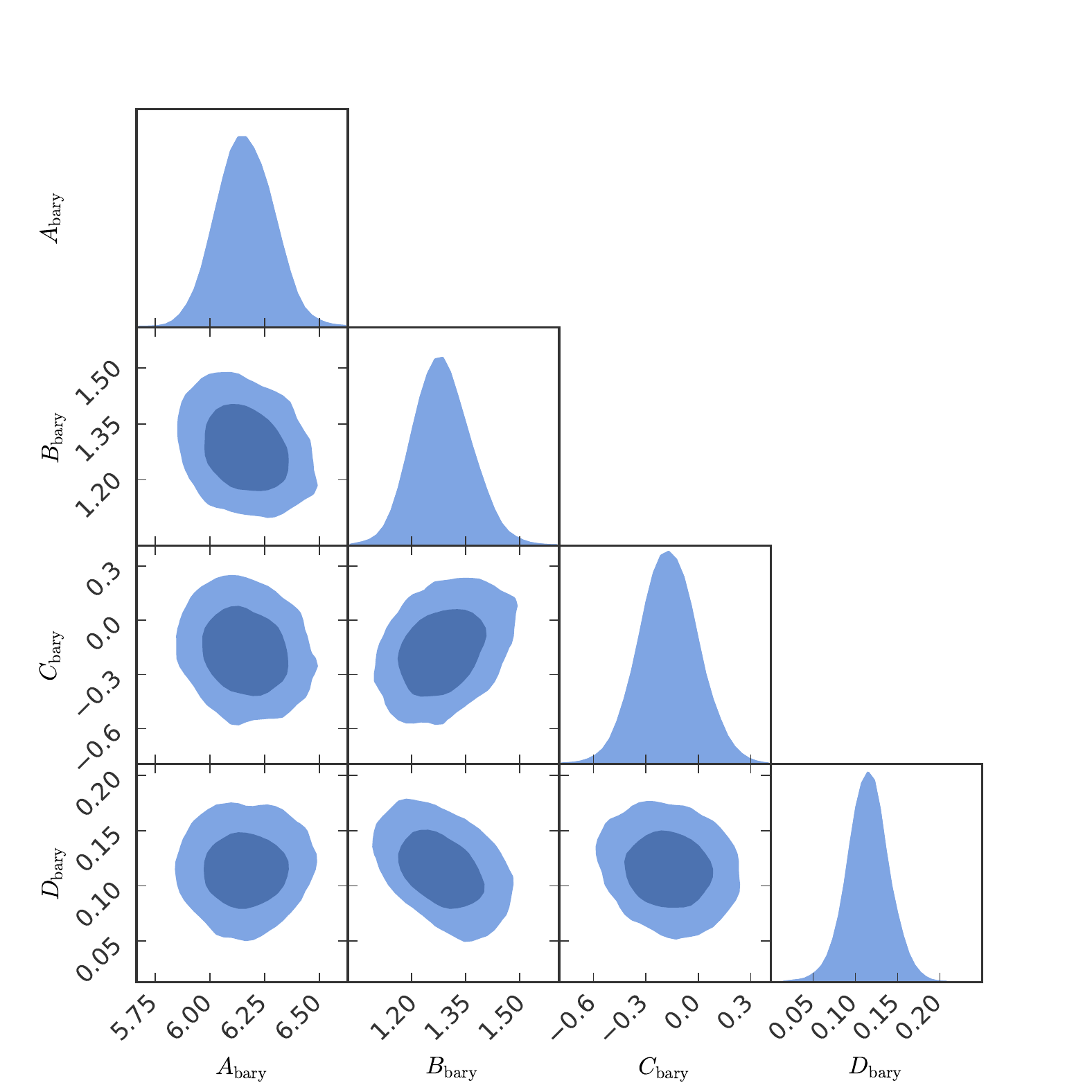}
}
\resizebox{0.47\textwidth}{!}{
\includegraphics[scale=1.0]{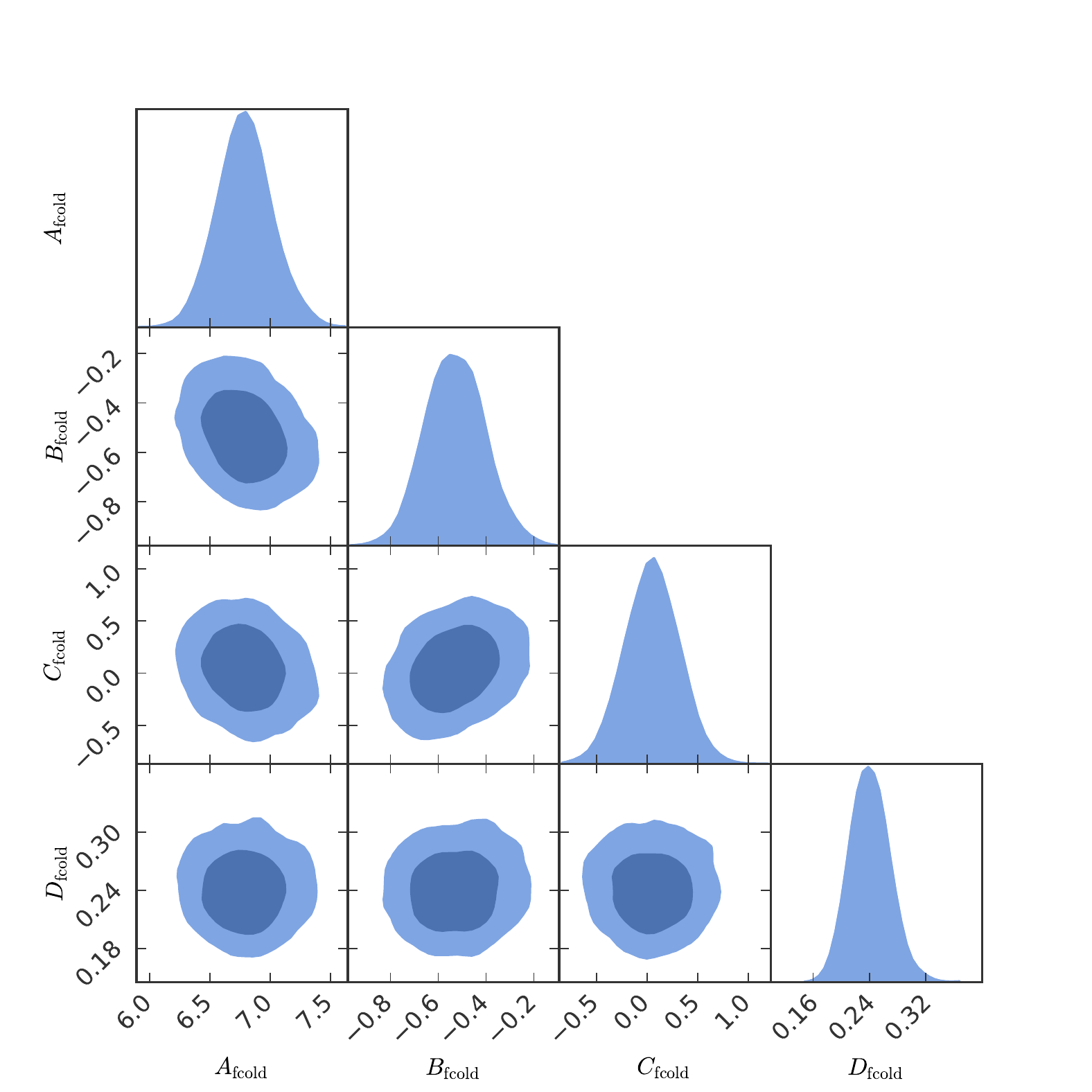}
}
\caption{
The single and joint parameter posterior likelihood distributions for the scaling relations stellar mass to halo mass (the upper-left panel), ICM mass to halo mass (the upper-right panel), baryonic mass to halo mass (the lower-left panel), and cold baryon fraction to halo mass (the lower-right panel).
The normalizations \Astar, \Agas, \Abary\ and \Acold\ are in the units of $10^{12}\Msun$, $10^{13}\Msun$, $10^{13}\Msun$ and $10^{-2}$ , respectively.
These plots are generated using the \texttt{pygtc} package \citep{pygtc16}.
}
\label{fig:triangles}
\end{figure*}
%

%
%

\section{Scaling Relation Form and Fitting Method}
\label{sec:method}

In this work, we use the following functional form to describe the scaling relation between the observable $\mathcal{X}$, the halo mass and the redshift:
\begin{equation}
\label{eq:sr_functional_form}
\mathcal{X} = A_{\mathcal{X}}
\left( \frac{\Mfiveoo}{\MPIV} \right)^{B_{\mathcal{X}}}
\left( \frac{f_{\redshift}(\redshift)}{f_{\redshift}(\ZPIV)} \right)^{C_{\mathcal{X}}} \, 
\end{equation}
with log-normal intrinsic scatter in observable at fixed mass $D_{\mathcal{X}} \equiv \sigma_{\ln \mathcal{X}|M_{500}}$, where $A_{\mathcal{X}}$ is the normalization at the pivot mass \MPIV\ and redshift \ZPIV, $B_{\mathcal{X}}$ and $C_{\mathcal{X}}$ are the power law indices of the mass and redshift trends, respectively, and the notation $\mathcal{X}$ runs over \Mstar, \Mgas, \Mbary\ and \fcold.
The function $f_{\redshift}$ describes the functional form of the redshift trend. 
We use two functional forms for $f_{\redshift}$ in each scaling relation: the first one is $f_{\redshift} \equiv E(\redshift)$, which is conventionally used in the community of X-ray cluster cosmology and implies that the redshift evolution of the observable at fixed mass is 
cosmology
dependent.  The second form is $f_{\redshift}\equiv(1+\redshift)$, which is a direct observable and has no cosmological sensitivity.
In addition, we adopt the pivot mass and redshift $\MPIV = 4.8\times10^{14}\Msun$ and $\ZPIV = 0.6$ throughout this work, because they are the median values of mass and redshift for our cluster sample.

\subsection{Fitting procedure}
\label{sec:fitting_procedure}

We fit the scaling relations in a Bayesian framework, which accounts for the Eddington bias, Malmquist bias and the selection function of our cluster sample.
The likelihood adopted in this work has also been used in several previous studies \cite[][Bulbul et. al. in preparation]{liu15a, chiu16c} and has been tested using large mocks, demonstrating that this likelihood can recover unbiased input parameters.
We defer the reader to the earlier references for more detail, and provide here only a briefly description of the likelihood.

This likelihood is specifically designed to obtain the targeted observable $\mathcal{X}$ to halo mass relation (e.g., equation~\ref{eq:sr_functional_form}) of a sample of clusters selected using another observable (e.g., the SZE observable $\xi$ used in this work).  
In this likelihood, we explore the parameter space of the targeted scaling relation $r_{\mathcal{X}}$ while fixing the cosmological parameters and the scaling relation $r_{\mathrm{SZ}}$ that is used to infer cluster halo masses.

We explicitly write down the likelihood as follows.
We first evaluate---for the $i$-th cluster at redshift $\redshift_{i}$---the probability $\mathcal{L}_{i}(r_{\mathcal{X}})$ of observing the observable $\mathcal{X}_{i}$ given the scaling relations ($r_{\mathcal{X}}$  and $r_{\mathrm{SZ}}$) and the selection observable $\xi_{i}$ that is used for inferred cluster mass (see equation~(\ref{eq:zeta2mass}) and equation~(\ref{eq:xi2zeta})), i.e.,
\begin{equation}
\label{eq:l15_likelihood1}
\begin{split}
\mathcal{L}_{i}(r_{\mathcal{X}}) &= 
P(\mathcal{X}_{i} | \xi_{i}, z_{i}, r_{\mathcal{X}}, r_{\mathrm{SZ}} )    \\
              &= 
\frac{
\int\mathrm{d}\Mfiveoo\ P(\mathcal{X}_{i}, \xi_{i}| z_{i}, r_{\mathcal{X}}, r_{\mathrm{SZ}} )\ n(\Mfiveoo, z_{i})
}{
\int\mathrm{d}\Mfiveoo\ P( \xi_{i}| z_{i}, r_{\mathcal{X}}, r_{\mathrm{SZ}} ) \ n(\Mfiveoo, z_{i})} \, ,
\end{split}
\end{equation}
\noindent where $n(\Mfiveoo, z_{i})$ is the mass function, for which the shape is fixed because we do not vary the cosmological parameters.
The \citet{tinker08} mass function is used for calculating $n(\Mfiveoo, z_{i})$, but for the mass range in this SPT cluster sample, using the more accurate mass functions extracted from hydrodynamical simulations would have no observable impact \citep{bocquet16}.
The best-fit scaling relation parameters $r_{\mathcal{X}}$ are then obtained by maximizing the sum of the log-likelihoods of the $N_\mathrm{cl}$ clusters,
\begin{equation}
\label{eq:full_like}
\ln \mathcal{L}(r_{\mathcal{X}}) = \sum_{i=1}^{N_{\mathrm{cl}}}~\ln \mathcal{L}_{i}(r_{\mathcal{X}}).
\end{equation}
We note that our scaling relation analysis does not include correlated scatter between SZE based halo masses and the ICM and stellar mass measurements.  In other recent analyses of the SPT cluster sample, no evidence for correlated scatter has emerged, and therefore adding an additional correlation coefficient would not impact our results.  Specifically, in \citet{dehaan16} a correlation coefficient $\rho_{SZ,Y}$ is included in the analysis but not well constrained by the data and has a value consistent with zero.  In Figure~7 of \citet{dietrich17} the correlation coefficients describing the correlated scatter among the SZE, X-ray and weak lensing mass proxies all have large uncertainties and values that are consistent with zero.  This is not to say that there is no correlated scatter among these observables, but it is proof that with this sample any correlated scatter that is present is too weak to be measured or to have an impact on our fit parameters.
Concretely, the extracted ICM mass scatter at fixed halo mass from the \citet{dietrich17} analysis is measured to be $0.106^{+0.041}_{-0.020}$, fully consistent with the results from our analysis that we present in Table~\ref{tab:sr_params}.

We use the python package \texttt{emcee} \citep{foreman13} to explore the parameter space of $r_{\mathcal{X}}$.
The intrinsic scatter and measurement uncertainties of $\xi_{i}$ for each cluster are taken into account while evaluating equation~(\ref{eq:full_like}). 
We apply flat priors for $r_{\mathcal{X}}$ during the likelihood maximization.
Precisely speaking, we adopt the flat priors of $B_{\mathcal{X}}$ in the range $(0.1, 3.5)$, $C_{\mathcal{X}}$ in $(-4,4)$, and $D_{\mathcal{X}}$ in $(10^{-3}, 1.5)$ for all scaling relations (i.e., $\mathcal{X}$ runs over \Mstar, \Mgas, \Mbary\ and \fcold) except that we use the flat prior of $(-1.5,0.0)$ for the mass trend of \fcold.
For the normalization $A_{\mathcal{X}}$, we apply a flat prior of 
$\mathcal{U}(10^{11}, 10^{13})\times\Msun$,
$\mathcal{U}(10^{12}, 10^{14})\times\Msun$, 
$\mathcal{U}(10^{12}, 10^{14})\times\Msun$, and 
$\mathcal{U}(10^{-2}, 10^{-1})$
for the observable $\mathcal{X}$ as \Mstar, \Mgas, \Mbary\ and \fcold,  respectively,
where the notation of $\mathcal{U}$ denotes a uniform interval.
All measurement uncertainties of \Mstar, \Mgas, \Mbary, and \fcold\ are taken to be Gaussian.

%
%

\subsection{Cluster halo mass \Mfiveoo\ systematic uncertainties}
\label{sec:sys_mass}

The uniformity of our sizable sample of galaxy clusters that have been selected through their SZE signatures over a wide redshift range of $0.25\lesssim\redshift\lesssim1.25$ represents one of the major strengths of this work.  Moreover, we analyze the multi-wavelength datasets in the same manner for every system, and in doing so we further avoid systematic uncertainties that can creep in with different treatments by a variety of codes and authors.
These two elements of our current analysis enable a reduction in systematic uncertainties in comparison to many previous studies. 
Nevertheless, halo mass related systematics remain.

We quantify the impact of systematic uncertainties due to remaining uncertainties in the SZE observable to mass scaling relation (i.e., equation~(\ref{eq:params_sz}).  These systematic uncertainties correspond to the variation of the best-fit SZE observable to mass relation that are impacted by the number count constraints in the full likelihood analysis including the variation of the cosmological parameters.
As quantified in \cite{dehaan16},
the $1\sigma$ uncertainties of the parameters $\left(\Asz,\Bsz,\Csz,\Dsz\right)$, which are fully marginalized over other nuisance parameters when performing a full likelihood analysis, 
are $\left(0.91, 0.08, 0.32, 0.07\right)$.

We separately vary the best-fit parameters of  \Asz, \Bsz, \Csz\ and \Dsz\ by their corresponding $+1\sigma$ and $-1\sigma$ uncertainties, and then re-run the likelihood fitting code to calculate the resulting difference in the best-fit parameters quoted in Table~\ref{tab:sr_params}.  Given the lack of evidence for covariance among the parameters of the SZE--mass relation, we ignore the correlation among \Asz, \Bsz, \Csz\ and \Dsz. That is, the resulting difference of $A_{\Mstar}$, for instance, is calculated as $0.5 \times \left| A_{\Mstar}|_{_{(\Asz+\sigma_{\Asz})}} - A_{\Mstar}|_{_{(\Asz-\sigma_{\Asz})}} \right|$, and this is the same for other parameters and observable--mass relations.  These differences serve as the systematic uncertainties that appear in Table~\ref{tab:sr_params}.  The resulting systematic uncertainties are smaller than or comparable with the statistical uncertainties except in the case of the normalization parameter $A_{\mathcal{X}}$.  As will be shown, including the systematic uncertainties increases the total error budget of $A_{\Mstar}$, $A_{\Mgas}$, $A_{\Mbary}$ and $A_{\fcold}$ by a factor of $\approx2$, $\approx5.6$, $\approx5.1$ and $\approx1.6$, respectively.   On the other hand, the systematic uncertainties are subdominant for mass and redshift trends, and thus do not change the overall interpretation that significant infall from surrounding environments must be taking place.

We present the resulting best fit parameters followed by their statistical uncertainties and estimated systematic uncertainties for each scaling relations in Table~\ref{tab:sr_params}.  In the discussion of the results in the text we combine the statistical and systematic uncertainties in quadrature and present only this combined estimate of the total uncertainty.

%
%

\section{Results}
\label{sec:scalingrelation}

In this section, we aim to derive the scaling relations of galaxy clusters describing the quantitative relationship between the baryon content in its various forms and the cluster halo mass and redshift.
Specifically, we focus on (1) the stellar mass to halo mass and redshift relation \Mstar--\Mfiveoo-\redshift, (2) the ICM mass to halo mass and redshift relation \Mgas--\Mfiveoo-\redshift, (3) the baryonic mass to halo mass and redshift relation \Mbary--\Mfiveoo-\redshift, and (4) the fraction of cold collapsed baryons to halo mass and redshift relation \fcold--\Mfiveoo-\redshift.

We estimate the halo masses \Mfiveoo\ and the ICM masses \Mgas\ of the 91 SPT clusters using their SZE observables and uniform \CHANDRA\ X-ray followup imaging, respectively. 
A subset of 84 clusters out of the full sample is imaged in the optical as part of DES and in the NIR with \Spitzer\ and \WISE, enabling us to obtain the stellar masses \Mstar\ of these 84 systems.
As a result, the scaling relations we present contain only 84 SZE-selected clusters except in the case of the ICM mass to halo mass relation, where we are able to use the full sample.

In the following subsections, we 
present our results.

\begin{table}
\centering
\caption{
The best-fit parameters of the observable $\mathcal{X}$ to halo mass and redshift scaling relations (equation~\ref{eq:sr_functional_form}).
Columns $A_{\mathcal{X}}$, $B_{\mathcal{X}}$, $C_{\mathcal{X}}$ and $D_{\mathcal{X}}$ are, respectively, the normalization, the power law index of the mass trends and redshift trends, and the log-normal intrinsic scatter of the observable at fixed mass and redshift.  $\mathcal{X}$ is used to represent \Mstar, \Mgas, \Mbary\ and \fcold.
The units of the normalization $A_{\mathcal{X}}$ are in $10^{12}\Msun$, $10^{13}\Msun$, $10^{13}\Msun$ and $10^{-2}$ for $\mathcal{X}$ of \Mstar, \Mgas, \Mbary\ and \fcold, respectively.
Along with the best-fit value, we present the statistical and systematic uncertainties in that order. The systematic uncertainties reflect the underlying uncertainties on the SZE mass--observable relation as described in detail in Section~\ref{sec:sys_mass}.  When quoting parameter uncertainties in the text we combine these statistical and systematic uncertainties in quadrature.
The case of $f_{\redshift} \equiv (1 + \redshift)$ is in the first tier of the table, followed by the case of $f_{\redshift} \equiv E(\redshift)$.
}
\label{tab:sr_params}
\resizebox{0.45\textwidth}{!}{
\begin{tabular}{lcccc}
\hline\hline
$\mathcal{X}$ &$A_{\mathcal{X}}$ &$B_{\mathcal{X}}$ &$C_{\mathcal{X}}$ &$D_{\mathcal{X}}$  \\[6pt] \hline
\multicolumn{5}{c}{$f_{\redshift} \equiv (1 + \redshift)$ in equation~(\ref{eq:sr_functional_form})} \\[6pt] \hline
\Mstar  &\numAstaronez $\pm0.25$ &\numBstaronez $\pm0.03$ &\numCstaronez $\pm0.11$ &\numDstaronez $\pm0.01$ \\[6pt] \hline
\Mgas   &\numAgasonez $\pm0.61$ &\numBgasonez $\pm0.05$ &\numCgasonez $\pm0.17$ &\numDgasonez $\pm0.03$ \\[6pt] \hline
\Mbary  &\numAbaryonez $\pm0.61$ &\numBbaryonez $\pm0.05$ &\numCbaryonez $\pm0.17$ &\numDbaryonez $\pm0.03$ \\[6pt] \hline
\fcold   &\numAcoldonez $\pm0.28$ &\numBcoldonez $\pm0.02$ &\numCcoldonez $\pm0.05$ &\numDcoldonez $\pm0.01$ \\[6pt] \hline\hline
\multicolumn{5}{c}{$f_{\redshift} \equiv E(\redshift)$ in equation~(\ref{eq:sr_functional_form})} \\[6pt] \hline
\Mstar  &\numAstarez $\pm0.25$ &\numBstarez $\pm0.03$ &\numCstarez $\pm0.11$ &\numDstarez $\pm0.01$ \\[6pt] \hline
\Mgas   &\numAgasez $\pm0.61$ &\numBgasez $\pm0.05$ &\numCgasez $\pm0.17$ &\numDgasez $\pm0.03$ \\[6pt] \hline
\Mbary  &\numAbaryez $\pm0.61$ &\numBbaryez $\pm0.05$ &\numCbaryez $\pm0.17$ &\numDbaryez $\pm0.03$ \\[6pt] \hline
\fcold   &\numAcoldez $\pm0.28$ &\numBcoldez $\pm0.02$ &\numCcoldez $\pm0.05$ &\numDcoldez $\pm0.01$ \\[6pt] \hline\hline
\end{tabular}
}
\end{table}
\begin{figure*}
\centering
\vskip-0.1in
\resizebox{!}{0.5\textheight}{
\includegraphics[scale=1.0]{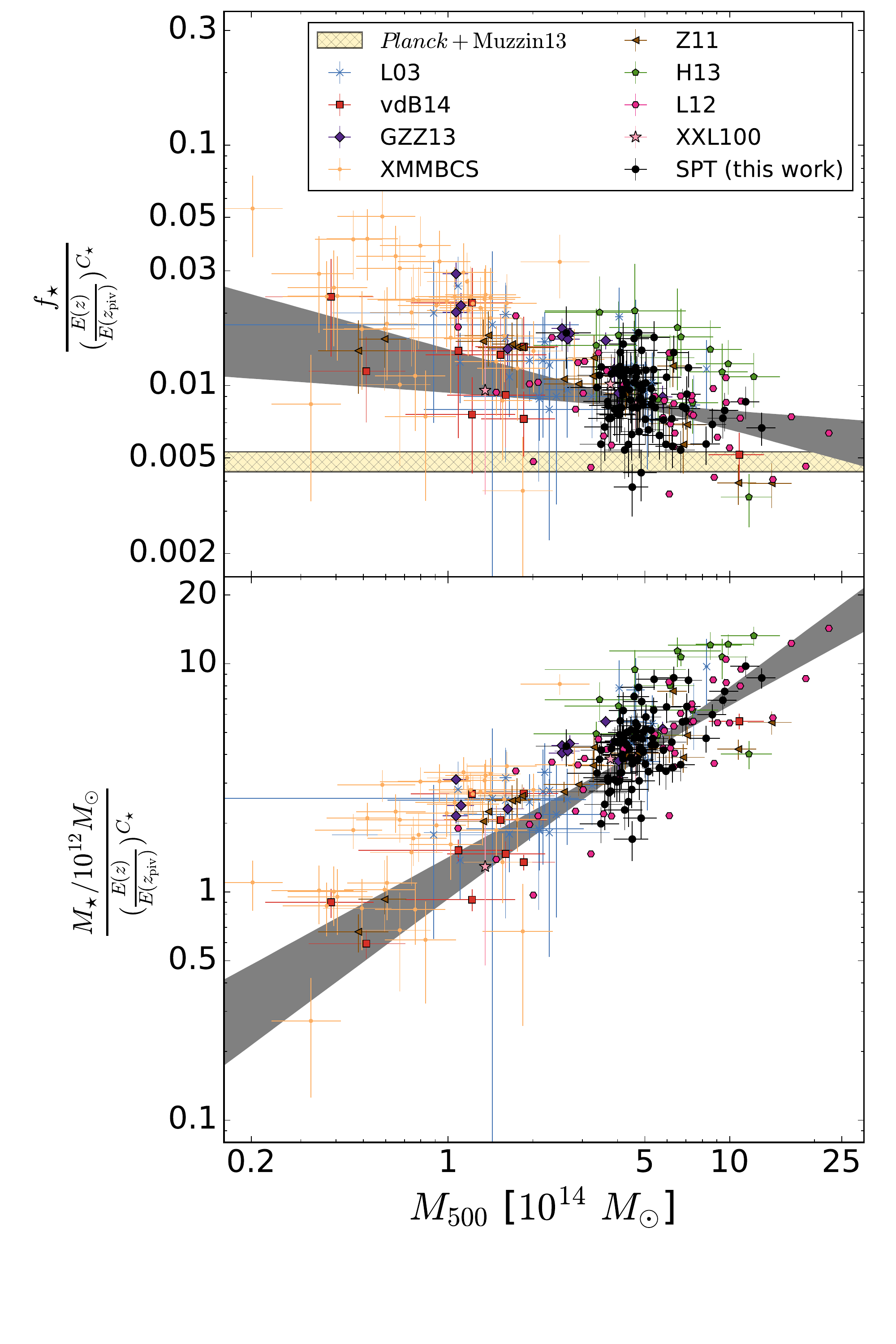}
}
\resizebox{!}{0.5\textheight}{
\includegraphics[scale=1.0]{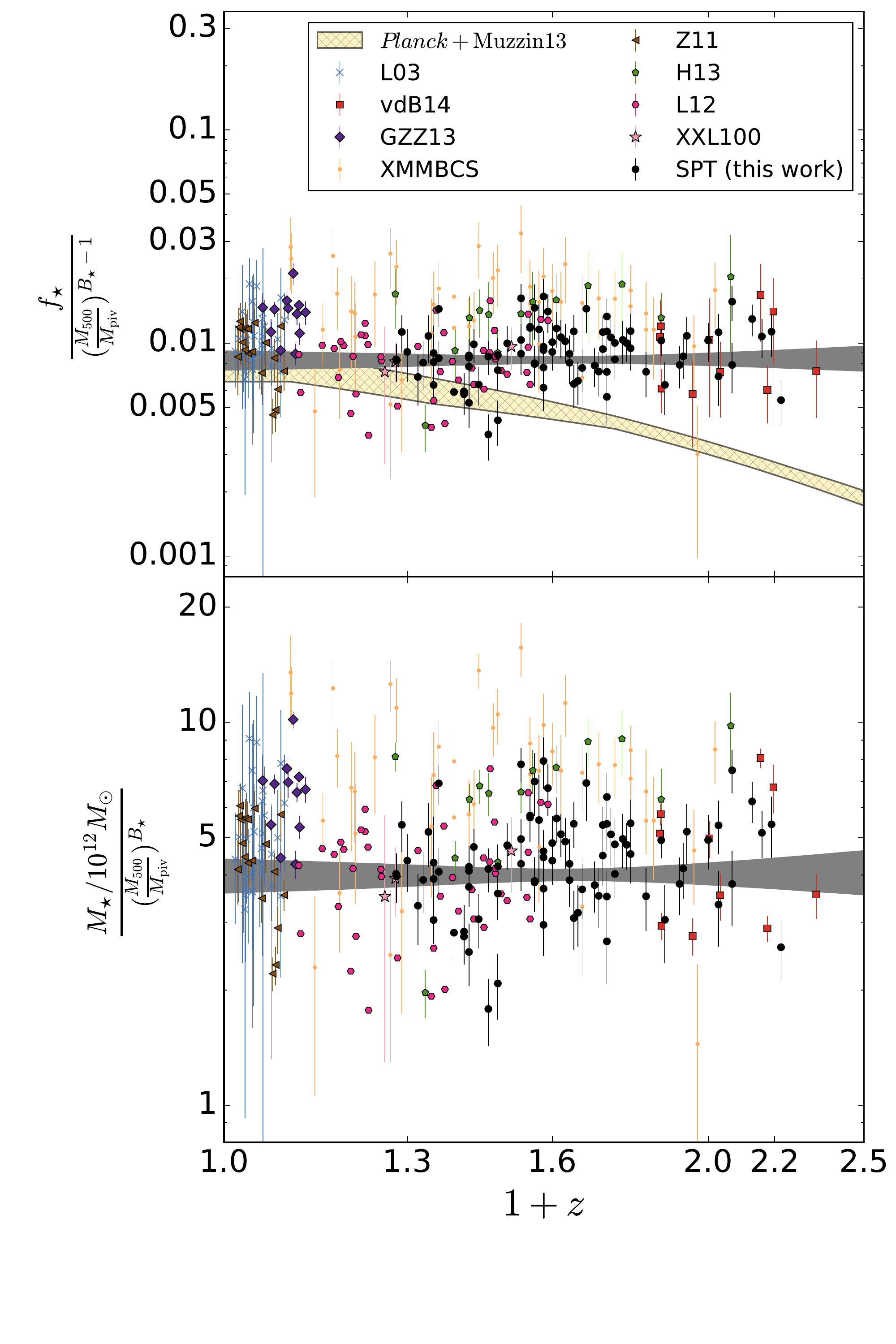}
}
\vskip-0.40in
\caption{
The scaling relations of stellar mass \Mstar\ (lower half) and stellar mass fraction \fstar\ (upper half) based on 84 SPT clusters.
The left and right panels show the mass and redshift trends with respect to the pivot mass $\MPIV=4.8\times10^{14}\Msun$ and redshift $\ZPIV=0.6$, respectively.
The black points are the measurements of the SPT clusters, while the comparison samples are color-coded as shown in the legend.  The cosmic value of the stellar mass fraction, derived from the combination of the field stellar mass density \citep{muzzin13,cole01,bell03,baldry12} and the CMB cosmological constraints  \citep{planck16cosmological}, is indicated by the yellow bar.
The grey areas indicate the fully marginalized $1\sigma$ confidence regions of the best-fit scaling relation extracted from the SPT clusters only.
}
\label{fig:xvpmstar_sr}
\end{figure*}

\subsection{Stellar mass to halo mass relation}
\label{sec:stellar_to_halo_mass_relation}

In this work, we obtain the stellar mass to halo mass scaling relation based on the 84 clusters selected by the SPT at $0.25\lesssim\redshift\lesssim1.25$.
The resulting scaling relation is
\begin{eqnarray}
\label{eq:mstar_sr_oneplusz}
\Mstar &=& (\numAstaronezsys)\times10^{12}\Msun\times\nonumber\\
&&\left( \frac{\Mfiveoo}{4.8\times10^{14}\Msun} \right)^{\numBstaronezsys}
\left( \frac{1 + \redshift}{1 + 0.6} \right)^{\numCstaronezsys} \, 
\end{eqnarray}
with the log-normal intrinsic scatter of $\numDstaronezsys$.
Full results with both forms of the redshift evolution are shown in Table~\ref{tab:sr_params}.
The fully marginalized posteriors and covariance of these parameters appear in Fig.~\ref{fig:triangles}.

The best-fit scaling relation and the derived \Mstar\ are shown in Fig.~\ref{fig:xvpmstar_sr}, where we also show the stellar mass fraction defined by
\[
\fstar \equiv \frac{\Mstar}{\Mfiveoo} \, .
\]
In Fig.~\ref{fig:xvpmstar_sr}, we present \Mstar\ and \fstar\ as functions of cluster mass \Mfiveoo\ and redshift \redshift.  Because the constraints of two kinds of the scaling relations are very similar, we only show the case for $f_{\redshift} = E(z)$ in this figure.
The mass trends of \Mstar\ (the lower panel) and \fstar\ (the upper panel) with respect to the pivot \ZPIV\ are contained in the left panel, while their redshift trends at the pivot mass \MPIV\ are shown in the right panel.
To present the mass trends at the characteristic redshift $\ZPIV=0.6$, we normalize \Mstar\ and \fstar\ to the pivot redshift \ZPIV\ (i.e., dividing them by the best-fit redshift trend $\left(E(\redshift) / E(\ZPIV)\right)^{\Cstar}$).
Similarly, we remove the mass trends to highlight the redshift trends by dividing the \Mstar\ and \fstar\ by an appropriate factor (i.e. $\left( \Mfiveoo/4.8\times10^{14}\Msun\right)^{\Bstar}$ for \Mstar).

The derived scaling relation suggests a strong mass trend $\Bstar = \numBstaronezsys$, while the redshift trend is statistically consistent with zero ($\Cstar = \numCstaronezsys$) with a large uncertainty.
The normalization $\Astar =(\numAstaronezsys)\times10^{12}\Msun$ implies a stellar mass fraction \fstar\ of $(0.83\pm0.06)$\percent\ at the pivot mass $\MPIV = 4.8\times10^{14}\Msun$ and the pivot redshift $\ZPIV = 0.6$.
Our results, based on an approximately mass-limited sample of clusters, suggest that the stellar mass content is well-established and not evolving in massive clusters with $\Mfiveoo\gtrsim4\times10^{14}\Msun$ at $0.25\lesssim\redshift\lesssim1.25$.
Switching from $f_{\redshift} \equiv E(\redshift)$ to $f_{\redshift} \equiv (1 + \redshift)$ provides a similar scenario.

We compare our SPT results to the previous work, as also shown in Fig.~\ref{fig:xvpmstar_sr}.
The comparison samples are 
(1) \citet[][L03]{lin03b}, where they measured the \Mgas\ and \Mstar\ of 27 nearby clusters at $\redshift\lesssim0.1$,
(2) \citet[][Z11]{zhang11}, where a sample of 19 clusters selected by their X-ray fluxes was studied, 
(3) \citet[][L12]{lin12}, where a census of baryon content using 94 clusters at $0 < \redshift < 0.6$ was conducted,
(4) \citet[][GZZ13]{gonzalez13}, where they studied baryon fractions of 12 clusters at $\redshift \approx 0.1$, 
(5) \citet[][H13]{hilton13}, where the stellar content of a sample of 14 SZE-selected clusters was measured,
(6) \citet[][vdB14]{burg14}, where they measured the stellar masses of a sample of 10 low-mass clusters selected in NIR at high redshift ($\redshift\approx1$),
(7) the \XMMBCS\ sample from \citet{chiu16c}, where they used uniform NIR imaging deriving the stellar masses of 46 X-ray selected galaxy groups, and
(8) the latest results from XXL100---the 100 brightest galaxy clusters or groups selected by the XXL survey \citep{eckert16}, for which we only use a subset of 34 clusters with available measurements of the ICM and stellar masses (see their Table~1).

For a fair comparison, we need to account for various systematic differences among the comparison samples.
For example, a different initial mass function (e.g., the \citet{salpeter55} model) of stellar population synthesis used in inferring stellar masses results in a factor $\approx2$ higher estimations than the ones derived using the \citet{chabrier03} mass function, which we use here.
Also, it has been demonstrated and quantified in \cite{bocquet15} that the cluster masses inferred by X-ray---usually based on the assumption of hydrostatic equilibrium in the state of ICM---are biased low by  
$\approx12\percent$
as compared to our SZE derived masses. 
Additionally, the cluster masses inferred from cluster velocity dispersions are $\approx4\percent$ higher
than the SZE derived masses.
Therefore, we apply the corrections to the comparison samples.
Specifically, we multiply $0.76$ ($0.58$) to the stellar mass fractions of the samples in L03, L12 and GZZ13 (Z11 and H13) to bring their stellar masses into the mass floor determined by the \citet{chabrier03} initial mass function.
To account for the systematic shifts in cluster masses, we also multiply a factor of $1.12$ ($0.96$) to the \Mfiveoo\ estimates in L03, L12, GZZ13 and the \XMMBCS\ samples (Z11, H13 and vdB14), resulting in another correction of a factor 
$1.04$ ($0.98$) to the \Mstar\ estimation due to the changing \Rfiveoo\  due to the updated \Mfiveoo.
For the XXL100 sample, we also multiply a factor of $0.96$ ($0.98$) to the \Mfiveoo\ (\Mstar) estimates based on their reported systematics in mass\footnote{In \citet{eckert16}, the ratio of the weak lensing mass to the one inferred by SZE is $\approx0.96$ based on their reported values.}.
We stress that the best-fit (grey) region is the fit only to the SPT sample and is extrapolating to the mass and redshift ranges sampled by the comparison samples.

As seen in Fig.~\ref{fig:xvpmstar_sr}, the SPT clusters are consistent with all the comparison samples in the context of mass and redshift trends---showing that (1) 
higher mass clusters have lower stellar mass fractions,
with the stellar mass fraction \fstar\ decreasing from $\approx3\percent$ at $\Mfiveoo\approx5\times10^{13}\Msun$ to $\approx0.5\percent$ at $\Mfiveoo\approx2\times10^{15}\Msun$ as ${\Mfiveoo}^{-0.20\pm0.12}$, and (2) the stellar mass at the 
typical cluster mass $4.8\times10^{14}\Msun$ does not vary with redshift to within the uncertainties such that the stellar mass fraction $\fstar$ is $\approx0.8\percent$ out to redshift $\redshift\approx1.3$.
The mass slope ($\Bstar=\numBstaronezsys$) of the SPT clusters is statistically consistent with L03 ($0.74\pm0.09$), Z11 ($0.61\pm0.09$), L12 ($0.71\pm0.04$), H13 ($1.11\pm0.4$), GZZ13 ($0.52\pm0.04$) and the \XMMBCS\ sample ($0.69\pm0.15$).
It is worth mentioning that our sample of SPT clusters uniformly samples the high-mass end in a wide redshift range $0.25\lesssim\redshift\lesssim1.25$, providing a direct constraint on the redshift trends out to $\redshift\approx1.25$, for which the best-fit redshift trend ($\Cstar=\numCstaronezsys$) is in good agreement with L12 ($-0.06\pm0.22$) and the \XMMBCS\ sample ($-0.04\pm0.47$) at the low-mass end.
We note that there are residual systematics in the normalization of \Mstar\ on the order of $\lesssim40\percent$ depending on the comparison samples.
However, this does not change the qualitative picture significantly, given the large intrinsic scatter and measurement uncertainties, which are comparable to the systematic uncertainties here.
Therefore, we conclude that the stellar mass shows a strong correlation with cluster mass as $\approx{\Mfiveoo}^{\numBstaronezsys}$ and no statistically significant redshift trend out to $\redshift\approx1.3$.

We compare the stellar mass fraction \fstar\ in the environment of galaxy clusters to the cosmic stellar mass fraction, which is inferred from the ratio of the stellar mass density in the field to the mean matter density.
Specifically, we use the evolution of the stellar mass densities (in comoving volume with the unit of $\Msun/\mathrm{Mpc}^3$) measured from the COSMOS/UltraVISTA survey \citep{muzzin13} at $0.2<\redshift<1.5$ together with the ones estimated at $\redshift<0.2$ \citep[from][]{cole01,bell03,baldry12}, and convert them into stellar mass fraction as a function of redshift by dividing by the matter density estimated by the cosmological parameters determined by \PLANCK\  \citep{planck16cosmological}.
Note that we linearly interpolate the stellar mass densities between the measurements at the adjacent redshift bins in \cite{cole01}, \cite{bell03}, \cite{baldry12} and \cite{muzzin13}.
We show the cosmic stellar mass fraction with yellow bars in Fig.~\ref{fig:xvpmstar_sr}, where the mass trend of the cosmic stellar mass fraction (in the left panel) is normalized at redshift $\ZPIV=0.6$ (same as the clusters), and the redshift trend (in the right panel) shows the evolution of the stellar mass fraction in the field.
As seen in the upper-left panel of Fig.~\ref{fig:xvpmstar_sr}, the stellar mass per unit halo mass in the cluster environment is significantly higher than the cosmic value at the characteristic redshift $\ZPIV=0.6$.
In the upper-right panel, the stellar mass fraction \fstar\ in the environment of galaxy clusters remains approximately constant with redshift, while the stellar mass per unit total matter in the field grows significantly by about an order of magnitude since redshift $\redshift\approx1.5$.
This clearly suggests that, based on the decreasing mass trend of the stellar mass fraction to halo mass relation without a significant redshift trend, massive clusters cannot form by simply accreting clusters with lower masses.   In such a scenario the stellar mass fraction \fstar\ of high mass clusters would be indistinguishable from low mass clusters.
Instead, a significant amount of infall from the lower density 
surrounding structures, which have substantially lower stellar mass fractions, must
contribute to the matter assembly of galaxy clusters such that the stellar mass fraction \fstar\ remains roughly constant 
at fixed mass over cosmic time.
That is, the infall from the surrounding environments must be in balance with the matter accretion from low mass galaxy clusters or groups to maintain the approximately constant stellar mass per hosting mass in the environment of clusters.
We return to this discussion in Section~\ref{sec:infall}.

\begin{figure*}
\centering
\vskip-0.10in
\resizebox{!}{0.5\textheight}{
\includegraphics[scale=1.0]{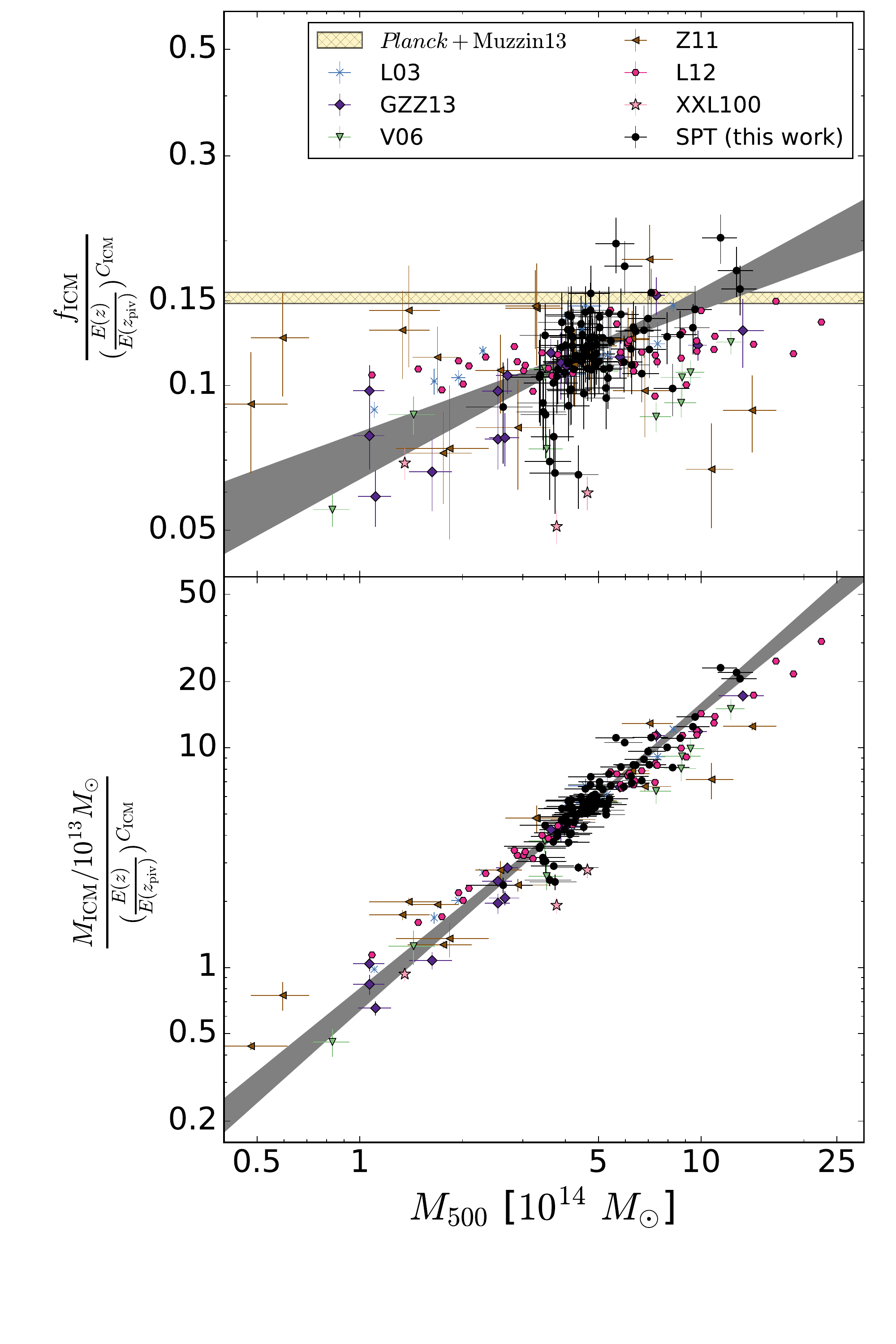}
}
\resizebox{!}{0.5\textheight}{
\includegraphics[scale=1.0]{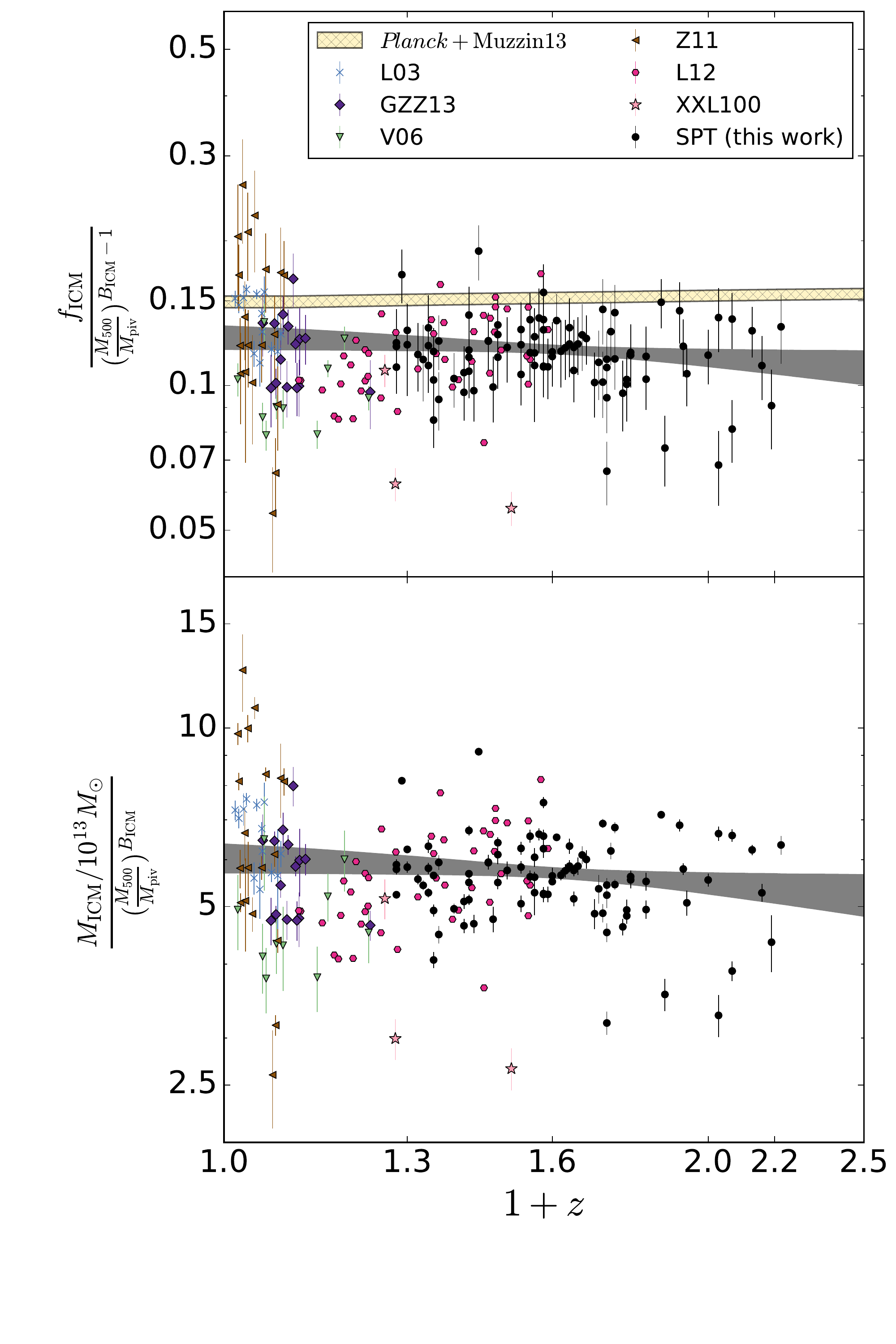}
}
\vskip-0.40in
\caption{
The scaling relations of the ICM mass \Mgas\ (below) and ICM mass fraction \fgas\ (above) to halo mass are based on 91 SPT clusters.
The SPT clusters and the comparison samples are plotted in the same manner as in Fig.~\ref{fig:xvpmstar_sr}, while the cosmic value, derived from the combination of the field stellar mass density \citep{muzzin13,cole01,bell03,baldry12} and the CMB cosmological constraints  \citep{planck16cosmological}, is indicated by the yellow bar.
Note that the yellow bars represent the cosmic mean gas fraction rather than the ICM fraction we expect within clusters. The grey areas indicate the fully marginalized $1\sigma$ confidence regions of the best-fit scaling relation extracted from the SPT clusters only.
}
\label{fig:xvpmgas_sr}
\end{figure*}

\subsection{ICM mass to halo mass relation}
\label{sec:icm_to_halo_mass_relation}

The ICM mass to halo mass scaling relation is obtained using the full sample of 91 clusters at redshift $0.25\lesssim\redshift\lesssim1.25$.
The best-fit parameters are shown in Table~\ref{tab:sr_params}, and the parameter covariances can be seen in Fig.~\ref{fig:triangles}.
We present the best-fit scaling relations and our measurements of \Mgas\ in Fig.~\ref{fig:xvpmgas_sr}, where we also show the results for the ICM mass fraction \fgas\ defined by
\[
\fgas \equiv \frac{\Mgas}{\Mfiveoo} \, .
\]
In Fig.~\ref{fig:xvpmgas_sr}, we normalize \Mgas\ and \fgas\ in the same way as in Section~\ref{sec:stellar_to_halo_mass_relation} to allow a clear picture of the mass and redshift trends.

The best-fit scaling relation is
\begin{eqnarray}
\label{eq:mgas_sr_oneplusz}
\Mgas &=& (\numAgasonezsys)\times 10^{13}\Msun \times \nonumber \\
&&\left( \frac{\Mfiveoo}{4.8\times10^{14}\Msun} \right)^{\numBgasonezsys}
\left( \frac{1 + \redshift}{1 + 0.6} \right)^{\numCgasonezsys} \, 
\end{eqnarray}
with log-normal intrinsic scatter of $\numDgasonezsys$.
The resulting mass and redshift trend parameters are $\Bgas = \numBgasonezsys$ and $\Cgas = \numCgasonezsys$, respectively, indicating a highly significant mass trend but a redshift trend that is statistically consistent with zero out to redshift $\redshift\approx1.25$. 
The best-fit normalization \Agas\ is $(\numAgasonezsys)\times10^{13}\Msun$, implying that the  
typical ICM mass fraction \fgas\ is $(12\pm1.3)\percent$ at the pivot mass $\MPIV\equiv4.8\times10^{14}\Msun$ and redshift $\ZPIV\equiv0.6$.

Similar to the stellar mass to halo mass scaling relation, we compare our results to those from previous studies.
We include \citet[][V06]{vikhlinin06}---where they studied the X-ray scaling relations of 13 relaxed clusters at low redshift $\redshift\lesssim0.3$---in this comparison.
To remove the known systematics raised from deriving cluster masses \Mfiveoo\ in different ways, we again multiply the halo masses by a factor of $1.12$ ($0.96$) in the samples of L03, V06, L12 and GZZ13 (Z11, XXL100), and this correspondingly results in a factor of $1.04$ ($0.98$) change to the \Mgas\ estimates due to the change in \Rfiveoo.
It has been demonstrated that the ICM mass determination is more robust as compared to other X-ray observables (e.g., temperature), and no strong systematics exist between values obtained using different X-ray telescopes \citep[e.g.,][]{martino14,schellenberger15}---therefore, we do not apply observatory based systematic corrections to the ICM mass estimations in the comparison samples.

We show the comparison samples in Fig.~\ref{fig:xvpmgas_sr}.
The mass trend parameter ($\Bgas = \numBgasez$) for the SPT clusters is statistically consistent with most comparison samples---Z11 ($1.38\pm0.36$), L12 ($1.13\pm0.03$), GZZ13 ($1.26\pm0.03$), and XXL100 ($1.21^{+0.11}_{-0.10}$) but is in some tension with another sample WtG16 ($1.04\pm0.05$) \citep{mantz16}.
As clearly seen in the left panel, the ICM mass \Mgas\ is a strong function of cluster mass \Mfiveoo\ increasing as $\propto{\Mfiveoo}^{\numBgasonezsys}$, which implies that the \fgas\ increases as $\appropto{\Mfiveoo}^{0.33\pm0.07}$ from $\approx7\percent$ at $\Mfiveoo\approx10^{14}\Msun$ to $\approx15\percent$ at $\Mfiveoo\approx10^{15}\Msun$.
The departure of the mass trend parameter for our SPT cluster sample from 1 (i.e. no mass trend, the self-similar expectation) is statistically significant at the $\gtrsim4.5\sigma$ level.
This mass-dependent ICM mass fraction \fgas\ was first noted in a homogeneous analysis of a large, local cluster sample by \cite{mohr99}, and was also observed later in various sizable samples \citep[e.g.,][]{vikhlinin06,vikhlinin09a}.  
Under the assumption that the temperature to mass relation is approximately self-similar ($T_{\mathrm{X}}\approx {\Mfiveoo}^{\frac{2}{3}}$), the constraints on the mass trend parameter of the \fgas\ to mass relation would be $\fgas\appropto{\Mfiveoo}^{0.32\pm0.08}$ \citep{mohr99} and $\fgas\appropto{\Mfiveoo}^{0.15\pm0.02}$ \citep{vikhlinin09a}\footnote{Note that we re-fit the data using the functional form of $\fgas \approx {\Mfiveoo}^{B}$ instead of quoting the original value obtained with the functional form of $\fgas \approx B \log(\Mfiveoo)$ used in \cite{vikhlinin09a}. In addition, we confirm that we can recover their mass slope using their functional form.}.  Both of these studies show inconsistency with the self-similar expectation (i.e. constant gas fraction) at $\gtrsim4\sigma$ significance.  
A mass dependent \fgas\ has long been suggested as the underlying cause of the non-self similar slopes of the luminosity--temperature \citep{david93,mushotzky97} and the X-ray isophotal size--temperature relations \citep{mohr97}.

Interestingly, as can be seen in the right panel, the ICM mass at the 
typical mass $\MPIV = 4.8\times10^{14}\Msun$ is $\approx(\numAgasonezsys)\times10^{13}\Msun$, and it shows no statistically significant redshift trend ($\propto (1+z)^{\numCgasonezsys}$) out to $\redshift\approx1.3$.
Our results obtained from the SPT clusters together with the comparison samples suggest that (1) the ICM mass inside the massive clusters shows strong mass-dependent behavior with the ICM mass fraction \fgas\ increasing with cluster mass (trend significant at $\gtrsim4.5\sigma$ level in current analysis), and (2) the ICM content of galaxy clusters (with $\Mfiveoo\gtrsim10^{14}\Msun$) has not changed significantly within \Rfiveoo\ since redshift $\redshift\approx1.25$.

Similarly to Section~\ref{sec:stellar_to_halo_mass_relation}, we also compare the ICM mass fraction \fgas\ of galaxy clusters to the cosmic value. 
To derive the cosmic value, we calculate the total baryon fraction (the baryonic mass per total mass) from the cosmological parameters determined by \PLANCK\  \citep{planck16cosmological}, and then subtract the cosmic stellar mass fraction (see Section~\ref{sec:stellar_to_halo_mass_relation}).
We show the cosmic value by the yellow bars in Fig.~\ref{fig:xvpmgas_sr} in the same manner as in Fig.~\ref{fig:xvpmstar_sr}.
As seen in Fig.~\ref{fig:xvpmgas_sr}, the ICM mass per total mass in galaxy clusters is a strong function of halo mass and is all significantly lower than the cosmic value ($\approx0.15$) for all but the most massive clusters over the full redshift range probed.
To have \fgas\ remain roughly constant with redshift, the balance between the infall from the surrounding environments and accretion of cluster and group scale subhalos must exist during cluster formation, which is qualitatively consistent with the picture implied by the stellar mass fraction (see Section~\ref{sec:stellar_to_halo_mass_relation}).
We will discuss this scenario in detail in Section~\ref{sec:infall}.

\begin{figure*}
\centering
\vskip-0.1in
\resizebox{!}{0.5\textheight}{
\includegraphics[scale=1.0]{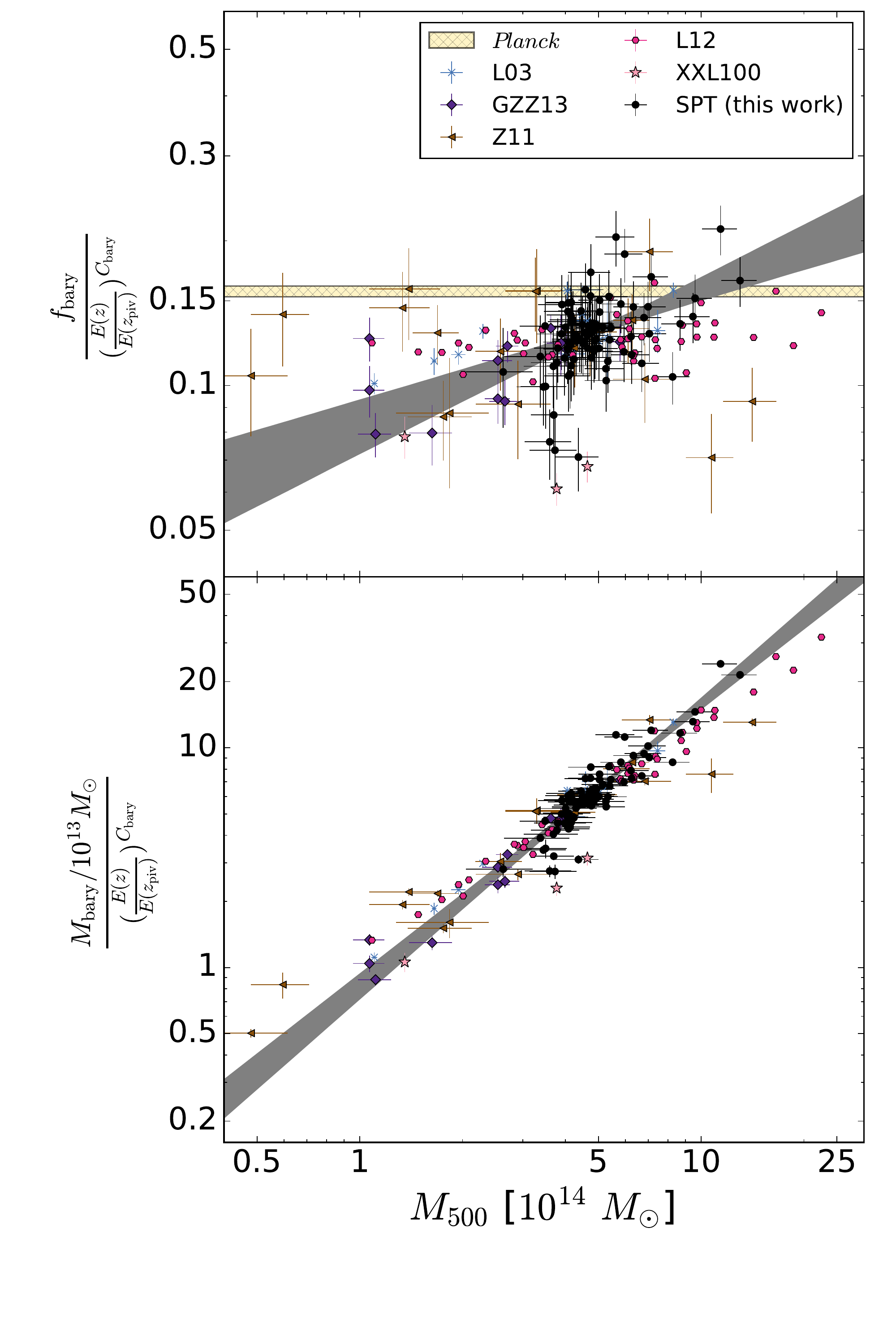}}
\resizebox{!}{0.5\textheight}{
\includegraphics[scale=1.0]{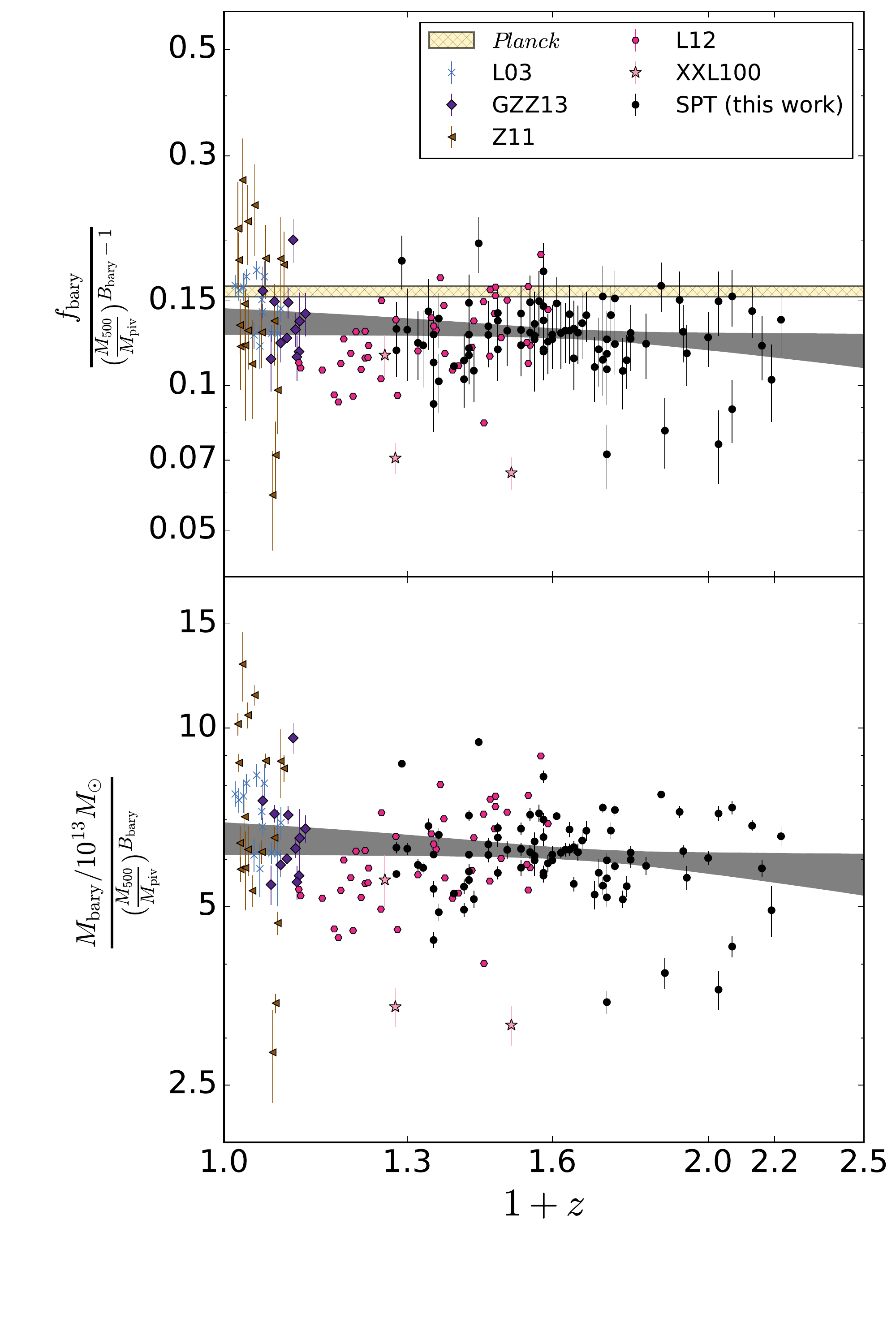}}
\vskip-0.4in
\caption{
The scaling relations of the baryonic mass \Mbary\ (below) and baryonic mass fraction \fbary\ (above) to halo mass are based on 84 SPT clusters.
The SPT clusters and the comparison samples are plotted in the same manner as in Fig.~\ref{fig:xvpmstar_sr}, while the cosmic value of the baryon fraction, derived from the CMB cosmological constraints \citep{planck16cosmological}, is indicated by the yellow bar.
The grey areas indicate the fully marginalized $1\sigma$ confidence regions of the best-fit scaling relation extracted from the SPT clusters only.
}
\label{fig:xvpmbary_sr}
\end{figure*}

\subsection{Baryonic mass to halo mass relation}
\label{sec:bary_to_halo_mass_relation}

The total baryonic mass \Mbary\ is estimated as the sum of the ICM and stellar masses ($\Mbary\equiv\Mgas + \Mstar$).
The baryonic mass to halo mass scaling relation is obtained using the subsample of 84 clusters with \Mstar\ measurements at redshift $0.25\lesssim\redshift\lesssim1.25$.
The best-fit parameters and their joint confidence constraints are in Table~\ref{tab:sr_params} and Fig.~\ref{fig:triangles}, respectively.
We present the best-fit scaling relation and our measurements of baryonic mass \Mbary\ and baryon fraction \fbary,
\[
\fbary \equiv \frac{\Mbary}{\Mfiveoo} \, ,
\] 
in Fig.~\ref{fig:xvpmbary_sr}, where we normalize \Mgas\ and \fgas\ in the same way as in Section~\ref{sec:stellar_to_halo_mass_relation} to disentangle the mass and redshift trends.

The best-fit scaling relation is
\begin{eqnarray}
\label{eq:mbary_sr_oneplusz}
\Mbary &=&(\numAbaryonezsys)\times10^{13}\Msun \times \nonumber \\
&&\left( \frac{\Mfiveoo}{4.8\times10^{14}\Msun} \right)^{\numBbaryonezsys}
\left( \frac{1 + \redshift}{1 + 0.6} \right)^{\numCbaryonezsys} \, 
\end{eqnarray}
with log-normal intrinsic scatter of $\numDbaryonezsys$. 
The resulting mass and redshift trend parameters are $\Bbary = \numBbaryonezsys$ and $\Cbary = \numCbaryonezsys$, respectively.
The best-fit normalization \Abary\ is $(\numAbaryonezsys)\times10^{13}\Msun$, suggesting that the 
typical total baryonic mass fraction \fbary\ is about $(12.8\pm1.29)\percent$ at the pivot mass $\MPIV\equiv4.8\times10^{14}\Msun$ and redshift $\ZPIV\equiv0.6$.
The general picture is the same as for the ICM mass to halo mass relation (see Section~\ref{sec:icm_to_halo_mass_relation}): adding the stellar mass to the ICM mass flattens the mass trend by $\approx0.5\sigma$ (from $\Bgas\approx1.32$ to $\Bbary\approx1.29$) and results in an increase in the normalization of $\approx9\percent$ (from $\Agas\approx5.7\times10^{13}\Msun$ to $\Abary\approx6.2\times10^{13}\Msun$).

After applying the corrections to remove the known systematics (see Section~\ref{sec:stellar_to_halo_mass_relation} and Section~\ref{sec:icm_to_halo_mass_relation}), we also compare our results to previous work (L03, Z11, L12, GZZ13).
The mass slope \Bbary\ of the SPT clusters is \numBbaryonezsys, which is statistically consistent with L03 ($1.148\pm0.04$), Z11 ($1.22\pm0.57$) and GZZ13 ($1.16\pm0.04$).
Again, no significant redshift trend is observed ($\Cbary=\numCbaryonezsys$, see the right panel in Fig.~\ref{fig:xvpmbary_sr}).
We stress that our SPT sample provides unique access to the baryon content of galaxy clusters out to redshift $\redshift\approx1.25$ and therefore constrains the redshift trends directly for the first time based on a uniformly selected, approximately mass-limited sample with homogeneously estimated masses.

We compare the baryon fractions of galaxy clusters to the cosmic baryon fraction, 
\begin{equation}
\frac{\Omega_{\mathrm{b}}}{\OmegaM} \approx 0.157 \pm 0.004 \, ,
\label{eq:cosmicbaryonfraction}
\end{equation}
determined by the cosmological parameters estimated by \PLANCK\  \citep{planck16cosmological} in Fig.~\ref{fig:xvpmbary_sr}.
The baryon fraction provides a cleaner measure in the context of matter accretion than stellar or ICM mass fractions, because the total mass of baryons per unit mass is expected to be invariant over cosmic time and is less subject to ICM cooling and star formation, which result in the interchange between cold and hot baryonic components.
We show the cosmic baryon fraction with the yellow bars in Fig.~\ref{fig:xvpmbary_sr} in the same manner as in previous subsections.
At the typical mass scale of $\MPIV=4.8\times10^{14}\Msun$ (see the upper-right panel of Fig.~\ref{fig:xvpmbary_sr}), the baryon fraction of clusters is constantly lower than the cosmic baryon fraction since the highest redshifts probed here $\redshift\approx1.25$.
On this mass scale, the baryon depletion factor is 
\begin{equation}
\label{eq:depletion_factor}
\mathcal{D} \equiv 1 - \frac{\fbary}{ \Omega_{\mathrm{b}}/\OmegaM } = 0.18\pm0.02.
\end{equation}
over the full redshift range we probe.  We return to a discussion of the depletion factor in Section~\ref{sec:sys_mass}.

A clear mass trend of total baryon fraction in galaxy clusters that is significantly lower than the cosmic fraction since redshift $\redshift\approx1.25$ again reinforces the picture that the infall from lower density regions outside collapsed halos must be in rough balance with the infall from material within the virial regions of lower mass galaxy clusters and groups. We will return to this discussion in Section~\ref{sec:infall}.

\subsection{Cold collapsed baryonic fraction to halo mass relation}
\label{sec:fcold_to_halo_mass_relation}

\begin{figure}
\hskip-0.1in
\resizebox{!}{0.45\textwidth}{
\includegraphics[scale=1.0]{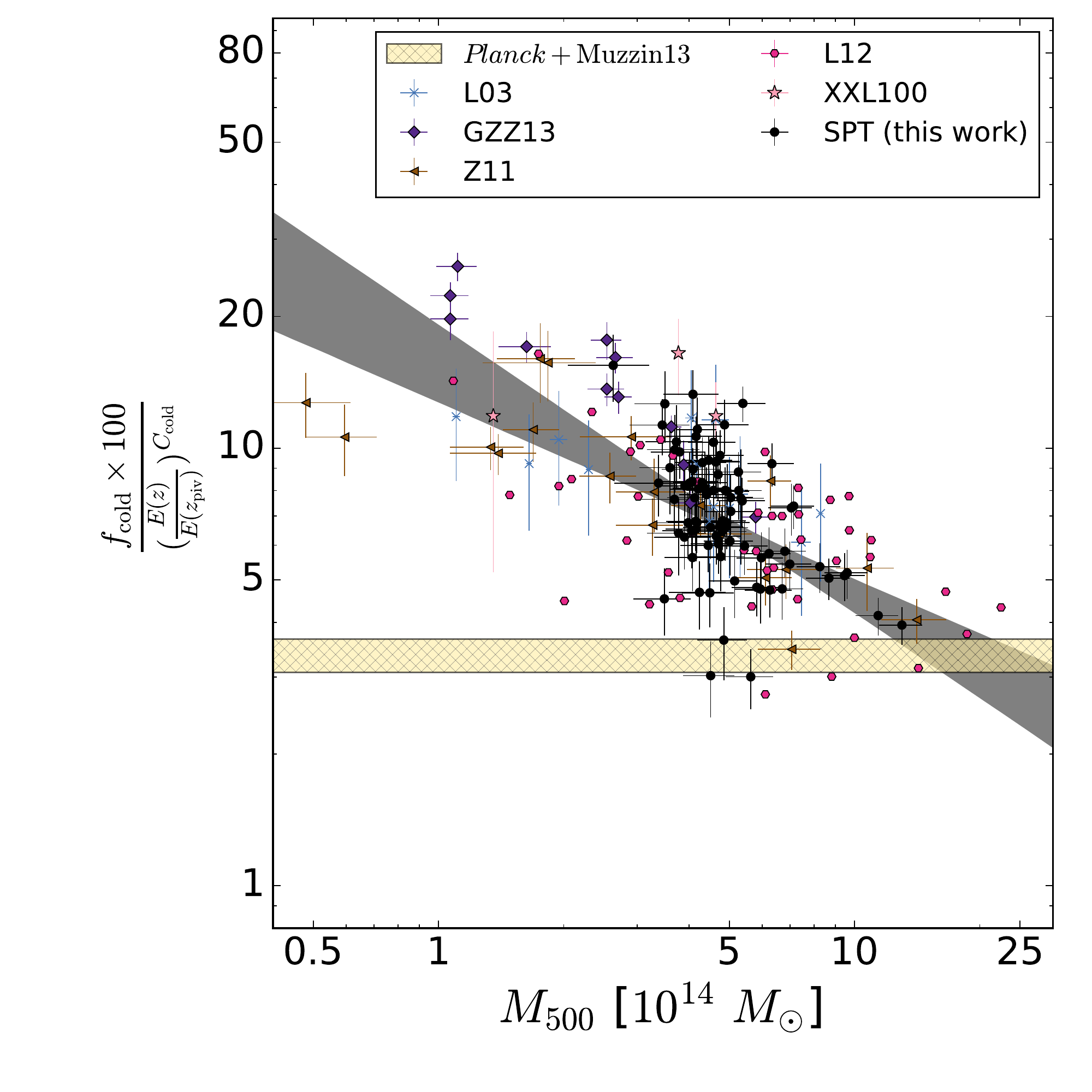}}
\resizebox{!}{0.45\textwidth}{
\includegraphics[scale=1.0]{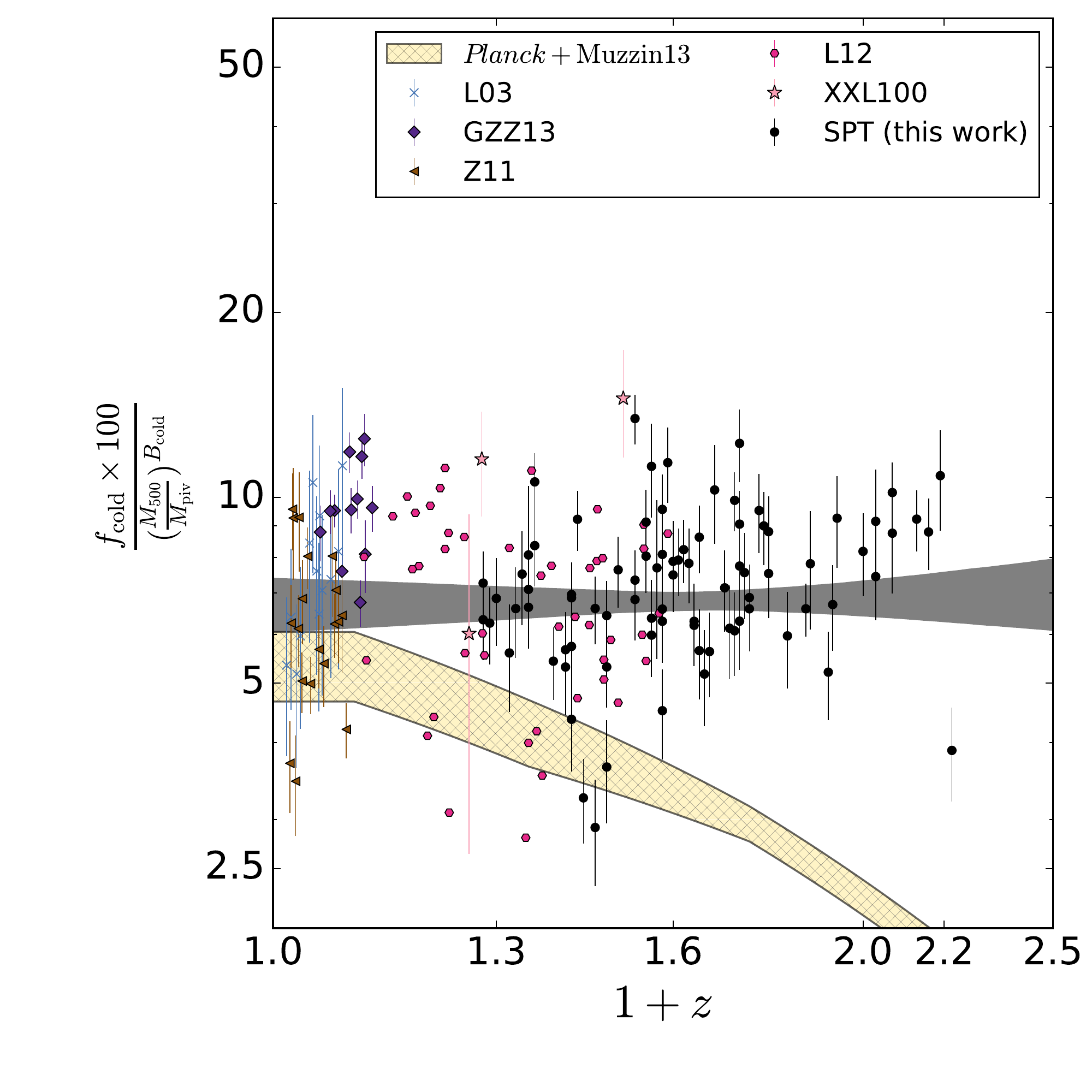}}
\vskip-0.2in
\caption{
The scaling relation of the cold baryonic fraction \fcold\ to halo mass is based on 84 SPT clusters.
The SPT clusters and the comparison samples are plotted in the same manner as in Fig.~\ref{fig:xvpmstar_sr}, while the cosmic value of the cold baryon fraction, derived from the combination of the field stellar mass density \citep{muzzin13,cole01,bell03,baldry12} and the CMB cosmological constraints  \citep{planck16cosmological}, is indicated by the yellow bar.
The grey areas indicate the fully marginalized $1\sigma$ confidence regions of the best-fit scaling relation extracted from the SPT clusters only.
}
\label{fig:xvpfcold_sr}
\end{figure}

The cold collapsed baryonic fraction (or cold fraction),
\[
\fcold \equiv \frac{\Mstar}{\Mbary} = \frac{\Mstar}{\Mgas + \Mstar} \, ,
\]
is a particularly interesting measurement because it involves only the ICM and stellar mass estimations, and is expected to be less subject to any systematics that impact the cluster halo mass \Mfiveoo.
Moreover, the cold fraction is defined as the fraction of baryonic mass in stars, which could also serve as a constraint on the integrated star formation rate inside galaxy clusters.
The cold baryonic fraction to halo mass scaling relation is obtained based on the subsample of 84 clusters at redshift $0.25\lesssim\redshift\lesssim1.25$ with \Mstar\ measurements.
The best-fit parameters,  the joint parameter constraints and the measurements are presented in Table~\ref{tab:sr_params}, Fig.~\ref{fig:triangles} and Fig.~\ref{fig:xvpfcold_sr}, respectively.
Again, we normalize \fcold\ in Fig.~\ref{fig:xvpfcold_sr} in the same way as in Section~\ref{sec:stellar_to_halo_mass_relation} to present the mass and redshift trends independently.

The resulting scaling relation is
\begin{eqnarray}
\label{eq:fcold_sr_oneplusz}
\fcold \times100 &=& \left(\numAcoldonezsys \right)\percent
\left( \frac{\Mfiveoo}{4.8\times10^{14}\Msun} \right)^{\numBcoldonezsys}  \nonumber \\
&&\left( \frac{1 + \redshift}{1 + 0.6} \right)^{\numCcoldonezsys} \, 
\end{eqnarray}
with the log-normal intrinsic scatter $\numDcoldonez$.
The best-fit mass and redshift trend parameters are$\Bcold = \numBcoldonezsys$ and $\Ccold = \numCcoldonezsys$, respectively.
The normalization \Acold\ indicates that the fraction of baryonic mass in stars is (\numAcoldonezsys)\percent\ at the pivot mass $4.8\times10^{14}\Msun$ and redshift $\ZPIV = 0.6$. 
The cold fraction is a strong function of mass, falling with $\approx4.5\sigma$ statistical significance such that at the pivot redshift $\ZPIV = 0.6$, \fcold\ decreases from 
$\approx10\percent$ to $\lesssim3\percent$ as the cluster halo mass increases by a factor of $\approx10$ from $\Mfiveoo\approx2\times10^{14}\Msun$ to $\Mfiveoo\gtrsim2\times10^{15}\Msun$.
The cold fraction shows no statistically significant evidence of a redshift trend.

We compare our results to previous work in Fig.~\ref{fig:xvpfcold_sr}.
The results of the SPT clusters are in good agreement with previous work\footnote{We derive the mass slopes of the comparison samples based on their reported mass slopes of \Bbary\ and \Bstar\ (i.e., $\Bcold\approx\Bstar - \Bbary$)}: the mass slope ($\Bcold=\numBcoldonez$) is statistically consistent with L03 and L12 ($-0.41\pm0.10$), Z11 ($-0.61\pm0.58$) and GZZ13 ($-0.64\pm0.06$). 
As seen in Fig.~\ref{fig:xvpfcold_sr}, \fcold\ for the SPT clusters behaves in a manner that is consistent with the comparison samples that extend to lower mass.  That is, the cold fraction decreases from $\approx20\percent$ at $\Mfiveoo\approx10^{14}\Msun$ to $\lesssim3\percent$ at $\Mfiveoo\gtrsim2\times10^{15}\Msun$.
In addition, this cold fraction has little dependence on cluster redshift.
This suggests that cold fractions of galaxy clusters have been well-established since redshift $\redshift\approx1.25$ and are determined primarily by the host halo mass.  
In addition, the integrated star formation rates for the components that make up massive galaxy clusters are significantly suppressed relative to those for the components of lower mass galaxy clusters.

\begin{figure}
\centering
\resizebox{0.3\textheight}{!}{
\includegraphics[scale=1.0]{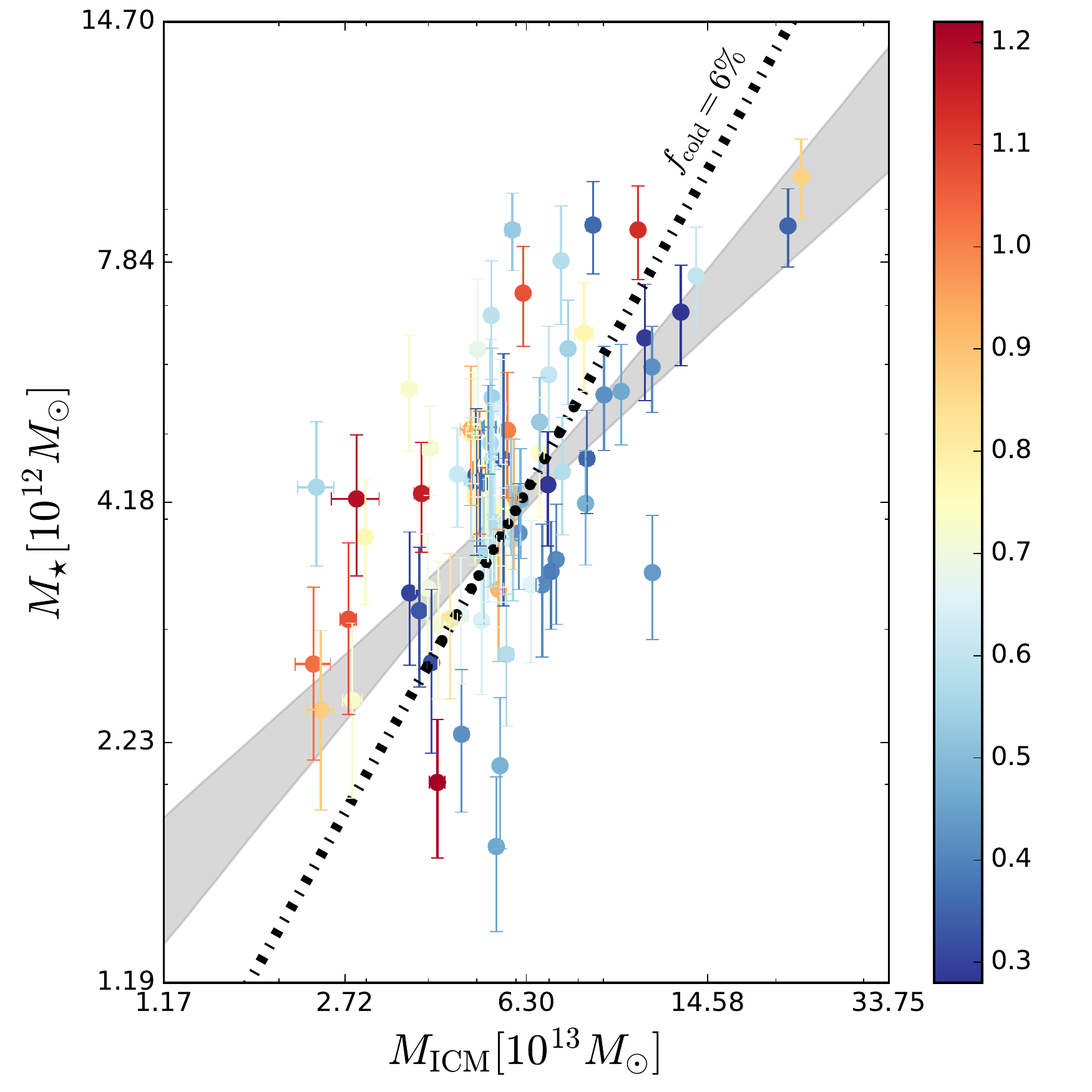}
}
\caption{
The relation between the stellar mass \Mstar\ and the ICM mass \Mgas\ among the 84 clusters.
This relation is approximated as 
$\Mstar = 4\times10^{12}\Msun\times\left(  \frac{\Mgas}{5.7\times10^{13}\Msun}\right)^{0.60}$ by assuming no redshift trends for \Mstar-to-\Mfiveoo\ and \Mgas-to-\Mfiveoo\ relations.
Each cluster is color-coded by the redshift, while the confidence levels (grey area) of the \Mstar-to-\Mgas\ relation are fully marginalized and extracted from the MCMC chains of the \Mstar-to-\Mfiveoo\ and \Mgas-to-\Mfiveoo\ relations.
The dashed line indicates the cold baryon fraction of $6\percent$ at the pivot mass and redshift.
}
\label{fig:mstar_mgas}
\end{figure}

This picture is further illustrated in Fig.~\ref{fig:mstar_mgas}, where we show the stellar mass as a function of ICM mass.
In addition, we also show the constant cold baryon fraction at $6\percent$ as the dashed line in Fig~\ref{fig:mstar_mgas}.
We derive the stellar mass to ICM mass relation directly from equations~(\ref{eq:mstar_sr_oneplusz}) and (\ref{eq:mgas_sr_oneplusz}), assuming that the there is no redshift trend.  This relation can be approximated as
\begin{equation}
\label{eq:mstar_mgas}
\Mstar = 4\times10^{12}\Msun\times\left(  \frac{\Mgas}{5.7\times10^{13}\Msun}\right)^{0.60\pm0.10}
\end{equation}
We also further extract the fully marginalized 1$\sigma$ confidence region for equation~(\ref{eq:mstar_mgas}) directly from the MCMC chains from fitting the stellar mass to halo mass (equation~(\ref{eq:mstar_sr_oneplusz})) and ICM mass to halo mass (equation~(\ref{eq:mgas_sr_oneplusz})) relations;  this is indicated as the grey area in Fig.~\ref{fig:mstar_mgas}.
The tilt of the stellar mass to ICM mass relation from a constant cold baryon fraction reflects the presence of the mass-dependent integrated star formation rate.
It is worth mentioning that this result may be an indication that stellar mass growth is strongly affected by merger-triggered star formation in clusters of galaxies, resulting in an inverse halo mass dependence \citep[e.g.,][]{brodwin13}.

We also compare the cold fraction of galaxy clusters to the cosmic value inferred from the stellar mass and the total baryon fraction of the field.
We obtain the cosmic cold fraction by dividing the stellar mass fraction obtained from the field luminosity function by the total baryon fraction for the Universe.  This is shown with the yellow bars in Fig.~\ref{fig:xvpfcold_sr}.
In the upper panel, the cosmic cold fraction is at $\approx3.8\pm0.7\percent$ at the characteristic redshift $\ZPIV=0.6$, which is significantly lower than the cold fraction of galaxy clusters across the mass range ($\Mfiveoo\lesssim10^{15}\Msun$).
In the lower panel, the cosmic cold fraction grows significantly from 
$\lesssim1\percent$ at $\redshift\gtrsim0.7$ to $\approx5.5\percent$ at $\redshift\approx0$, 
which follows from the significant growth of the cosmic stellar mass fraction seen as the yellow bar in the upper-right panel of Fig.~\ref{fig:xvpmstar_sr}.
Therefore, the same scenario where there is significant infall from the field for massive clusters is also supported by the results from the cold baryon fraction analysis.

\begin{figure}
\vskip-0.20in
\centering
\resizebox{0.5\textwidth}{!}{
\includegraphics[scale=1.0]{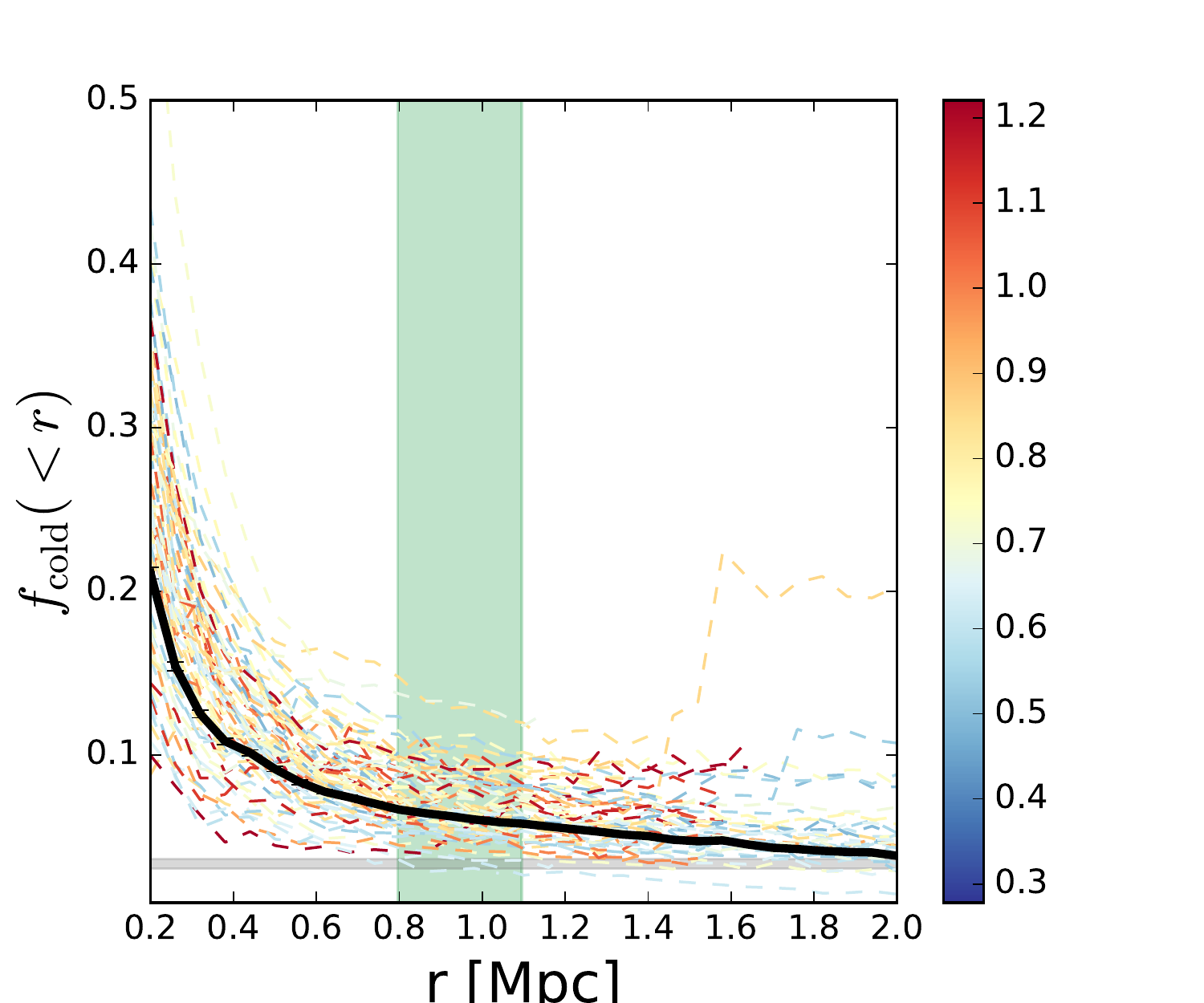}
}
\vskip-0.1in
\caption{
The average cold baryon fraction $\fcold(<r)$ within a radius $r$.
Individual systems are shown with dashed-lines that are color-coded by the redshift according to the color bar on the right, while the mean of these clusters is shown in black.
The average cold baryon fraction $\fcold(<r)$ falls with increasing radius, but shows no redshift dependence.
The mean \Rfiveoo\ (with the RMS of the sample) is plotted as the green area, centered at $\approx0.9$~Mpc.
The grey, horizontal bar indicates the cosmic value of the cold baryon fraction at the pivot redshift $\ZPIV = 0.6$.
}
\label{fig:fcold_rmpc}
\end{figure}

Another interesting measurement is the cold fraction as a function of radius within the cluster, because it is less subject to the systematics associated with the overall halo mass scale.
To make this measurement, we repeat the analysis in Sections~\ref{sec:icmmass} and \ref{sec:stellarmass},
i.e. the determination of ICM and stellar masses, 
for radii between $0.2$~Mpc and $2$~Mpc with a step of $50$~kpc.  The enclosed cold fraction $\fcold(<r)$ as a function radius appears in Fig.~\ref{fig:fcold_rmpc}. 
In Fig.~\ref{fig:fcold_rmpc}, individual cluster profiles are color-coded according to their redshift, and we only use measurements with the masking correction $f_{\mathrm{mask}} < 1.2$. 
The stacked $\fcold(<r)$ profile of all clusters is shown as the black line, while the typical \Rfiveoo\ with its standard deviation is independently plotted as the green region.
We can see that the \fcold\ is clearly a function of radius that monotonically decreases from $\approx20\percent$ in the central regions ($r<0.2$~Mpc) to the $\lesssim6\percent$ beyond $\approx1$~Mpc, which is typically the scale of \Rfiveoo\ for the SPT clusters.

\section{Discussion}
\label{sec:discussion}

In this section we discuss a toy model that provides a scenario by which the behavior we have presented could emerge from hierarchical structure formation.  Thereafter, we explore the impact of adopting a Planck CMB anisotropy based cosmology prior. 

\subsection{Sources of Accretion onto Halos}
\label{sec:infall}

In Section~\ref{sec:scalingrelation}, we have presented the baryon content of galaxy clusters from a large, approximately mass-limited cluster sample ($\Mfiveoo\gtrsim4\times10^{14}\Msun$) and the variations of these baryonic quantities with halo mass and redshift.  Extrapolations of the halo mass trends from our sample to lower masses appear to be in good agreement with results from previous studies that have focused on clusters and groups at lower mass scales ($\Mfiveoo\gtrsim5\times10^{13}\Msun$).
The emerging trends are that all baryonic components of clusters show strong, non-self similar dependences on cluster halo mass but that there is no statistically significant evidence of redshift trends.  The models we fit to the current dataset assume redshift and mass trends are separable, and indeed our data provide no evidence of tension with this model.
 
This result is quite remarkable.  Within the radius \Rfiveoo\ 
the observations we present suggest that the stellar and ICM mass components of galaxy clusters 
scale as a simple power law with the halo mass \Mfiveoo, and this scaling has not changed significantly since redshift $\redshift\approx1.25$.
Given that our halo, ICM and stellar masses are measured homogeneously over the full redshift range, this result cannot easily be attributed to systematic differences in the measurements at low and high redshift.  We caution that we can only constrain the redshift trends to within the stated measurement uncertainties, which are still considerable; with large, future samples the redshift and mass trends can be constrained more precisely, and presumably we will be able to resolve any underlying redshift trends in these baryonic quantities. 
 
As is clear in Fig.~\ref{fig:xvpmgas_sr}, the universal ICM mass fraction is higher than that of all but the most massive clusters.  That implies an increasing ICM mass fraction as one moves outside the virial regions of these systems and into their infall regions.  The rough constancy of ICM mass fraction at fixed mass with cosmic time 
(as seen in the upper-right panel of Fig.~\ref{fig:xvpmgas_sr}) can then be used to provide a constraint on the combination of (1) entropy injection from AGN, star formation and other sources together with (2) the mix of material falling in from subclusters and that falling in from the low density regions surrounding the parent halo 

The same picture of mass trends and weak redshift trends is also suggested by the stellar mass fraction \fstar, although in this case the trends with mass are in the opposite sense: the stellar mass fraction falls with halo mass on cluster scales.  One cannot construct massive clusters through the accretion of lower mass subclusters in this case, because as the main cluster grows in mass the stellar mass fraction must fall, and it is not possible to effect this change with lower mass subclusters that have larger stellar mass fractions.  In the case of the stellar mass fraction, entropy injection doesn't impact the mass or redshift trend.  Thus, we can conclude that material accreted from the lower density infall regions surrounding the parent cluster is having a balancing effect that keeps the stellar mass fraction roughly constant over cosmic time.  

Similar arguments can be made for the baryon fraction \fbary\ and the cold baryon fraction \fcold, although in both those cases the ICM plays a major role, and so entropy injection and infall from surrounding regions can both contribute to the observed mass and redshift trends.   

We have discussed the impact of infall from the low density regions or ``field'' for structure formation on cluster scales already in \cite{chiu16a} in a study of 14 SPT clusters in combination with a heterogeneous set of results from the literature, and a similar picture emerged in \cite{chiu16c} where we studied the stellar mass fractions of a larger sample of 46  low mass galaxy groups uniformly selected in the XMM-BCS X-ray survey \citep{suhada12,desai12} at redshifts $0.1\lesssim\redshift\lesssim1$.
The significant improvements of our current analysis are that---for the first time---(1) we conduct a study of an approximately ``mass-limited'' sample with a much larger size ($91$ clusters) that has been uniformly selected by the SZE signatures of the clusters over a wide redshift range $0.25\lesssim\redshift\lesssim1.25$, (2) for each cluster, we study the full baryonic mass (the ICM and stellar content) and the halo mass \Mfiveoo\ using consistent methods on uniform multi-wavelength datasets (optical, NIR, X-ray and mm-wavelength), which dramatically reduces the systematic uncertainties that were present in the previous work, and (3) we derive the scaling relations in a Bayesian framework that fully accounts for various selection biases, especially those associated with the intrinsic scatter of the SZE $\zeta$ to mass relation that we use to infer cluster mass \Mfiveoo.
With increased statistical power and reduced systematic uncertainties, we are in a position to more quantitatively constrain the contributions to cluster growth from lower density regions that lie outside the 
virial region, a region that is often referred to as the field.

\begin{figure}
\vskip-0.20in
\centering
\resizebox{0.5\textwidth}{!}{
\includegraphics[scale=0.8]{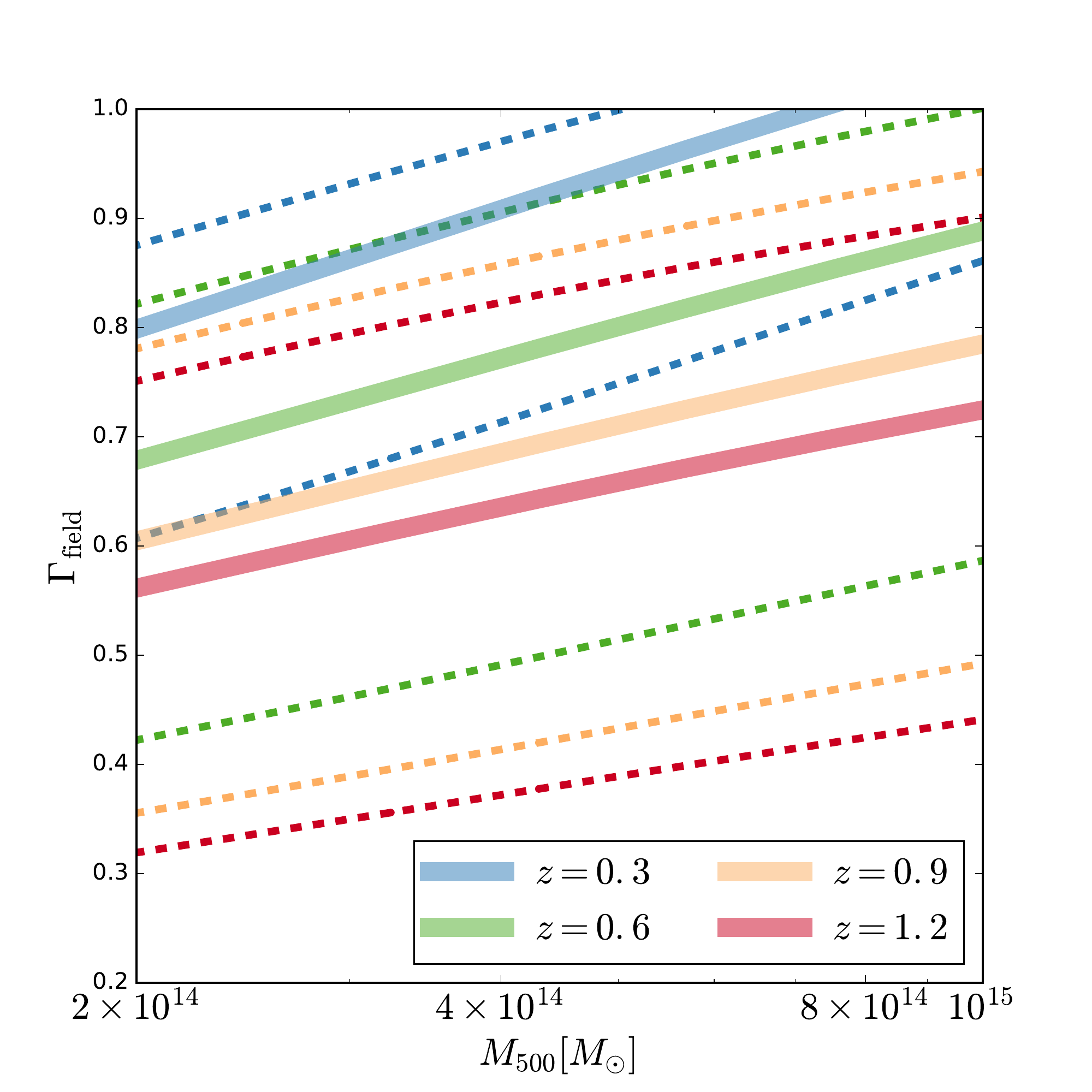}
}
\vskip-0.1in
\caption{
Estimates of the fraction $\Gamma_{\mathrm{fld}}$ of the cluster halo mass contributed by material outside the radius \Rfiveoo\ or in halos of mass 
$\Mfiveoo<10^{13}\Msun$ as a function of the cluster mass. 
Results are shown for redshifts $\redshift=0, 0.3, 0.6, 1.2$, which are color-coded as shown in the lower-right corner.
The associated $1\sigma$ lower and upper bounds are presented by the dashed lines. 
These field infall fractions are estimated from the mass and redshift trends of the stellar mass to halo mass scaling relation together with a simple infall model discussed in Section~\ref{sec:infall} and Appendix~\ref{sec:infall_equation}. 
}
\label{fig:infall}
\end{figure}

We construct a toy model to help us explore the infall of a variety of components with different properties in the Appendix~\ref{sec:infall_equation}.
In short, assuming that the stellar mass to halo mass and redshift relation provides a good description for the stellar mass inside the dense virial regions of halos (within \Rfiveoo), the mass trend parameter \Bstar\ of the stellar mass to halo mass relation provides a constraint on the fractional contributions of infalling galaxies from different sources.
As an example, if we adopt a model where there is infall from only two sources---the low density surrounding infall region of the parent halo and the high density virial regions of infalling subhalos of mass \Mfiveoosub, denoted by subscripts $\mathrm{fld}$ and $\mathrm{sub}$, respectively---we have 
\begin{equation}
\label{eq:infall}
\Bstar = 
\Gamma_{\mathrm{fld}}
\frac{ f_{\mathrm{fld}}(\redshift) }{  
\fstar(\Mfiveoo, \redshift) } + 
\left(1 - \Gamma_{\mathrm{fld}} \right)
\frac{f_\mathrm{sub}(\Mfiveoosub,\redshift) }{  
\fstar(\Mfiveoo, \redshift) }
\, ,
\end{equation}
where $\Gamma_{\mathrm{fld}} = \Gamma_{\mathrm{fld}}(\Mfiveoo, \redshift)$
is the fraction of the total infall from the low density region with the stellar mass fraction $f_{\mathrm{fld}}$, the subhalos make up the rest of the infall with stellar mass fraction $f_\mathrm{sub}$, and the cluster being accreted upon has the stellar mass fraction $\fstar(\Mfiveoo, \redshift)$ which we have measured as a function of mass and redshift.

For a given redshift \redshift, we can approximate the characteristic stellar mass fraction $\left<f_\mathrm{sub}(<\Mfiveoo,\redshift)\right>$ of the subhalos that fall into a cluster of mass \Mfiveoo\ by integrating the stellar mass fraction to mass relation over the mass function of halos $n({\Mfiveoosub}, \redshift)$ with masses less than the mass of the target halo:
\begin{equation}
\label{eq:weighted_fbary}
\left<f_\mathrm{sub}(<\Mfiveoo,\redshift)\right> = \frac{ 
\int\dif {\Mfiveoosub}\, \fstar({\Mfiveoosub}, \redshift) \Mfiveoosub n({\Mfiveoosub}, \redshift) 
}{
\int \dif {\Mfiveoosub}\, \Mfiveoosub n({\Mfiveoosub}, \redshift) 
} \,  .
\end{equation}
In equation~(\ref{eq:weighted_fbary}), we integrate ${\Mfiveoosub}$ from $10^{13}\Msun$
to half the mass of the parent halo $\Mfiveoo/2$ and use our measured stellar mass to halo mass relation over this whole range.  This gives mean stellar mass fractions of 
$\left<f_\mathrm{sub}(<\Mfiveoo,\redshift)\right>\approx1.4\percent$ for a cluster with $\Mfiveoo=6\times10^{14}\Msun$ over the redshift range of interest.
With $f_{\mathrm{fld}}(\redshift)$ and $\left<f_\mathrm{sub}(<\Mfiveoo,\redshift)\right>$, we can constrain the fraction of the material  $\Gamma_\mathrm{fld}$ in a halo of mass \Mfiveoo\ that originated outside the virial regions of halos with masses $10^{13}\Msun$ and greater:
\begin{equation}
\label{eq:gamma}
\Gamma_{\mathrm{fld}}(\Mfiveoo, \redshift) = \frac{
\fstar(\Mfiveoo, \redshift) \Bstar - \left<f_\mathrm{sub}(<\Mfiveoo,\redshift)\right> 
}
{
f_\mathrm{fld}(\redshift) - \left<f_\mathrm{sub}(<\Mfiveoo,\redshift)\right>
} \,  .
\end{equation}
This toy model ignores the timescale over which a cluster forms, using the measured stellar mass fractions for the subhalos and the field at redshift \redshift\ to estimate  $\Gamma_\mathrm{fld}$.     In Fig.~\ref{fig:infall} we plot the behavior of the low density infall fraction as a function of mass for clusters at four different redshifts.   The character of the solution is as expected---that the fraction of mass contributed from the low density regions surrounding a cluster grows with cluster mass.  At $\redshift=0$ the low density infall fraction $\Gamma_{\mathrm{fld}}$ reaches $100\percent$.  At this mass and redshift, the cluster stellar mass fraction falls below the field stellar mass fraction, and within the toy model it is no longer possible to form clusters of such low stellar mass fraction.  Accounting for the timescale over which clusters form, including an estimate of the intracluster light fraction (ICL) that could vary with mass and allowing for uncertainties both in our cluster stellar mass fractions and the field stellar mass fractions would all provide flexibility to explain the low stellar mass fractions of the highest mass clusters at redshift $\redshift=0$.  
It is also worth mentioning that we assume no in-situ star formation in clusters in this simple toy model.
A more accurate exploration of this scenario would require hydrodynamical simulations that include star formation and AGN feedback at resolutions adequate to form and maintain galaxy populations.

It is interesting to mention the recent result from \cite{lin17}, where they studied the growth of stellar mass in galaxy clusters by pairing the progenitors and descendants from $\redshift\approx1$ to $\redshift\approx0.3$.
They found a redshift-independent correlation of $\Mstar\propto\Mfiveoo^{0.7}$ that is in good agreement with our results, suggesting that the scaling derived in this work is also a good description for the evolution of baryon content from progenitors to descendants of galaxy clusters.

To conclude, our measured mass trend parameter in the stellar mass to halo mass relation together with a lack of strong redshift evolution implies a cluster mass and redshift dependent infall from the regions outside the \Rfiveoo\ regions of massive halos during cluster formation.  This trend is qualitatively similar to that noted in simulations by \citet{mcgee09}.  Moreover, if we use the ICM mass or baryon mass to halo mass and redshift relations within this simple model we infer a much steeper mass dependence in the infall fraction $\Gamma_\mathrm{fld}$ than in the case of the stellar mass.  This is as expected, because a strong trend of increasing ICM mass fraction (or baryon fraction) with mass could also be impacted by entropy injection by AGN, star formation or other sources.  Therefore, the departures from self-similarity in the ICM mass (and baryon mass) to halo mass relation is presumably due to a combination of entropy injection and the mass dependent infall fraction from the lower density areas surrounding the clusters.  
Thus, 
the measured mass trends without strong redshift scaling
in, for example, the baryon fraction, stellar mass fraction or ICM mass fraction provides evidence for the presence of the so-called ``missing'' baryonic mass in the Universe.

\subsection{Impact of Adopting a \PLANCK\ CMB Cosmology Prior}
\label{sec:planckpriors}

As described in Section~\ref{sec:totalmass}, we estimate the cluster mass \Mfiveoo\ using the SZE observable $\xi$ and redshift together with the best fit SZE observable to mass scaling relation from a previous analysis \citep{dehaan16} that used the SPT cluster counts together with $Y_{\mathrm{X}}$ measurements and an externally calibrated $Y_{\mathrm{X}}$ to mass scaling relation.  As noted in \cite{bocquet15}, the SZE observable to mass scaling relation parameters can shift when including the cosmology priors from the \PLANCK\ CMB anisotropy dataset.  As quantified in \cite{dehaan16}, analysis of the \PLANCK\ CMB measurements and the SPT cluster dataset within a flat \LCDM\ cosmology gives SZE observable to mass scaling relation parameters:
\[
r_{\mathrm{SZ}} = \left( 3.53, 1.66, 0.73, 0.20 \right) \, .
\]
These parameters imply cluster masses \Mfiveoo\ that are typically $22\percent$ larger in comparison to the default masses used in this work.

To quantify the impact from the \PLANCK\ CMB priors on our cluster halo mass calibration, we redo the analysis with these new halo masses, because \Mfiveoo\ determines \Rfiveoo\ ($22\percent$ in mass corresponds to $7\percent$ in radius), which is the radius within which we must extract the ICM and stellar masses.   The resulting scaling relations with new masses suggest that the normalizations at $\Mfiveoo=4.8\times10^{14}\Msun$ of \Astar, \Agas\ and \Abary\ decrease by $\approx12\percent$, $\approx23\percent$ and $\approx23\percent$, respectively---therefore---resulting in an increase of \Acold\ by $\approx12\percent$.  Apart from the normalization, the mass trend, redshift trend and intrinsic scatter are all consistent with the previous results within the quoted uncertainties and are therefore not sensitive to the $22\percent$ shift in halo mass.

Interestingly, the baryon fraction of galaxy clusters with the new \Mfiveoo\ estimates at the median mass and redshift becomes $\fbary\approx\left(10.8\pm0.2\right)\percent$.  That is, after including the \PLANCK\ CMB constraints in the cluster mass calibration, the depletion factor of our sample (at their median mass \MPIV) is shifted  from $0.18\pm0.02$ to
\[
\mathcal{D} = 0.30\pm0.03 \, .
\]
This shift is statistically significant and is strongly driven by the change of the absolute mass scale of \Mfiveoo\ due to the inclusion of the \PLANCK\ CMB constraints.  Various simulation based studies indicate that the depletion factor within \Rfiveoo\ ranges from $\approx10\percent$ to $\approx20\percent$ \citep[e.g.,][]{mccarthy11,planelles12,barnes17}, which is in good agreement with our baseline result $\mathcal{D} = 0.18\pm0.02$, but is strongly inconsistent with our result when including the \PLANCK\ CMB based cosmology prior. This result implies that the cosmological framework anchored by the \PLANCK\ CMB constraints predicts cluster masses that are too high to be consistent with simulations.  As already mentioned, we disfavor the adoption of the \PLANCK\ cosmology prior, because directly measured weak gravitational lensing \citep{gruen14,dehaan16,dietrich17,stern18} and dynamical mass measurements \citep{bocquet15,capasso17} are in better agreement with the masses emerging from the standard SPT cluster analysis results from \citet{dehaan16} employed in this paper.

%
%

\section{Conclusions}
\label{sec:xvp_conclusion}

We estimate the halo mass \Mfiveoo\ and the ICM mass \Mgas\ using the SZE observable and X-ray observations, respectively, of 91 SZE selected galaxy clusters from the SPT-SZ survey.  In addition, we derive the stellar mass \Mstar\ of 84 of those clusters using SED fitting of the galaxies with $griz$ optical imaging from DES combined with NIR photometry either from \WISE\ (\Wone\Wtwo) or from \Spitzer\ (\IRACone\IRACtwo).  This sample spans a mass range from $\approx2.5\times10^{14}\Msun$ to $\approx1.3\times10^{15}\Msun$ (with median of $4.8\times10^{14}\Msun$) and a redshift range from $\redshift=0.278$ to $\redshift=1.22$ (median $\redshift=0.58$).
The well understood selection, the sample size and uniform analysis applied over a wide redshift range combine to allow us to study the baryon content of galaxy clusters and trends with mass and redshift with lower systematic uncertainties than previous work.  All measurements are extracted at a consistent radius \Rfiveoo, allowing us to self-consistently compare the baryon components within the same portion of the virial region, regardless of cluster mass or redshift.

We use these measurements to study four different observable--halo mass--redshift scaling relations, where the four observables are the stellar mass, the ICM mass, the baryonic mass and the cold baryon fraction.  Our fits are carried out using a Bayesian framework that includes selection effects and accounts for both measurement uncertainties and intrinsic scatter.  Fitting for these scaling relations allows us to constrain the variation of the stellar mass fraction, ICM mass fraction and baryonic fraction as a function of halo mass and redshift.  

Our results indicate that the baryon content within \Rfiveoo\ strongly depends on the cluster halo mass such that the stellar mass, ICM mass, baryonic mass and the cold fraction scale with \Mfiveoo\ as ${\Mfiveoo}^{\numBstaronezsys}$, ${\Mfiveoo}^{\numBgasonezsys}$, ${\Mfiveoo}^{\numBbaryonezsys}$ and ${\Mfiveoo}^{\numBcoldonezsys}$, respectively.
On the other hand, we do not observe significant redshift trends in any of these scaling relations, where if the observable at a fixed mass scales as $(1+z)^\gamma$ we measure $\gamma=\numCstaronezsys$, $\numCgasonezsys$, $\numCbaryonezsys$ and $\numCcoldonezsys$ for stars, ICM, baryons and cold fractions, respectively.

Our observations imply that at the 
typical mass $\MPIV = 4.8\times10^{14}\Msun$ and redshift $\ZPIV=0.6$ of our sample, the stellar mass fraction, ICM mass fraction, baryonic mass fraction and the cold baryon fraction are  
\fstar=$(0.83\pm0.06)\percent$,
$\fgas=(12.0\pm1.3)\percent$,  
$\fbary=(12.8\pm1.29)\percent$ and
$\fcold=(\numAcoldonezsys)$\percent. 
The logarithmic intrinsic scatters in these observables at fixed mass are found to be
$\numDstaronezsys$,
$\numDgasonezsys$,
$\numDbaryonezsys$ and
$\numDcoldonezsys$, respectively.
We carefully quantify the systematic uncertainties associated with the determination of the halo masses and find that these are too small to affect our mass or redshift trends in an important way.  

Our measurements are in good agreement with previous work in the mass and redshift ranges where they overlap.  
The extrapolation of our measured scaling relations to lower mass is broadly consistent with measurements available in the literature.  Our results demonstrate that the baryon content of galaxy clusters is determined by the halo mass and has not changed much in the past 9~Gyr of cosmic evolution.

We also compare our results to the cosmic values estimated from the cosmological parameters \citep{planck16cosmological} and the analysis of the stellar mass function in the field from the COSMOS/UltraVISTA survey \citep{muzzin13}.
We find that the baryon content of clusters deviates significantly from these cosmic values, with stellar mass fractions (ICM mass fractions) being larger than (smaller than) the cosmic values for all but the most massive galaxy clusters ($\gtrsim8\times10^{14}\Msun$).

The strong mass trends and weak redshift trends of these baryonic fractions of the clusters provide an indication that galaxy clusters cannot grow solely in a self-similar fashion.  Specifically, while ICM mass fractions that increase with halo mass can also be introduced through entropy injection from AGN feedback, star formation or other sources \citep[e.g.][]{ponman99}, this entropy injection would not impact the stellar mass fraction decrease with halo mass that we observe.  Thus, our measurements suggest that higher mass cluster halos contain a larger fraction of material that has been accreted from regions outside the high density, virial regions of clusters and groups.  In these surrounding low density regions the  characteristic stellar mass fractions (or baryon fractions) lie below (above) the mass dependent fractions we measure in the virial regions ($r<\Rfiveoo$ in our analysis).  Furthermore, the infall from the surrounding low density regions must be in balance with the infall from the lower mass halos such that the baryon content of galaxy clusters remains approximately the same at a fixed mass scale since redshift $\redshift\approx1.25$ \citep{chiu16a}.  Our results provide a direct indication of the presence and the impact of the so-called ``missing baryons'' in our Universe.  If the baryons were not present at the expected densities outside collapsed halos, then there would be no way of explaining in a hierarchical structure formation model how baryonic scaling relations vary steeply with mass and still remain roughly constant in time.

To explore this ansatz, we construct a toy model that we use to estimate the fraction of material that has been accreted from the field (i.e. outside the virial regions or in halos with masses $\Mfiveoo<10^{13}\Msun$) based on the derived scaling relations together with the cosmic values in the field.  Results from our toy model suggest that the fraction of infalling material from low density regions increases with cluster halo mass, varying from $\approx60\percent$ at $\Mfiveoo=10^{14}\Msun$ to $\gtrsim75\percent$ at $\Mfiveoo=8\times10^{14}\Msun$ at intermediate redshift.  Interestingly, recent structure formation simulations that include AGN feedback produce high mass halo populations that exhibit strong mass scaling and weak redshift evolution in ICM mass and stellar mass scaling relations \citep{wu15, truong16, barnes17, lebrun17} that are qualitatively consistent with the behavior we find in the SPT selected sample.   With further hydrodynamical simulations of the high mass cluster population, we expect to be able to carry out a detailed analysis of the physical implications of observed and predicted scaling relations and their redshift trends since $z\approx1$.

We emphasize that, as stated explicitly above, the uncertainties on the redshift and mass trends of the baryonic scaling relations are still considerable even with our large sample; therefore, even larger samples (e.g., including the galaxy clusters identified in DES, SPT-3G and eROSITA surveys at the low mass end) will be needed in future studies to resolve the precise scale of the variations in baryonic content of galaxy clusters over cosmic time.

%
%

\section*{Acknowledgements}
\label{sec:acknowledgements}

I-Non Chiu thanks Chien-Ho Lin, Keiichi Umetsu, Yen-Ting Lin and James Chan for support and useful discussions during this work.
We thank the anonymous referee for the constructive comments that lead to the improvement of this paper.
We acknowledge the support by the DFG Cluster of Excellence ``Origin and Structure of the Universe'', the DLR award 50 OR 1205 that supported I. Chiu during his PhD project, and the Transregio program TR33 ``The Dark Universe''.   
The data processing has been carried out on the computing facilities of the Computational Center for Particle and Astrophysics (C2PAP), located at the Leibniz Supercomputer Center (LRZ) in Garching.

The South Pole Telescope is supported by the National Science Foundation through grant PLR-1248097. Partial support is also provided by the NSF Physics Frontier Center grant PHY-1125897 to the Kavli Institute of Cosmological Physics at the University of Chicago, the Kavli Foundation and the Gordon and Betty Moore Foundation grant GBMF 947.  

This paper has gone through internal review by the DES collaboration.  Funding for the DES Projects has been provided by the U.S. Department of Energy, the U.S. National Science Foundation, the Ministry of Science and Education of Spain, the Science and Technology Facilities Council of the United Kingdom, the Higher Education Funding Council for England, the National Center for Supercomputing  Applications at the University of Illinois at Urbana-Champaign, the Kavli Institute of Cosmological Physics at the University of Chicago, the Center for Cosmology and Astro-Particle Physics at the Ohio State University, the Mitchell Institute for Fundamental Physics and Astronomy at Texas A\&M University, Financiadora de Estudos e Projetos, Funda{\c c}{\~a}o Carlos Chagas Filho de Amparo {\`a} Pesquisa do Estado do Rio de Janeiro, Conselho Nacional de Desenvolvimento Cient{\'i}fico e Tecnol{\'o}gico and the Minist{\'e}rio da Ci{\^e}ncia, Tecnologia e Inova{\c c}{\~a}o, the Deutsche Forschungsgemeinschaft and the Collaborating Institutions in the Dark Energy Survey. 

The Collaborating Institutions are Argonne National Laboratory, the University of California at Santa Cruz, the University of Cambridge, Centro de Investigaciones Energ{\'e}ticas, Medioambientales y Tecnol{\'o}gicas-Madrid, the University of Chicago, University College London, the DES-Brazil Consortium, the University of Edinburgh, the Eidgen{\"o}ssische Technische Hochschule (ETH) Z{\"u}rich, Fermi National Accelerator Laboratory, the University of Illinois at Urbana-Champaign, the Institut de Ci{\`e}ncies de l'Espai (IEEC/CSIC), the Institut de F{\'i}sica d'Altes Energies, Lawrence Berkeley National Laboratory, the Ludwig-Maximilians Universit{\"a}t M{\"u}nchen and the associated Excellence Cluster Universe, the University of Michigan, the National Optical Astronomy Observatory, the University of Nottingham, The Ohio State University, the University of Pennsylvania, the University of Portsmouth, SLAC National Accelerator Laboratory, Stanford University, the University of Sussex, Texas A\&M University, and the OzDES Membership Consortium.

The DES data management system is supported by the National Science Foundation under Grant Number AST-1138766. The DES participants from Spanish institutions are partially supported by MINECO under grants AYA2012-39559, ESP2013-48274, FPA2013-47986, and Centro de Excelencia Severo Ochoa SEV-2012-0234.  Research leading to these results has received funding from the European Research Council under the European Union's Seventh Framework Programme (FP7/2007-2013) including ERC grant agreements  240672, 291329, and 306478.

This work is based in part on archival data obtained with the \Spitzer\ Space Telescope, which is operated by the Jet Propulsion Laboratory, California Institute of Technology under a contract with NASA.  
This publication makes use of data products from the Wide-field Infrared Survey Explorer \citep{2010AJ....140.1868W}, which is a joint project of the University of California, Los Angeles, and the Jet Propulsion Laboratory/California Institute of Technology, funded by the National Aeronautics and Space Administration. 
This publication also makes use of data products from NEOWISE, which is a project of the Jet Propulsion Laboratory/California Institute of Technology, funded by the Planetary Science Division of the National Aeronautics and Space Administration. 
This research has made use of data obtained from the Chandra Data Archive and the Chandra Source Catalog, and software provided by the Chandra X-ray Center (CXC) in the application packages CIAO, ChIPS, and Sherpa. 
This research made use of ds9, a tool for data visualization supported by the Chandra X-ray Science Center (CXC) and the High Energy Astrophysics Science Archive Center (HEASARC) with support from the JWST Mission office at the Space Telescope Science Institute for 3D visualization. 
This paper includes data gathered with the Blanco 4~m telescope, located at the Cerro Tololo Inter-American Observatory in Chile, which is part of the U.S. National Optical Astronomy Observatory, which is operated by the Association of Universities for Research in Astronomy (AURA), under contract with the NSF.
This work includes data products from observations made with ESO Telescopes at the La Silla Paranal Observatory under ESO programme ID 179.A-2005 and data products produced by TERAPIX and the Cambridge Astronomy Survey Unit on behalf of the UltraVISTA consortium.
This work made use of the IPython package \citep{PER-GRA:2007}, SciPy \citep{jones_scipy_2001}, TOPCAT, an interactive graphical viewer and editor for tabular data \citep{2005ASPC..347...29T}, matplotlib, a Python library for publication quality graphics \citep{Hunter:2007}, Astropy, a community-developed core Python package for Astronomy \citep{2013A&A...558A..33A}, Sherpa \citep{2001SPIE.4477...76F}, XSPEC \citep{1996ASPC..101...17A}, NumPy \citep{van2011numpy}.

%
%

\bibliographystyle{mn2e}
\bibliography{literature}

%
%

\appendix

\section{Toy Model for Accretion}
\label{sec:infall_equation}

We provide the derivation of equation~(\ref{eq:infall}) for the toy model that describes the connection of the mass slope of the scaling relation to the materials that fall into the main halo.  For simplicity, we consider the stellar mass \Mstar\ to halo mass scaling relation here.

For a halo of mass \Mhalo\ and stellar mass \Mstar\ at redshift \redshift, the process of accretion or infall onto the halo will increase the halo and stellar masses by $\dif\Mhalo$ and $\dif\Mstar$, respectively, over a period $\dif\redshift$ in the formation history.
Then, the stellar mass fraction \fstar\ after the accretion of matter could be written as follows.
\begin{equation}
\label{eq:derive_infall_1}
\fstar(\Mhalo + \dif\Mhalo, \redshift + \dif\redshift) = 
\frac{\Mstar+ \dif\Mstar}{\Mhalo + \dif\Mhalo } \, .
\end{equation}
In the limit of small changes in mass, equation~(\ref{eq:derive_infall_1}) could be approximated as
\begin{equation}
\label{eq:derive_infall_2}
\fstar(\Mhalo + \dif\Mhalo, \redshift + \dif\redshift) \approx 
\fstar(\Mhalo, \redshift) \left(
1 + 
\frac{\dif\Mstar}{\Mstar} - 
\frac{\dif\Mhalo}{\Mhalo}
\right)
\, .
\end{equation}
This accretion of surrounding material could consist of other halos of a given mass and even the material outside the virial regions of these halos.  We can express $\dif\Mstar$ by summing the stellar mass infall from each of these components $i$ with the stellar fraction of $f_{\mathrm{\star},i}$, where each component contributes a fraction $\Gamma_{i}(\Mhalo, \redshift)$ of the infalling halo mass $\dif\Mhalo$:
\begin{equation}
\label{eq:derive_infall_3}
\dif\Mstar = \dif\Mhalo
\sum_{i \in \mathrm{infall}} 
\Gamma_{i}(\Mhalo, \redshift)  f_{\mathrm{\star},i}(\redshift)
\, ,
\end{equation}
where the fractions sum to unity
\begin{equation}
\label{eq:derive_infall_35}
\sum_{i \in \mathrm{infall}} 
\Gamma_{i}(\Mhalo, \redshift)  = 1
\, .
\end{equation}
Substituting equation~(\ref{eq:derive_infall_3}) and $\fstar(\Mhalo, \redshift) = \frac{\Mstar}{ \Mhalo }$ in equation~(\ref{eq:derive_infall_2}), we have 
\begin{eqnarray}
\label{eq:derive_infall_4}
{\fstar(\Mhalo + \dif \Mhalo, \redshift + \dif\redshift) \over \fstar(\Mhalo, \redshift)}
=& 1+
\frac{ \dif\Mhalo \sum_{i \in \mathrm{infall}}\Gamma_{i}(\Mhalo, \redshift)  f_{\mathrm{\star},i}(\redshift) }
{\Mhalo~\fstar(\Mhalo, \redshift)} -\frac{\dif \Mhalo}{\Mhalo}
 \nonumber \\
=&1+\frac{\dif \Mhalo}{\Mhalo}
\left(\frac{ \sum_{i \in \mathrm{infall}} \Gamma_{i}(\Mhalo, \redshift)  f_{\mathrm{\star},i}(\redshift) }{\fstar(\Mhalo, \redshift) } 
-1\right) 
\end{eqnarray}
This expression can be further rewritten as
\begin{eqnarray}
\label{eq:derive_infall_5}
\frac{\Mhalo}{\dif \Mhalo}
\left(
\frac{
\fstar(\Mhalo + \dif \Mhalo, \redshift + \dif\redshift)  - 
\fstar(\Mhalo, \redshift)
}{
\fstar(\Mhalo, \redshift)
}
\right) + 1\nonumber \\ 
= 
\sum_{i \in \mathrm{infall}} 
\Gamma_{i}(\Mhalo, \redshift)  
\frac{ f_{\mathrm{\star},i}(\redshift) }{  
\fstar(\Mhalo, \redshift) }
\, .
\end{eqnarray}
The first term on the left side of equation~(\ref{eq:derive_infall_5}) is simply the logarithmic derivative, and so we can write
\begin{equation}
\label{eq:derive_infall_6}
\frac{
\dif \ln \fstar(\Mhalo, \redshift)
}{
\dif \ln \Mhalo
} + 1 = 
\sum_{i \in \mathrm{infall}} 
\Gamma_{i}(\Mhalo, \redshift)  
\frac{ f_{\mathrm{\star},i}(\redshift) }{  
\fstar(\Mhalo, \redshift) }
\, .
\end{equation}
Note that the left hand side of equation~(\ref{eq:derive_infall_6}) is just the power law index $\Bstar$ of the mass trend in the stellar mass to halo mass relation.  That is, 
\begin{equation}
\label{eq:derive_infall_7}
\Bstar = 
\sum_{i \in \mathrm{infall}} 
\Gamma_{i}(\Mhalo, \redshift)  
\frac{ f_{\mathrm{\star},i}(\redshift) }{  
\fstar(\Mhalo, \redshift) }
\, .
\end{equation}
Thus, the slope parameter $\Bstar$ of the mass trend of the stellar mass to halo mass relation reflects the stellar mass fractions of the components that are accreted onto the halo.
If one can quantify 
(1) the mass trend parameter $\Bstar$ of the stellar mass to halo mass relation, and 
(2) the stellar mass fractions $f_{\mathrm{\star},i}(\redshift)$ of the infalling components $i$ and of the main halo $\fstar(\Mhalo, \redshift)$,
then---based on Equation~\ref{eq:derive_infall_7}---the fraction $\Gamma_{i}(\Mhalo, \redshift)$ of the infalling mass from the various components $i$ could be determined.

Consider a particularly simple case where there are only two components that fall into clusters---material within the virial regions of subhalos and material outside the virial regions of halos, which we refer to as the field.  Denoting these using $\mathrm{sub}$ and $\mathrm{fld}$, respectively, we can write
equation~(\ref{eq:derive_infall_7}) as
\begin{equation}
\label{eq:derive_infall_8}
\Bstar = 
\Gamma_{\mathrm{fld}}(\Mhalo, \redshift)  
\frac{ f_{\mathrm{fld}}(\redshift) }{  
\fstar(\Mhalo, \redshift) } + 
(1 - \Gamma_{\mathrm{fld}}(\Mhalo, \redshift) )
\frac{ f_{\mathrm{sub}}(\redshift) }{  
\fstar(\Mhalo, \redshift) }
\, .
\end{equation}
In the extreme case that stellar mass fraction is constant in halos regardless of hosting mass (i.e., $\Bstar = 1$ and $f_{\mathrm{sub}}(\redshift) = \fstar(\Mhalo, \redshift)$), then equation~(\ref{eq:derive_infall_7}) becomes
\[
\Gamma_{\mathrm{fld}}(\Mhalo, \redshift)
\left(
\frac{ f_{\mathrm{fld}}(\redshift) }{  
\fstar(\Mhalo, \redshift) } - 1
\right) = 0 \, ,
\]
which must imply no infall from the field (i.e., $\Gamma_{\mathrm{fld}}(\Mhalo, \redshift) = 0$) if the field stellar mass fraction is different from the halo stellar mass fraction $f_{\mathrm{fld}}(\redshift) \neq \fstar(\Mhalo, \redshift)$.
Because we have measured the stellar mass fractions as a function of mass and redshift on the cluster mass scale out to $\redshift\approx1.25$, we can estimate the contribution from infalling subhalos at any redshift by adopting the halo mass function at that redshift $\frac{dn}{dM}(\Mhalo,\redshift)$.  This together with the field stellar mass fraction, which can serve as  a reference for the mean stellar fraction outside halo virial regions, then allows us to estimate the fraction of accreted mass coming from virial regions of subhalos and the fraction coming from the lower density regions that lie outside virial regions---a region that is often referred to as the field.

\paragraph*{Affiliations:}  
\scriptsize
$^{1}$ Academia Sinica Institute of Astronomy and Astrophysics (ASIAA), 11F of AS/NTU Astronomy-Mathematics Building, No.1, Sec. 4, Roosevelt Rd, Taipei 10617, Taiwan\\
$^{2}$ Faculty of Physics, Ludwig-Maximilians-Universit\"at, Scheinerstr. 1, 81679 Munich, Germany\\
$^{3}$ Excellence Cluster Universe, Boltzmannstr.\ 2, 85748 Garching, Germany\\
$^{4}$ Max Planck Institute for Extraterrestrial Physics, Giessenbachstrasse, 85748 Garching, Germany\\
$^{5}$ Kavli Institute for Astrophysics and Space Research, Massachusetts Institute of Technology, 77 Massachusetts Avenue, Cambridge, MA 02139\\
$^{6}$ Argonne National Laboratory, High-Energy Physics Division, 9700 S. Cass Avenue, Argonne, IL, USA 60439\\
$^{7}$ Department of Physics, IIT Hyderabad, Kandi, Telangana 502285, India\\
$^{8}$ Harvard-Smithsonian Center for Astrophysics, 60 Garden Street, Cambridge, MA 02138\\
$^{9}$ Physics Department, University of California, Davis, CA 95616\\
$^{10}$ Fermi National Accelerator Laboratory, Batavia, IL 60510-0500\\
$^{11}$ Department of Astronomy and Astrophysics, University of Chicago, 5640 South Ellis Avenue, Chicago, IL 60637\\
$^{12}$ Kavli Institute for Cosmological Physics, University of Chicago, 5640 South Ellis Avenue, Chicago, IL 60637\\
$^{13}$ Department of Physics and Astronomy, University of Missouri, 5110 Rockhill Road, Kansas City, MO 64110\\
$^{14}$ Cerro Tololo Inter-American Observatory, National Optical Astronomy Observatory, Casilla 603, La Serena, Chile\\
$^{15}$ Department of Physics and Electronics, Rhodes University, PO Box 94, Grahamstown, 6140, South Africa\\
$^{16}$ Department of Physics \& Astronomy, University College London, Gower Street, London, WC1E 6BT, UK\\
$^{17}$ Fermi National Accelerator Laboratory, P. O. Box 500, Batavia, IL 60510, USA\\
$^{18}$ Kavli Institute for Astrophysics and Space Research, Massachusetts Institute of Technology, 77 Massachusetts Avenue, Cambridge, MA 02139\\
$^{19}$ Sorbonne Universit\'es, UPMC Univ Paris 06, UMR 7095, Institut d'Astrophysique de Paris, F-75014, Paris, France\\
$^{20}$ CNRS, UMR 7095, Institut d'Astrophysique de Paris, F-75014, Paris, France\\
$^{21}$ Argonne National Laboratory, High-Energy Physics Division, 9700 S. Cass Avenue, Argonne, IL, USA 60439\\
$^{22}$ Kavli Institute for Cosmological Physics, University of Chicago, 5640 South Ellis Avenue, Chicago, IL 60637\\
$^{23}$ Excellence Cluster Universe, Boltzmannstr.\ 2, 85748 Garching, Germany\\
$^{24}$ Faculty of Physics, Ludwig-Maximilians-Universit\"at, Scheinerstr. 1, 81679 Munich, Germany\\
$^{25}$ Department of Astronomy and Astrophysics, University of Chicago, 5640 South Ellis Avenue, Chicago, IL 60637\\
$^{26}$ Observat\'orio Nacional, Rua Gal. Jos\'e Cristino 77, Rio de Janeiro, RJ - 20921-400, Brazil\\
$^{27}$ Laborat\'orio Interinstitucional de e-Astronomia - LIneA, Rua Gal. Jos\'e Cristino 77, Rio de Janeiro, RJ - 20921-400, Brazil\\
$^{28}$ Institut de F\'{\i}sica d'Altes Energies (IFAE), The Barcelona Institute of Science and Technology, Campus UAB, 08193 Bellaterra (Barcelona) Spain\\
$^{29}$ Institute of Space Sciences, IEEC-CSIC, Campus UAB, Carrer de Can Magrans, s/n,  08193 Barcelona, Spain\\
$^{30}$ Kavli Institute for Particle Astrophysics \& Cosmology, P. O. Box 2450, Stanford University, Stanford, CA 94305, USA\\
$^{31}$ Department of Physics and Astronomy, University of Pennsylvania, Philadelphia, PA 19104, USA\\
$^{32}$ Department of Physics, California Institute of Technology, Pasadena, CA 91125, USA\\
$^{33}$ Jet Propulsion Laboratory, California Institute of Technology, 4800 Oak Grove Dr., Pasadena, CA 91109, USA\\
$^{34}$ Department of Astronomy, University of Michigan, Ann Arbor, MI 48109, USA\\
$^{35}$ Department of Physics, University of Michigan, Ann Arbor, MI 48109, USA\\
$^{36}$ Instituto de Fisica Teorica UAM/CSIC, Universidad Autonoma de Madrid, 28049 Madrid, Spain\\
$^{37}$ Department of Astronomy and Astrophysics, The Pennsylvania State University, 201 Old Main, University Park, Pennsylvania 16802\\
$^{38}$ Department of Astronomy, University of Florida, Gainesville, FL 32611\\
$^{39}$ SLAC National Accelerator Laboratory, Menlo Park, CA 94025, USA\\
$^{40}$ Department of Astronomy, University of Illinois, 1002 W. Green Street, Urbana, IL 61801, USA\\
$^{41}$ National Center for Supercomputing Applications, 1205 West Clark St., Urbana, IL 61801, USA\\
$^{42}$ Universit\"ats-Sternwarte, Fakult\"at f\"ur Physik, Ludwig-Maximilians Universit\"at M\"unchen, Scheinerstr. 1, 81679 M\"unchen, Germany\\
$^{43}$ Department of Physics, University of Montreal, Montreal, QC H3C 3J7, Canada\\
$^{44}$ Department of Physics, The Ohio State University, Columbus, OH 43210, USA\\
$^{45}$ Center for Cosmology and Astro-Particle Physics, The Ohio State University, Columbus, OH 43210, USA\\
$^{46}$ Astronomy Department, University of Washington, Box 351580, Seattle, WA 98195, USA\\
$^{47}$ Santa Cruz Institute for Particle Physics, Santa Cruz, CA 95064, USA\\
$^{48}$ Harvard-Smithsonian Center for Astrophysics, 60 Garden Street, Cambridge, MA 02138, USA\\
$^{49}$ Australian Astronomical Observatory, North Ryde, NSW 2113, Australia\\
$^{50}$ Argonne National Laboratory, 9700 South Cass Avenue, Lemont, IL 60439, USA\\
$^{51}$ Departamento de F\'isica Matem\'atica, Instituto de F\'isica, Universidade de S\~ao Paulo, CP 66318, S\~ao Paulo, SP, 05314-970, Brazil\\
$^{52}$ George P. and Cynthia Woods Mitchell Institute for Fundamental Physics and Astronomy, and Department of Physics and Astronomy, Texas A\&M University, College Station, TX 77843,  USA\\
$^{53}$ Department of Astrophysical Sciences, Princeton University, Peyton Hall, Princeton, NJ 08544, USA\\
$^{54}$ Instituci\'o Catalana de Recerca i Estudis Avan\c{c}ats, E-08010 Barcelona, Spain\\
$^{55}$ Center for Astrophysics and Space Astronomy, Department of Astrophysical and Planetary Science, University of Colorado, Boulder, C0 80309, USA\\
$^{56}$ NASA Postdoctoral Program Senior Fellow, NASA Ames Research Center, Moffett Field, CA 94035, USA\\
$^{57}$ School of Physics, University of Melbourne, Parkville, VIC 3010, Australia\\
$^{58}$ Department of Physics and Astronomy, Pevensey Building, University of Sussex, Brighton, BN1 9QH, UK\\
$^{59}$ Centro de Investigaciones Energ\'eticas, Medioambientales y Tecnol\'ogicas (CIEMAT), Madrid, Spain\\
$^{60}$ Department of Astronomy, University of Michigan , 1085 S. University Ave., Ann Arbor, MI\\
$^{61}$ School of Physics and Astronomy, University of Southampton,  Southampton, SO17 1BJ, UK\\
$^{62}$ Instituto de F\'isica Gleb Wataghin, Universidade Estadual de Campinas, 13083-859, Campinas, SP, Brazil\\
$^{63}$ LSST, 950 North Cherry Avenue, Tucson, AZ 85719\\
$^{64}$ Computer Science and Mathematics Division, Oak Ridge National Laboratory, Oak Ridge, TN 37831\\
$^{65}$ Max Planck Institute for Extraterrestrial Physics, Giessenbachstrasse, 85748 Garching, Germany\\

\end{document}

%% file: authorlist.tex
\author[Chiu et al.]{
\parbox{\textwidth}{
\Large
I.~Chiu$^{1,2,3}$,
J.~J.~Mohr$^{2,3,4}$,
M.~McDonald$^{5}$,
S.~Bocquet$^{6}$,
S.~Desai$^{7}$,
M.~Klein$^{2,4}$,
H.~Israel$^{2}$,
M.~L.~N.~Ashby$^{8}$,
A.~Stanford$^{9}$,
B.~A.~Benson$^{10,11,12}$,
M.~Brodwin$^{13}$,
T.~M.~C.~Abbott$^{14}$,
F.~B.~Abdalla$^{15,16}$,
S.~Allam$^{17}$,
J.~Annis$^{17}$,
M.~Bayliss$^{18}$,
A.~Benoit-L{\'e}vy$^{19,16,20}$,
E.~Bertin$^{20,19}$,
L.~Bleem$^{21,22}$,
D.~Brooks$^{16}$,
E.~Buckley-Geer$^{17}$,
E.~Bulbul$^{18}$,
R.~Capasso$^{23,24}$,
J.~E.~Carlstrom$^{25,22}$,
A.~Carnero~Rosell$^{26,27}$,
J.~Carretero$^{28}$,
F.~J.~Castander$^{29}$,
C.~E.~Cunha$^{30}$,
C.~B.~D'Andrea$^{31}$,
L.~N.~da Costa$^{27,26}$,
C.~Davis$^{30}$,
H.~T.~Diehl$^{17}$,
J.~P.~Dietrich$^{24,23}$,
P.~Doel$^{16}$,
A.~Drlica-Wagner$^{17}$,
T.~F.~Eifler$^{32,33}$,
A.~E.~Evrard$^{34,35}$,
B.~Flaugher$^{17}$,
J.~Garc\'ia-Bellido$^{36}$,
G.~Garmire$^{37}$,
E.~Gaztanaga$^{29}$,
D.~W.~Gerdes$^{34,35}$,
A.~Gonzalez$^{38}$,
D.~Gruen$^{30,39}$,
R.~A.~Gruendl$^{40,41}$,
J.~Gschwend$^{27,26}$,
N.~Gupta$^{23,24,42}$,
G.~Gutierrez$^{17}$,
J.~Hlavacek-L.$^{43}$,
K.~Honscheid$^{44,45}$,
D.~J.~James$^{46}$,
T.~Jeltema$^{47}$,
R.~Kraft$^{48}$,
E.~Krause$^{30}$,
K.~Kuehn$^{49}$,
S.~Kuhlmann$^{50}$,
N.~Kuropatkin$^{17}$,
O.~Lahav$^{16}$,
M.~Lima$^{27,51}$,
M.~A.~G.~Maia$^{26,27}$,
J.~L.~Marshall$^{52}$,
P.~Melchior$^{53}$,
F.~Menanteau$^{40,41}$,
R.~Miquel$^{28,54}$,
S.~Murray$^{48}$,
B.~Nord$^{17}$,
R.~L.~C.~Ogando$^{26,27}$,
A.~A.~Plazas$^{33}$,
D.~Rapetti$^{55,56,23,24}$,
C.~L.~Reichardt$^{57}$,
A.~K.~Romer$^{58}$,
A.~Roodman$^{39,30}$,
E.~Sanchez$^{59}$,
A.~Saro$^{24}$,
V.~Scarpine$^{17}$,
R.~Schindler$^{39}$,
M.~Schubnell$^{35}$,
K.~Sharon$^{60}$,
R.~C.~Smith$^{14}$,
M.~Smith$^{61}$,
M.~Soares-Santos$^{17}$,
F.~Sobreira$^{27,62}$,
B.~Stalder$^{63}$,
C.~Stern$^{24,23}$,
V.~Strazzullo$^{24}$,
E.~Suchyta$^{64}$,
M.~E.~C.~Swanson$^{41}$,
G.~Tarle$^{35}$,
V.~Vikram$^{50}$,
A.~R.~Walker$^{14}$,
J.~Weller$^{42,23,65}$,
Y.~Zhang$^{17}$
}
\vspace{0.4cm}
\\
\parbox{\textwidth}{
Affiliations are shown in the end of this paper.
}
}

%% file: xvp.master.tex
    SPT-CL~J0000$-$5748    &$ 0.702 $    &$ 0.2518 $    &$ -57.8094 $     &$ 0.2501 $    &$ -57.8093 $     & $griz[3.6][4.5]$    \\ 
    SPT-CL~J0013$-$4906    &$ 0.406 $    &$ 3.3309 $    &$ -49.1160 $     &$ 3.3306 $    &$ -49.1099 $     & $grizW1W2$    \\ 
    SPT-CL~J0014$-$4952    &$ 0.752 $    &$ 3.6912 $    &$ -49.8800 $     &$ 3.7041 $    &$ -49.8851 $     & $grizW1W2$    \\ 
    SPT-CL~J0033$-$6326    &$ 0.597 $    &$ 8.4720 $    &$ -63.4429 $     &$ 8.4710 $    &$ -63.4449 $     & $grizW1W2$    \\ 
    SPT-CL~J0037$-$5047    &$ 1.026 $    &$ 9.4476 $    &$ -50.7876 $     &$ 9.4478 $    &$ -50.7890 $     & $griz[3.6][4.5]$    \\ 
    SPT-CL~J0040$-$4407    &$ 0.350 $    &$ 10.2085 $    &$ -44.1340 $     &$ 10.2080 $    &$ -44.1307 $     & $grizW1W2$    \\ 
    SPT-CL~J0058$-$6145    &$ 0.826 $    &$ 14.5829 $    &$ -61.7693 $     &$ 14.5842 $    &$ -61.7669 $     & $griz[3.6][4.5]$    \\ 
    SPT-CL~J0102$-$4603    &$ 0.722 $    &$ 15.6737 $    &$ -46.0652 $     &$ 15.6779 $    &$ -46.0710 $     & $griz[3.6][4.5]$    \\ 
    SPT-CL~J0102$-$4915    &$ 0.870 $    &$ 15.7350 $    &$ -49.2667 $     &$ 15.7407 $    &$ -49.2720 $     & $griz[3.6][4.5]$    \\ 
    SPT-CL~J0123$-$4821    &$ 0.620 $    &$ 20.7931 $    &$ -48.3567 $     &$ 20.7956 $    &$ -48.3563 $     & $griz[3.6][4.5]$    \\ 
    SPT-CL~J0142$-$5032    &$ 0.730 $    &$ 25.5464 $    &$ -50.5403 $     &$ 25.5401 $    &$ -50.5410 $     & $griz[3.6][4.5]$    \\ 
    SPT-CL~J0151$-$5954    &$ 1.035 $    &$ 27.8584 $    &$ -59.9076 $     &$ 27.8540 $    &$ -59.9123 $     & $griz[3.6][4.5]$    \\ 
    SPT-CL~J0156$-$5541    &$ 1.221 $    &$ 29.0405 $    &$ -55.6976 $     &$ 29.0381 $    &$ -55.7029 $     & $griz[3.6][4.5]$    \\ 
    SPT-CL~J0200$-$4852    &$ 0.498 $    &$ 30.1403 $    &$ -48.8752 $     &$ 30.1421 $    &$ -48.8712 $     & $grizW1W2$    \\ 
    SPT-CL~J0212$-$4657    &$ 0.655 $    &$ 33.1094 $    &$ -46.9495 $     &$ 33.0986 $    &$ -46.9537 $     & $griz[3.6][4.5]$    \\ 
    SPT-CL~J0217$-$5245    &$ 0.343 $    &$ 34.2947 $    &$ -52.7514 $     &$ 34.3122 $    &$ -52.7604 $     & $grizW1W2$    \\ 
    SPT-CL~J0232$-$5257    &$ 0.556 $    &$ 38.1977 $    &$ -52.9556 $     &$ 38.2058 $    &$ -52.9531 $     & $grizW1W2$    \\ 
    SPT-CL~J0234$-$5831    &$ 0.415 $    &$ 38.6777 $    &$ -58.5240 $     &$ 38.6761 $    &$ -58.5236 $     & $grizW1W2$    \\ 
    SPT-CL~J0236$-$4938    &$ 0.334 $    &$ 39.2495 $    &$ -49.6343 $     &$ 39.2569 $    &$ -49.6360 $     & $grizW1W2$    \\ 
    SPT-CL~J0243$-$5930    &$ 0.635 $    &$ 40.8638 $    &$ -59.5166 $     &$ 40.8628 $    &$ -59.5172 $     & $griz[3.6][4.5]$    \\ 
    SPT-CL~J0252$-$4824    &$ 0.421 $    &$ 43.1946 $    &$ -48.4136 $     &$ 43.2083 $    &$ -48.4162 $     & $grizW1W2$    \\ 
    SPT-CL~J0256$-$5617    &$ 0.580 $    &$ 44.1044 $    &$ -56.2977 $     &$ 44.1056 $    &$ -56.2978 $     & $griz[3.6][4.5]$    \\ 
    SPT-CL~J0304$-$4401    &$ 0.458 $    &$ 46.0659 $    &$ -44.0329 $     &$ 46.0701 $    &$ -44.0255 $     & $grizW1W2$    \\ 
    SPT-CL~J0304$-$4921    &$ 0.392 $    &$ 46.0664 $    &$ -49.3570 $     &$ 46.0673 $    &$ -49.3571 $     & $grizW1W2$    \\ 
    SPT-CL~J0307$-$5042    &$ 0.550 $    &$ 46.9603 $    &$ -50.7045 $     &$ 46.9605 $    &$ -50.7012 $     & $grizW1W2$    \\ 
    SPT-CL~J0307$-$6225    &$ 0.579 $    &$ 46.8275 $    &$ -62.4352 $     &$ 46.8195 $    &$ -62.4465 $     & $grizW1W2$    \\ 
    SPT-CL~J0310$-$4647    &$ 0.709 $    &$ 47.6343 $    &$ -46.7847 $     &$ 47.6354 $    &$ -46.7856 $     & $griz[3.6][4.5]$    \\ 
    SPT-CL~J0324$-$6236    &$ 0.730 $    &$ 51.0488 $    &$ -62.5984 $     &$ 51.0511 $    &$ -62.5988 $     & $griz[3.6][4.5]$    \\ 
    SPT-CL~J0330$-$5228    &$ 0.442 $    &$ 52.7226 $    &$ -52.4737 $     &$ 52.7374 $    &$ -52.4704 $     & $grizW1W2$    \\ 
    SPT-CL~J0334$-$4659    &$ 0.485 $    &$ 53.5501 $    &$ -46.9966 $     &$ 53.5457 $    &$ -46.9958 $     & $grizW1W2$    \\ 
    SPT-CL~J0346$-$5439    &$ 0.530 $    &$ 56.7320 $    &$ -54.6477 $     &$ 56.7308 $    &$ -54.6487 $     & $grizW1W2$    \\ 
    SPT-CL~J0348$-$4515    &$ 0.358 $    &$ 57.0701 $    &$ -45.2507 $     &$ 57.0712 $    &$ -45.2498 $     & $grizW1W2$    \\ 
    SPT-CL~J0352$-$5647    &$ 0.670 $    &$ 58.2403 $    &$ -56.7985 $     &$ 58.2397 $    &$ -56.7977 $     & $griz[3.6][4.5]$    \\ 
    SPT-CL~J0406$-$4805    &$ 0.737 $    &$ 61.7270 $    &$ -48.0853 $     &$ 61.7302 $    &$ -48.0826 $     & $grizW1W2$    \\ 
    SPT-CL~J0411$-$4819    &$ 0.424 $    &$ 62.8100 $    &$ -48.3214 $     &$ 62.7957 $    &$ -48.3277 $     & $grizW1W2$    \\ 
    SPT-CL~J0417$-$4748    &$ 0.581 $    &$ 64.3458 $    &$ -47.8140 $     &$ 64.3461 $    &$ -47.8132 $     & $griz[3.6][4.5]$    \\ 
    SPT-CL~J0426$-$5455    &$ 0.630 $    &$ 66.5207 $    &$ -54.9173 $     &$ 66.5171 $    &$ -54.9253 $     & $grizW1W2$    \\ 
    SPT-CL~J0438$-$5419    &$ 0.421 $    &$ 69.5775 $    &$ -54.3202 $     &$ 69.5734 $    &$ -54.3224 $     & $grizW1W2$    \\ 
    SPT-CL~J0441$-$4855    &$ 0.790 $    &$ 70.4503 $    &$ -48.9220 $     &$ 70.4497 $    &$ -48.9233 $     & $griz[3.6][4.5]$    \\ 
    SPT-CL~J0446$-$5849    &$ 1.186 $    &$ 71.5170 $    &$ -58.8294 $     &$ 71.5157 $    &$ -58.8304 $     & $griz[3.6][4.5]$    \\ 
    SPT-CL~J0449$-$4901    &$ 0.792 $    &$ 72.2741 $    &$ -49.0242 $     &$ 72.2819 $    &$ -49.0214 $     & $griz[3.6][4.5]$    \\ 
    SPT-CL~J0456$-$5116    &$ 0.562 $    &$ 74.1201 $    &$ -51.2777 $     &$ 74.1171 $    &$ -51.2764 $     & $griz[3.6][4.5]$    \\ 
    SPT-CL~J0509$-$5342    &$ 0.461 $    &$ 77.3383 $    &$ -53.7032 $     &$ 77.3393 $    &$ -53.7035 $     & $griz[3.6][4.5]$    \\ 
    SPT-CL~J0528$-$5300    &$ 0.768 $    &$ 82.0188 $    &$ -52.9961 $     &$ 82.0222 $    &$ -52.9981 $     & $griz[3.6][4.5]$    \\ 
    SPT-CL~J0533$-$5005    &$ 0.881 $    &$ 83.4018 $    &$ -50.0969 $     &$ 83.4033 $    &$ -50.0958 $     & $griz[3.6][4.5]$    \\ 
    SPT-CL~J0542$-$4100    &$ 0.642 $    &$ 85.7111 $    &$ -41.0019 $     &$ 85.7085 $    &$ -41.0001 $     & $grizW1W2$    \\ 
    SPT-CL~J0546$-$5345    &$ 1.066 $    &$ 86.6532 $    &$ -53.7604 $     &$ 86.6568 $    &$ -53.7586 $     & $griz[3.6][4.5]$    \\ 
    SPT-CL~J0551$-$5709    &$ 0.423 $    &$ 87.8954 $    &$ -57.1484 $     &$ 87.8931 $    &$ -57.1451 $     & $griz[3.6][4.5]$    \\ 
    SPT-CL~J0555$-$6406    &$ 0.345 $    &$ 88.8660 $    &$ -64.1058 $     &$ 88.8731 $    &$ -64.1068 $     & $--$    \\ 
    SPT-CL~J0559$-$5249    &$ 0.609 $    &$ 89.9357 $    &$ -52.8253 $     &$ 89.9301 $    &$ -52.8242 $     & $griz[3.6][4.5]$    \\ 
    SPT-CL~J0616$-$5227    &$ 0.684 $    &$ 94.1466 $    &$ -52.4555 $     &$ 94.1420 $    &$ -52.4525 $     & $griz[3.6][4.5]$    \\ 
    SPT-CL~J0655$-$5234    &$ 0.470 $    &$ 103.9721 $    &$ -52.5687 $     &$ 103.9760 $    &$ -52.5674 $     & $--$    \\ 
    SPT-CL~J2031$-$4037    &$ 0.342 $    &$ 307.9648 $    &$ -40.6220 $     &$ 307.9720 $    &$ -40.6252 $     & $--$    \\ 
    SPT-CL~J2034$-$5936    &$ 0.919 $    &$ 308.5371 $    &$ -59.6039 $     &$ 308.5390 $    &$ -59.6042 $     & $griz[3.6][4.5]$    \\ 
    SPT-CL~J2035$-$5251    &$ 0.528 $    &$ 308.7927 $    &$ -52.8554 $     &$ 308.7950 $    &$ -52.8564 $     & $griz[3.6][4.5]$    \\ 
    SPT-CL~J2043$-$5035    &$ 0.723 $    &$ 310.8243 $    &$ -50.5930 $     &$ 310.8230 $    &$ -50.5923 $     & $griz[3.6][4.5]$    \\ 
    SPT-CL~J2106$-$5844    &$ 1.132 $    &$ 316.5174 $    &$ -58.7426 $     &$ 316.5190 $    &$ -58.7411 $     & $griz[3.6][4.5]$    \\ 
    SPT-CL~J2135$-$5726    &$ 0.427 $    &$ 323.9111 $    &$ -57.4390 $     &$ 323.9060 $    &$ -57.4418 $     & $griz[3.6][4.5]$    \\ 
    SPT-CL~J2145$-$5644    &$ 0.480 $    &$ 326.4686 $    &$ -56.7470 $     &$ 326.4660 $    &$ -56.7482 $     & $griz[3.6][4.5]$    \\ 
    SPT-CL~J2146$-$4633    &$ 0.933 $    &$ 326.6456 $    &$ -46.5489 $     &$ 326.6470 $    &$ -46.5505 $     & $griz[3.6][4.5]$    \\ 
    SPT-CL~J2148$-$6116    &$ 0.571 $    &$ 327.1804 $    &$ -61.2788 $     &$ 327.1780 $    &$ -61.2795 $     & $griz[3.6][4.5]$    \\ 
    SPT-CL~J2218$-$4519    &$ 0.650 $    &$ 334.7461 $    &$ -45.3158 $     &$ 334.7470 $    &$ -45.3145 $     & $griz[3.6][4.5]$    \\ 
    SPT-CL~J2222$-$4834    &$ 0.652 $    &$ 335.7136 $    &$ -48.5770 $     &$ 335.7110 $    &$ -48.5764 $     & $griz[3.6][4.5]$    \\ 
    SPT-CL~J2232$-$5959    &$ 0.594 $    &$ 338.1428 $    &$ -59.9990 $     &$ 338.1410 $    &$ -59.9980 $     & $grizW1W2$    \\ 
    SPT-CL~J2233$-$5339    &$ 0.480 $    &$ 338.3233 $    &$ -53.6544 $     &$ 338.3150 $    &$ -53.6526 $     & $grizW1W2$    \\ 
    SPT-CL~J2236$-$4555    &$ 1.162 $    &$ 339.2196 $    &$ -45.9270 $     &$ 339.2230 $    &$ -45.9312 $     & $griz[3.6][4.5]$    \\ 
    SPT-CL~J2245$-$6206    &$ 0.580 $    &$ 341.2577 $    &$ -62.1185 $     &$ 341.2590 $    &$ -62.1272 $     & $griz[3.6][4.5]$    \\ 
    SPT-CL~J2248$-$4431    &$ 0.351 $    &$ 342.1875 $    &$ -44.5287 $     &$ 342.1830 $    &$ -44.5308 $     & $grizW1W2$    \\ 
    SPT-CL~J2258$-$4044    &$ 0.826 $    &$ 344.7062 $    &$ -40.7396 $     &$ 344.7010 $    &$ -40.7418 $     & $--$    \\ 
    SPT-CL~J2259$-$6057    &$ 0.750 $    &$ 344.7509 $    &$ -60.9590 $     &$ 344.7540 $    &$ -60.9595 $     & $griz[3.6][4.5]$    \\ 
    SPT-CL~J2301$-$4023    &$ 0.730 $    &$ 345.4692 $    &$ -40.3895 $     &$ 345.4700 $    &$ -40.3868 $     & $griz[3.6][4.5]$    \\ 
    SPT-CL~J2306$-$6505    &$ 0.530 $    &$ 346.7260 $    &$ -65.0902 $     &$ 346.7230 $    &$ -65.0882 $     & $griz[3.6][4.5]$    \\ 
    SPT-CL~J2325$-$4111    &$ 0.358 $    &$ 351.3023 $    &$ -41.1959 $     &$ 351.2990 $    &$ -41.2037 $     & $grizW1W2$    \\ 
    SPT-CL~J2331$-$5051    &$ 0.576 $    &$ 352.9610 $    &$ -50.8631 $     &$ 352.9630 $    &$ -50.8650 $     & $griz[3.6][4.5]$    \\ 
    SPT-CL~J2335$-$4544    &$ 0.547 $    &$ 353.7854 $    &$ -45.7396 $     &$ 353.7850 $    &$ -45.7391 $     & $grizW1W2$    \\ 
    SPT-CL~J2337$-$5942    &$ 0.775 $    &$ 354.3516 $    &$ -59.7061 $     &$ 354.3370 $    &$ -59.7109 $     & $griz[3.6][4.5]$    \\ 
    SPT-CL~J2341$-$5119    &$ 1.003 $    &$ 355.3009 $    &$ -51.3285 $     &$ 355.3010 $    &$ -51.3291 $     & $griz[3.6][4.5]$    \\ 
    SPT-CL~J2342$-$5411    &$ 1.075 $    &$ 355.6904 $    &$ -54.1838 $     &$ 355.6910 $    &$ -54.1847 $     & $griz[3.6][4.5]$    \\ 
    SPT-CL~J2344$-$4243    &$ 0.596 $    &$ 356.1839 $    &$ -42.7205 $     &$ 356.1830 $    &$ -42.7201 $     & $griz[3.6][4.5]$    \\ 
    SPT-CL~J2345$-$6405    &$ 0.937 $    &$ 356.2491 $    &$ -64.1001 $     &$ 356.2510 $    &$ -64.0927 $     & $griz[3.6][4.5]$    \\ 
    SPT-CL~J2352$-$4657    &$ 0.734 $    &$ 358.0684 $    &$ -46.9605 $     &$ 358.0680 $    &$ -46.9602 $     & $griz[3.6][4.5]$    \\ 
    SPT-CL~J2355$-$5055    &$ 0.320 $    &$ 358.9502 $    &$ -50.9283 $     &$ 358.9480 $    &$ -50.9277 $     & $griz[3.6][4.5]$    \\ 
    SPT-CL~J2359$-$5009    &$ 0.775 $    &$ 359.9321 $    &$ -50.1697 $     &$ 359.9280 $    &$ -50.1672 $     & $--$    \\ 
    SPT-CL~J0106$-$5943    &$ 0.348 $    &$ 16.6141 $    &$ -59.7200 $     &$ 16.6197 $    &$ -59.7201 $     & $grizW1W2$    \\ 
    SPT-CL~J2332$-$5053    &$ 0.560 $    &$ 353.0343 $    &$ -50.8911 $     &$ 353.0249 $    &$ -50.8849 $     & $griz[3.6][4.5]$    \\ 
    SPT-CL~J0232$-$4421    &$ 0.284 $    &$ 38.0711 $    &$ -44.3513 $     &$ 38.0773 $    &$ -44.3467 $     & $grizW1W2$    \\ 
    SPT-CL~J0235$-$5121    &$ 0.278 $    &$ 38.9345 $    &$ -51.3585 $     &$ 38.9387 $    &$ -51.3512 $     & $grizW1W2$    \\ 
    SPT-CL~J0516$-$5430    &$ 0.295 $    &$ 79.1479 $    &$ -54.5126 $     &$ 79.1556 $    &$ -54.5004 $     & $griz[3.6][4.5]$    \\ 
    SPT-CL~J0522$-$4818    &$ 0.296 $    &$ 80.5682 $    &$ -48.3039 $     &$ 80.5648 $    &$ -48.3049 $     & $grizW1W2$    \\ 
    SPT-CL~J0658$-$5556    &$ 0.296 $    &$ 104.6180 $    &$ -55.9448 $     &$ 104.6470 $    &$ -55.9491 $     & $--$    \\ 
    SPT-CL~J2011$-$5725    &$ 0.279 $    &$ 302.8611 $    &$ -57.4200 $     &$ 302.8620 $    &$ -57.4196 $     & $--$    \\ 

%% file: xvp.measurement.tex
    SPT-CL~J0000$-$5748    &$ 2.07 $    &$ 4.13 \pm 0.59 $    &$ 3.99 \pm 0.23 $    &$ 3.34 \pm 0.51 $     &$ 4.33 \pm 0.24 $    &$ 0.077 \pm 0.012 $    \\ 
    SPT-CL~J0013$-$4906    &$ 3.46 $    &$ 5.94 \pm 0.78 $    &$ 6.77 \pm 0.19 $    &$ 3.37 \pm 0.58 $     &$ 7.11 \pm 0.20 $    &$ 0.047 \pm 0.008 $    \\ 
    SPT-CL~J0014$-$4952    &$ 2.07 $    &$ 4.74 \pm 0.66 $    &$ 6.69 \pm 0.15 $    &$ 4.76 \pm 0.74 $     &$ 7.17 \pm 0.17 $    &$ 0.066 \pm 0.010 $    \\ 
    SPT-CL~J0033$-$6326    &$ 2.35 $    &$ 4.30 \pm 0.64 $    &$ 4.88 \pm 0.16 $    &$ 4.44 \pm 0.78 $     &$ 5.32 \pm 0.18 $    &$ 0.083 \pm 0.014 $    \\ 
    SPT-CL~J0037$-$5047    &$ 1.56 $    &$ 3.73 \pm 0.59 $    &$ 2.35 \pm 0.19 $    &$ 2.74 \pm 0.61 $     &$ 2.62 \pm 0.20 $    &$ 0.104 \pm 0.022 $    \\ 
    SPT-CL~J0040$-$4407    &$ 4.33 $    &$ 8.25 \pm 1.00 $    &$ 8.34 \pm 0.26 $    &$ 4.69 \pm 0.63 $     &$ 8.81 \pm 0.26 $    &$ 0.053 \pm 0.007 $    \\ 
    SPT-CL~J0058$-$6145    &$ 1.85 $    &$ 4.06 \pm 0.61 $    &$ 4.42 \pm 0.15 $    &$ 3.08 \pm 0.58 $     &$ 4.73 \pm 0.16 $    &$ 0.065 \pm 0.012 $    \\ 
    SPT-CL~J0102$-$4603    &$ 2.04 $    &$ 4.16 \pm 0.64 $    &$ 4.04 \pm 0.14 $    &$ 4.82 \pm 0.56 $     &$ 4.52 \pm 0.15 $    &$ 0.107 \pm 0.012 $    \\ 
    SPT-CL~J0102$-$4915    &$ 2.52 $    &$ 11.36 \pm 1.34 $    &$ 22.50 \pm 0.21 $    &$ 9.80 \pm 1.01 $     &$ 23.48 \pm 0.24 $    &$ 0.042 \pm 0.004 $    \\ 
    SPT-CL~J0123$-$4821    &$ 2.25 $    &$ 4.09 \pm 0.65 $    &$ 4.58 \pm 0.09 $    &$ 4.50 \pm 0.58 $     &$ 5.03 \pm 0.11 $    &$ 0.090 \pm 0.011 $    \\ 
    SPT-CL~J0142$-$5032    &$ 2.15 $    &$ 5.02 \pm 0.68 $    &$ 5.55 \pm 0.27 $    &$ 3.64 \pm 0.62 $     &$ 5.92 \pm 0.28 $    &$ 0.061 \pm 0.010 $    \\ 
    SPT-CL~J0151$-$5954    &$ 1.58 $    &$ 3.91 \pm 0.59 $    &$ 5.07 \pm 0.14 $    &$ 4.58 \pm 0.73 $     &$ 5.52 \pm 0.16 $    &$ 0.083 \pm 0.012 $    \\ 
    SPT-CL~J0156$-$5541    &$ 1.38 $    &$ 3.49 \pm 0.55 $    &$ 4.17 \pm 0.15 $    &$ 2.01 \pm 0.36 $     &$ 4.37 \pm 0.16 $    &$ 0.046 \pm 0.008 $    \\ 
    SPT-CL~J0200$-$4852    &$ 2.67 $    &$ 4.31 \pm 0.65 $    &$ 5.00 \pm 0.17 $    &$ 4.39 \pm 0.61 $     &$ 5.44 \pm 0.18 $    &$ 0.081 \pm 0.011 $    \\ 
    SPT-CL~J0212$-$4657    &$ 2.33 $    &$ 5.15 \pm 0.69 $    &$ 6.43 \pm 0.22 $    &$ 3.37 \pm 0.62 $     &$ 6.77 \pm 0.23 $    &$ 0.050 \pm 0.009 $    \\ 
    SPT-CL~J0217$-$5245    &$ 3.46 $    &$ 4.01 \pm 0.66 $    &$ 4.99 \pm 0.13 $    &$ 4.49 \pm 0.85 $     &$ 5.44 \pm 0.16 $    &$ 0.083 \pm 0.014 $    \\ 
    SPT-CL~J0232$-$5257    &$ 2.54 $    &$ 4.71 \pm 0.67 $    &$ 5.91 \pm 0.21 $    &$ 3.80 \pm 0.57 $     &$ 6.29 \pm 0.22 $    &$ 0.060 \pm 0.009 $    \\ 
    SPT-CL~J0234$-$5831    &$ 3.54 $    &$ 6.70 \pm 0.83 $    &$ 7.23 \pm 0.20 $    &$ 3.60 \pm 0.56 $     &$ 7.59 \pm 0.21 $    &$ 0.047 \pm 0.007 $    \\ 
    SPT-CL~J0236$-$4938    &$ 3.44 $    &$ 3.69 \pm 0.67 $    &$ 3.84 \pm 0.07 $    &$ 3.15 \pm 0.57 $     &$ 4.15 \pm 0.09 $    &$ 0.076 \pm 0.013 $    \\ 
    SPT-CL~J0243$-$5930    &$ 2.23 $    &$ 4.18 \pm 0.62 $    &$ 5.27 \pm 0.16 $    &$ 3.83 \pm 0.61 $     &$ 5.66 \pm 0.17 $    &$ 0.068 \pm 0.010 $    \\ 
    SPT-CL~J0252$-$4824    &$ 3.01 $    &$ 4.24 \pm 0.66 $    &$ 4.67 \pm 0.14 $    &$ 2.28 \pm 0.42 $     &$ 4.89 \pm 0.14 $    &$ 0.047 \pm 0.008 $    \\ 
    SPT-CL~J0256$-$5617    &$ 2.36 $    &$ 4.15 \pm 0.62 $    &$ 5.43 \pm 0.13 $    &$ 3.96 \pm 0.60 $     &$ 5.83 \pm 0.15 $    &$ 0.068 \pm 0.010 $    \\ 
    SPT-CL~J0304$-$4401    &$ 3.34 $    &$ 6.99 \pm 0.87 $    &$ 9.77 \pm 0.24 $    &$ 5.59 \pm 0.73 $     &$ 10.33 \pm 0.25 $    &$ 0.054 \pm 0.007 $    \\ 
    SPT-CL~J0304$-$4921    &$ 3.62 $    &$ 6.26 \pm 0.81 $    &$ 7.06 \pm 0.14 $    &$ 3.49 \pm 0.49 $     &$ 7.41 \pm 0.15 $    &$ 0.047 \pm 0.006 $    \\ 
    SPT-CL~J0307$-$5042    &$ 2.55 $    &$ 4.64 \pm 0.66 $    &$ 5.37 \pm 0.14 $    &$ 5.50 \pm 0.76 $     &$ 5.92 \pm 0.16 $    &$ 0.093 \pm 0.012 $    \\ 
    SPT-CL~J0307$-$6225    &$ 2.43 $    &$ 4.49 \pm 0.64 $    &$ 5.74 \pm 0.20 $    &$ 2.81 \pm 0.48 $     &$ 6.02 \pm 0.20 $    &$ 0.047 \pm 0.008 $    \\ 
    SPT-CL~J0310$-$4647    &$ 2.03 $    &$ 3.98 \pm 0.62 $    &$ 4.19 \pm 0.23 $    &$ 3.04 \pm 0.54 $     &$ 4.49 \pm 0.24 $    &$ 0.068 \pm 0.012 $    \\ 
    SPT-CL~J0324$-$6236    &$ 2.07 $    &$ 4.46 \pm 0.63 $    &$ 4.94 \pm 0.13 $    &$ 5.13 \pm 0.74 $     &$ 5.45 \pm 0.15 $    &$ 0.094 \pm 0.012 $    \\ 
    SPT-CL~J0330$-$5228    &$ 3.19 $    &$ 5.63 \pm 0.74 $    &$ 11.29 \pm 0.18 $    &$ 3.48 \pm 0.56 $     &$ 11.64 \pm 0.19 $    &$ 0.030 \pm 0.005 $    \\ 
    SPT-CL~J0334$-$4659    &$ 2.83 $    &$ 4.85 \pm 0.67 $    &$ 5.57 \pm 0.14 $    &$ 2.10 \pm 0.41 $     &$ 5.78 \pm 0.15 $    &$ 0.036 \pm 0.007 $    \\ 
    SPT-CL~J0346$-$5439    &$ 2.65 $    &$ 4.82 \pm 0.66 $    &$ 5.87 \pm 0.14 $    &$ 4.29 \pm 0.65 $     &$ 6.30 \pm 0.16 $    &$ 0.068 \pm 0.010 $    \\ 
    SPT-CL~J0348$-$4515    &$ 3.66 $    &$ 5.27 \pm 0.71 $    &$ 5.08 \pm 0.17 $    &$ 4.39 \pm 0.66 $     &$ 5.52 \pm 0.19 $    &$ 0.079 \pm 0.011 $    \\ 
    SPT-CL~J0352$-$5647    &$ 2.10 $    &$ 3.90 \pm 0.61 $    &$ 4.65 \pm 0.16 $    &$ 3.11 \pm 0.51 $     &$ 4.96 \pm 0.17 $    &$ 0.063 \pm 0.010 $    \\ 
    SPT-CL~J0406$-$4805    &$ 2.02 $    &$ 4.21 \pm 0.61 $    &$ 5.22 \pm 0.17 $    &$ 4.60 \pm 0.79 $     &$ 5.68 \pm 0.18 $    &$ 0.081 \pm 0.013 $    \\ 
    SPT-CL~J0411$-$4819    &$ 3.50 $    &$ 6.79 \pm 0.84 $    &$ 9.02 \pm 0.13 $    &$ 5.54 \pm 0.75 $     &$ 9.58 \pm 0.15 $    &$ 0.058 \pm 0.007 $    \\ 
    SPT-CL~J0417$-$4748    &$ 2.70 $    &$ 6.23 \pm 0.78 $    &$ 7.44 \pm 0.20 $    &$ 4.53 \pm 0.69 $     &$ 7.89 \pm 0.21 $    &$ 0.057 \pm 0.008 $    \\ 
    SPT-CL~J0426$-$5455    &$ 2.31 $    &$ 4.58 \pm 0.64 $    &$ 5.41 \pm 0.14 $    &$ 4.71 \pm 0.71 $     &$ 5.88 \pm 0.16 $    &$ 0.080 \pm 0.011 $    \\ 
    SPT-CL~J0438$-$5419    &$ 3.82 $    &$ 8.68 \pm 1.04 $    &$ 11.27 \pm 0.22 $    &$ 5.96 \pm 0.67 $     &$ 11.87 \pm 0.23 $    &$ 0.050 \pm 0.005 $    \\ 
    SPT-CL~J0441$-$4855    &$ 1.94 $    &$ 4.31 \pm 0.61 $    &$ 4.88 \pm 0.12 $    &$ 5.01 \pm 0.77 $     &$ 5.38 \pm 0.14 $    &$ 0.093 \pm 0.013 $    \\ 
    SPT-CL~J0446$-$5849    &$ 1.40 $    &$ 3.50 \pm 0.54 $    &$ 2.87 \pm 0.31 $    &$ 4.22 \pm 0.77 $     &$ 3.29 \pm 0.32 $    &$ 0.128 \pm 0.024 $    \\ 
    SPT-CL~J0449$-$4901    &$ 1.95 $    &$ 4.41 \pm 0.62 $    &$ 4.96 \pm 0.16 $    &$ 4.24 \pm 0.69 $     &$ 5.39 \pm 0.18 $    &$ 0.079 \pm 0.012 $    \\ 
    SPT-CL~J0456$-$5116    &$ 2.49 $    &$ 4.51 \pm 0.64 $    &$ 5.17 \pm 0.09 $    &$ 3.64 \pm 0.60 $     &$ 5.53 \pm 0.11 $    &$ 0.066 \pm 0.010 $    \\ 
    SPT-CL~J0509$-$5342    &$ 2.87 $    &$ 4.51 \pm 0.64 $    &$ 5.48 \pm 0.16 $    &$ 1.70 \pm 0.34 $     &$ 5.65 \pm 0.16 $    &$ 0.030 \pm 0.006 $    \\ 
    SPT-CL~J0528$-$5300    &$ 1.83 $    &$ 3.45 \pm 0.57 $    &$ 2.99 \pm 0.10 $    &$ 3.82 \pm 0.62 $     &$ 3.37 \pm 0.11 $    &$ 0.113 \pm 0.017 $    \\ 
    SPT-CL~J0533$-$5005    &$ 1.70 $    &$ 3.60 \pm 0.56 $    &$ 2.43 \pm 0.15 $    &$ 2.43 \pm 0.56 $     &$ 2.67 \pm 0.16 $    &$ 0.091 \pm 0.020 $    \\ 
    SPT-CL~J0542$-$4100    &$ 2.29 $    &$ 4.65 \pm 0.68 $    &$ 5.62 \pm 0.13 $    &$ 3.78 \pm 0.56 $     &$ 6.00 \pm 0.14 $    &$ 0.063 \pm 0.009 $    \\ 
    SPT-CL~J0546$-$5345    &$ 1.64 $    &$ 4.58 \pm 0.60 $    &$ 6.20 \pm 0.15 $    &$ 7.23 \pm 0.94 $     &$ 6.92 \pm 0.18 $    &$ 0.104 \pm 0.012 $    \\ 
    SPT-CL~J0551$-$5709    &$ 3.05 $    &$ 4.45 \pm 0.64 $    &$ 6.09 \pm 0.11 $    &$ 3.86 \pm 0.53 $     &$ 6.47 \pm 0.12 $    &$ 0.060 \pm 0.008 $    \\ 
    SPT-CL~J0555$-$6406    &$ 4.03 $    &$ 6.43 \pm 0.82 $    &$ 8.56 \pm 0.19 $    &$       --      $     &$       --      $    &$       --      $    \\ 
    SPT-CL~J0559$-$5249    &$ 2.44 $    &$ 5.04 \pm 0.67 $    &$ 6.99 \pm 0.09 $    &$ 5.84 \pm 0.79 $     &$ 7.57 \pm 0.12 $    &$ 0.077 \pm 0.010 $    \\ 
    SPT-CL~J0616$-$5227    &$ 2.06 $    &$ 4.19 \pm 0.47 $    &$ 5.02 \pm 0.18 $    &$ 6.24 \pm 1.26 $     &$ 5.65 \pm 0.22 $    &$ 0.111 \pm 0.020 $    \\ 
    SPT-CL~J0655$-$5234    &$ 2.83 $    &$ 4.53 \pm 0.67 $    &$ 4.42 \pm 0.22 $    &$       --      $     &$       --      $    &$       --      $    \\ 
    SPT-CL~J2031$-$4037    &$ 4.36 $    &$ 7.95 \pm 0.97 $    &$ 10.31 \pm 0.18 $    &$       --      $     &$       --      $    &$       --      $    \\ 
    SPT-CL~J2034$-$5936    &$ 1.73 $    &$ 4.08 \pm 0.58 $    &$ 5.54 \pm 0.13 $    &$ 3.33 \pm 0.57 $     &$ 5.87 \pm 0.14 $    &$ 0.057 \pm 0.009 $    \\ 
    SPT-CL~J2035$-$5251    &$ 2.76 $    &$ 5.39 \pm 0.73 $    &$ 5.90 \pm 0.19 $    &$ 8.53 \pm 0.86 $     &$ 6.76 \pm 0.21 $    &$ 0.126 \pm 0.012 $    \\ 
    SPT-CL~J2043$-$5035    &$ 2.03 $    &$ 4.17 \pm 0.64 $    &$ 5.72 \pm 0.10 $    &$ 4.01 \pm 0.66 $     &$ 6.12 \pm 0.12 $    &$ 0.066 \pm 0.010 $    \\ 
    SPT-CL~J2106$-$5844    &$ 1.83 $    &$ 7.14 \pm 0.85 $    &$ 10.56 \pm 0.21 $    &$ 8.53 \pm 1.04 $     &$ 11.41 \pm 0.23 $    &$ 0.075 \pm 0.009 $    \\ 
    SPT-CL~J2135$-$5726    &$ 3.20 $    &$ 5.26 \pm 0.70 $    &$ 5.29 \pm 0.19 $    &$ 5.09 \pm 0.59 $     &$ 5.80 \pm 0.20 $    &$ 0.088 \pm 0.010 $    \\ 
    SPT-CL~J2145$-$5644    &$ 3.03 $    &$ 5.82 \pm 0.75 $    &$ 8.28 \pm 0.17 $    &$ 4.17 \pm 0.62 $     &$ 8.70 \pm 0.18 $    &$ 0.048 \pm 0.007 $    \\ 
    SPT-CL~J2146$-$4633    &$ 1.82 $    &$ 4.89 \pm 0.66 $    &$ 5.94 \pm 0.13 $    &$ 4.22 \pm 0.71 $     &$ 6.36 \pm 0.15 $    &$ 0.066 \pm 0.011 $    \\ 
    SPT-CL~J2148$-$6116    &$ 2.38 $    &$ 4.07 \pm 0.61 $    &$ 5.33 \pm 0.12 $    &$ 4.88 \pm 1.53 $     &$ 5.82 \pm 0.20 $    &$ 0.084 \pm 0.024 $    \\ 
    SPT-CL~J2218$-$4519    &$ 2.28 $    &$ 4.70 \pm 0.66 $    &$ 5.60 \pm 0.13 $    &$ 5.35 \pm 0.73 $     &$ 6.14 \pm 0.15 $    &$ 0.087 \pm 0.011 $    \\ 
    SPT-CL~J2222$-$4834    &$ 2.28 $    &$ 4.77 \pm 0.66 $    &$ 5.12 \pm 0.15 $    &$ 3.07 \pm 0.54 $     &$ 5.42 \pm 0.16 $    &$ 0.057 \pm 0.009 $    \\ 
    SPT-CL~J2232$-$5959    &$ 2.45 $    &$ 4.87 \pm 0.69 $    &$ 5.35 \pm 0.16 $    &$ 6.82 \pm 1.05 $     &$ 6.04 \pm 0.19 $    &$ 0.113 \pm 0.016 $    \\ 
    SPT-CL~J2233$-$5339    &$ 2.85 $    &$ 4.80 \pm 0.69 $    &$ 6.13 \pm 0.23 $    &$ 4.21 \pm 0.60 $     &$ 6.55 \pm 0.23 $    &$ 0.064 \pm 0.009 $    \\ 
    SPT-CL~J2236$-$4555    &$ 1.46 $    &$ 3.80 \pm 0.56 $    &$ 3.88 \pm 0.13 $    &$ 4.28 \pm 0.61 $     &$ 4.30 \pm 0.15 $    &$ 0.099 \pm 0.013 $    \\ 
    SPT-CL~J2245$-$6206    &$ 2.47 $    &$ 4.75 \pm 0.67 $    &$ 7.40 \pm 0.16 $    &$ 7.87 \pm 1.21 $     &$ 8.18 \pm 0.20 $    &$ 0.096 \pm 0.013 $    \\ 
    SPT-CL~J2248$-$4431    &$ 5.04 $    &$ 13.05 \pm 1.54 $    &$ 21.15 \pm 0.22 $    &$ 8.62 \pm 0.88 $     &$ 22.01 \pm 0.24 $    &$ 0.039 \pm 0.004 $    \\ 
    SPT-CL~J2258$-$4044    &$ 2.00 $    &$ 5.18 \pm 0.68 $    &$ 5.47 \pm 0.19 $    &$       --      $     &$       --      $    &$       --      $    \\ 
    SPT-CL~J2259$-$6057    &$ 2.10 $    &$ 4.93 \pm 0.67 $    &$ 5.65 \pm 0.11 $    &$ 4.11 \pm 0.57 $     &$ 6.06 \pm 0.13 $    &$ 0.068 \pm 0.009 $    \\ 
    SPT-CL~J2301$-$4023    &$ 2.05 $    &$ 4.37 \pm 0.63 $    &$ 2.81 \pm 0.13 $    &$ 2.49 \pm 0.56 $     &$ 3.06 \pm 0.14 $    &$ 0.081 \pm 0.017 $    \\ 
    SPT-CL~J2306$-$6505    &$ 2.69 $    &$ 5.04 \pm 0.69 $    &$ 6.70 \pm 0.17 $    &$ 5.16 \pm 0.64 $     &$ 7.21 \pm 0.18 $    &$ 0.072 \pm 0.008 $    \\ 
    SPT-CL~J2325$-$4111    &$ 3.90 $    &$ 6.33 \pm 0.81 $    &$ 8.57 \pm 0.24 $    &$ 8.64 \pm 1.04 $     &$ 9.44 \pm 0.26 $    &$ 0.092 \pm 0.010 $    \\ 
    SPT-CL~J2331$-$5051    &$ 2.51 $    &$ 4.89 \pm 0.65 $    &$ 5.38 \pm 0.16 $    &$ 4.68 \pm 0.63 $     &$ 5.85 \pm 0.17 $    &$ 0.080 \pm 0.010 $    \\ 
    SPT-CL~J2335$-$4544    &$ 2.69 $    &$ 5.37 \pm 0.71 $    &$ 7.64 \pm 0.19 $    &$ 6.25 \pm 0.85 $     &$ 8.26 \pm 0.21 $    &$ 0.076 \pm 0.010 $    \\ 
    SPT-CL~J2337$-$5942    &$ 2.31 $    &$ 7.05 \pm 0.85 $    &$ 8.22 \pm 0.34 $    &$ 6.51 \pm 0.92 $     &$ 8.87 \pm 0.35 $    &$ 0.073 \pm 0.010 $    \\ 
    SPT-CL~J2341$-$5119    &$ 1.74 $    &$ 4.94 \pm 0.63 $    &$ 5.77 \pm 0.15 $    &$ 5.05 \pm 0.82 $     &$ 6.27 \pm 0.17 $    &$ 0.081 \pm 0.012 $    \\ 
    SPT-CL~J2342$-$5411    &$ 1.51 $    &$ 3.70 \pm 0.53 $    &$ 2.76 \pm 0.10 $    &$ 3.08 \pm 0.68 $     &$ 3.07 \pm 0.12 $    &$ 0.100 \pm 0.020 $    \\ 
    SPT-CL~J2344$-$4243    &$ 3.07 $    &$ 9.60 \pm 1.14 $    &$ 13.82 \pm 0.14 $    &$ 7.56 \pm 1.03 $     &$ 14.58 \pm 0.18 $    &$ 0.052 \pm 0.007 $    \\ 
    SPT-CL~J2345$-$6405    &$ 1.78 $    &$ 4.65 \pm 0.63 $    &$ 4.87 \pm 0.24 $    &$ 5.06 \pm 0.91 $     &$ 5.38 \pm 0.25 $    &$ 0.094 \pm 0.016 $    \\ 
    SPT-CL~J2352$-$4657    &$ 2.00 $    &$ 4.09 \pm 0.62 $    &$ 3.66 \pm 0.14 $    &$ 5.63 \pm 0.85 $     &$ 4.23 \pm 0.16 $    &$ 0.133 \pm 0.018 $    \\ 
    SPT-CL~J2355$-$5055    &$ 3.58 $    &$ 3.78 \pm 0.61 $    &$ 4.06 \pm 0.08 $    &$ 2.75 \pm 0.58 $     &$ 4.33 \pm 0.10 $    &$ 0.063 \pm 0.013 $    \\ 
    SPT-CL~J2359$-$5009    &$ 1.82 $    &$ 3.44 \pm 0.55 $    &$ 3.11 \pm 0.09 $    &$       --      $     &$       --      $    &$       --      $    \\ 
    SPT-CL~J0106$-$5943    &$ 3.76 $    &$ 5.33 \pm 0.73 $    &$ 5.67 \pm 0.13 $    &$ 4.68 \pm 1.49 $     &$ 6.13 \pm 0.20 $    &$ 0.076 \pm 0.022 $    \\ 
    SPT-CL~J2332$-$5053    &$ 2.08 $    &$ 2.63 \pm 0.58 $    &$ 2.38 \pm 0.20 $    &$ 4.35 \pm 0.81 $     &$ 2.82 \pm 0.22 $    &$ 0.154 \pm 0.027 $    \\ 
    SPT-CL~J0232$-$4421    &$ 5.35 $    &$ 9.46 \pm 1.13 $    &$ 12.88 \pm 0.17 $    &$ 6.88 \pm 0.90 $     &$ 13.57 \pm 0.19 $    &$ 0.051 \pm 0.006 $    \\ 
    SPT-CL~J0235$-$5121    &$ 4.53 $    &$ 5.44 \pm 0.74 $    &$ 6.95 \pm 0.15 $    &$ 4.38 \pm 0.65 $     &$ 7.39 \pm 0.17 $    &$ 0.059 \pm 0.008 $    \\ 
    SPT-CL~J0516$-$5430    &$ 4.45 $    &$ 5.97 \pm 0.76 $    &$ 10.90 \pm 0.13 $    &$ 6.43 \pm 0.97 $     &$ 11.54 \pm 0.17 $    &$ 0.056 \pm 0.008 $    \\ 
    SPT-CL~J0522$-$4818    &$ 3.67 $    &$ 3.38 \pm 0.73 $    &$ 3.67 \pm 0.08 $    &$ 3.30 \pm 0.57 $     &$ 4.00 \pm 0.10 $    &$ 0.083 \pm 0.013 $    \\ 
    SPT-CL~J0658$-$5556    &$ 5.71 $    &$ 12.71 \pm 1.51 $    &$ 22.69 \pm 0.17 $    &$       --      $     &$       --      $    &$       --      $    \\ 
    SPT-CL~J2011$-$5725    &$ 3.85 $    &$ 3.36 \pm 0.65 $    &$ 3.61 \pm 0.07 $    &$       --      $     &$       --      $    &$       --      $    \\ 